\documentclass[11pt]{scrartcl}  
\usepackage[left=1in, right=1in, top=1in, bottom=1in]{geometry}
\usepackage{mathptmx} 
\usepackage{multirow,graphicx,physics,hyperref,amsmath,amssymb,mathalfa,mathtools,colortbl,hhline}
\numberwithin{equation}{section}
\usepackage{hyperref}
\hypersetup{colorlinks=true, citecolor=blue}
\usepackage{array}

\usepackage{longtable,caption}
\newcolumntype{L}[1]{>{\raggedright\let\newline\\\arraybackslash\hspace{0pt}}m{#1}}
\newcolumntype{C}[1]{>{\centering\let\newline\\\arraybackslash\hspace{0pt}}m{#1}}
\newcolumntype{R}[1]{>{\raggedleft\let\newline\\\arraybackslash\hspace{0pt}}m{#1}}
\usepackage[framemethod=TikZ]{mdframed}
\mdfdefinestyle{sid}{%
    linecolor=black,
    outerlinewidth=1.0pt,
    roundcorner=7pt,
    innerrightmargin=15pt,
    innerleftmargin=15pt,
    backgroundcolor=gray!30
		}
\mdfdefinestyle{sid2}{%
    linecolor=black,
    outerlinewidth=1.0pt,
    roundcorner=7pt,
    innerrightmargin=15pt,
    innerleftmargin=15pt,
    backgroundcolor=blue!10
		}
\definecolor{Gray}{gray}{0.8}
\definecolor{MyBlue}{rgb}{0.0,0.0,0.9}
\definecolor{MyRed}{rgb}{0.0,0.9,0.0}
\colorlet{Bluee}{MyBlue!6}
\colorlet{Redd}{MyRed!1}
\DeclareMathAlphabet{\mathcal}{OMS}{cmsy}{m}{n} 

\usepackage{xcolor}
\colorlet{sectioncolor}{blue!20}
\colorlet{subsectioncolor}{orange!70}
\colorlet{subsubsectioncolor}{green!40}
\makeatletter
\renewcommand\sectionlinesformat[4]{%
  \colorbox{#1color}{%
    \parbox[t]{\dimexpr\textwidth-2\fboxsep\relax}{%
      \raggedsection\color{black}\@hangfrom{#3}{#4}%
}}}
\makeatother
\usepackage{blindtext}
\title{\textbf{Topological entanglement and number theory}}
\date{}
\author{\textbf{\emph{Siddharth Dwivedi}}\thanks{Email: siddharth.dwivedi@curaj.ac.in}\\\\ Department of Physics, School of Physical Sciences, \\ Central University of Rajasthan, Ajmer, 305817, India}
\begin{document}
\maketitle
\begin{abstract}
The recent developments in the study of topological multi-boundary entanglement in the context of 3d Chern-Simons theory (with gauge group $G$ and level $k$) suggest a strong interplay between entanglement measures and number theory. The purpose of this note is twofold. First, we introduce a $q$-deformed version of the Witten zeta function using the Chern-Simons theory at level $k$. We analyze the large $k$ limit of this function and show that it converges to an integer multiple of the classical Witten zeta function of $G$, where the integer multiple is precisely the order of the center of the group. This analysis provides an alternative way to compute the classical zeta functions, and we present some examples. Next, we study the quantum state associated with the $S^3$ complement of torus links of type $T_{p,p}$ and show that we can write the R\'enyi entropies at finite $k$ in terms of $q$-deformed Witten zeta functions. Using our first result, we obtain the $k \to \infty$ limit of the R\'enyi entropies and find that the entropies converge to finite values, which can be written in terms of the classical Witten zeta functions evaluated at positive integers. Since Witten zeta functions naturally appear in the symplectic volumes of moduli spaces of flat connections on Riemann surfaces, we give a geometric interpretation of the $k \to \infty$ limit of the R\'enyi and entanglement entropies in terms of these volumes. The results of this paper reveal an intriguing connection between topological entanglement, number-theoretic structures arising from Witten zeta functions, and the geometry of moduli spaces. 
\end{abstract}
\hypersetup{linkcolor=red}
\listoftables

\newpage
\hypersetup{linkcolor=blue}
\tableofcontents
\newpage
\section{Introduction} \label{sec1}
\begin{figure}[htbp]
	\centering
		\includegraphics[width=0.85\textwidth]{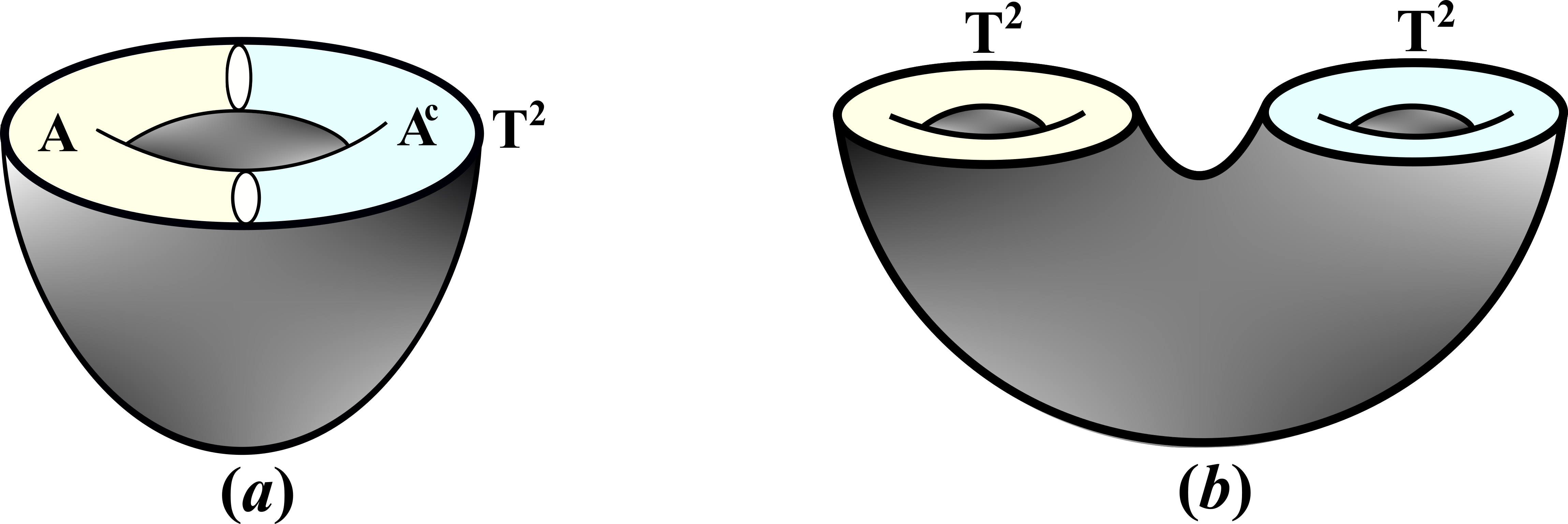}
	\caption{Two different set-ups to study topological entanglement: the manifold in ($a$) has a single $T^2$ boundary, which is bi-partitioned into spatially connected regions $A$ and $A^c$. The manifold in ($b$) has two disjoint $T^2$ boundaries.}
	\label{T2partition}
\end{figure}

Quantum entanglement \cite{Horodecki:2009zz} is an important fundamental aspect of the quantum system and is at the core of many revolutionary advances in fields of science and technology, including a wide range of applications in quantum information science.  
Analyzing and classifying the possible patterns of entanglement that can emerge in quantum field theory is an important question in quantum mechanics and quantum information theory. Though the study of entanglement in a generic QFT is difficult, a simple case where this could be done is for `topological quantum field theories' (TQFT's). A TQFT is a class of field theory in which the physical observables and the correlation functions do not depend on the spacetime metric. 
Such theories do not have local dynamics; therefore, all of the entanglement arises from
the topological properties of the underlying manifolds. For example, consider the manifold shown in the figure \ref{T2partition}($a$) having a torus boundary. The boundary is bi-partitioned into spatially connected sections $A$ and its complement $A^c$.  
When we trace out the region $A^c$, we are essentially calculating the `topological entanglement entropy', which is independent of the length or area of the region $A$ or $A^c$. Such entropies have been studied in \cite{Kitaev:2005dm,Levin:2006zz,Dong:2008ft} in the context of 2+1 dimensional Chern-Simons theory, the best understood TQFT \cite{Witten:1988hf}. Another approach was used in \cite{Balasubramanian:2016sro}, where the authors study the entanglement entropy of the states obtained by the path integral in Chern-Simons theory performed on a link complement\footnote{If a link $\mathcal{L}$  is embedded in $S^3$, then the link complement is a three-dimensional manifold which is obtained by removing a tubular neighborhood around $\mathcal{L}$ from $S^3$, i.e. $S^3 \backslash \mathcal{L} \equiv S^3 - \text{interior}(\mathcal{L}_{\text{tub}})$.} $S^3 \backslash \mathcal{L}$. In this case, we have two or more disjoint torus boundaries similar to the one shown in figure \ref{T2partition}($b$). The topological entanglement structure, in this case, can be obtained by tracing out one of the boundary components, which is termed `multi-boundary entanglement.' We refer the interested readers to \cite{Balasubramanian:2016sro,Dwivedi:2017rnj,Balasubramanian:2018por,Hung:2018rhg,Melnikov:2018zfn,Camilo:2019bbl,Dwivedi:2019bzh,Buican:2019evc,Zhou:2019ezk,Dwivedi:2020jyx,Dwivedi:2020rlo,Dwivedi:2021dix} for the recent developments in this study. 

An important aspect of any topological field theory, which is the motivation behind the study of multi-boundary entanglement, is the following decomposition of the Hilbert space associated with a boundary with multiple disjoint components:
\begin{equation}
\mathcal{H}_{\partial M} = \mathcal{H}_{\Sigma_1} \otimes \mathcal{H}_{\Sigma_2} \otimes \ldots \otimes \mathcal{H}_{\Sigma_n} ~,
\end{equation}
where the boundary of the manifold $M$ consists of disjoint components: $\partial M = \Sigma_1 \sqcup \Sigma_2 \sqcup \ldots \sqcup \Sigma_n$ and $\mathcal{H}_{\Sigma_i}$ denotes the Hilbert space associated with the $i^{\text{th}}$ component. The Chern-Simons path integral on such a manifold associates a quantum state $\ket{\Psi}$ to $M$ which lives in the Hilbert space $\mathcal{H}_{\partial M}$. The entanglement structure of $\ket{\Psi}$ can be studied by tracing out a subset of the Hilbert spaces. It was shown in \cite{Balasubramanian:2016sro} that when $M=S^3 \backslash \mathcal{L}$, the probability amplitudes of the associated state $\ket{\mathcal{L}}$ are the partition functions of $S^3$ in the presence of link $\mathcal{L}$ which are also the quantum link invariants of $\ket{\mathcal{L}}$ (see \cite{Witten:1988hf}).

Apart from the connection with knot theory, topological entanglement also has interesting number-theoretic properties, which can be seen in \cite{Dwivedi:2020rlo}. This motivation brings us to the present work, where we will study the R\'enyi entropies associated with torus link complements $S^3 \backslash T_{p,p}$ for various simple and semi-simple Lie groups. The manifold $S^3 \backslash T_{p,p}$ has $p$ number of disjoint torus boundaries, and the associated quantum state $\ket{T_{p,p}}$ lives in the tensor product of $p$ copies of $\mathcal{H}_{T^2}$. In particular, we investigate the number-theoretic properties of the entanglement measures associated with $\ket{T_{p,p}}$. Given the Chern-Simons theory with gauge group $G$ and level $k$, we show that the spectrum of the reduced density matrix in this case is given by the quantum dimensions of the integrable highest-weight representations associated with the affine group $G_k$. The important aspect of this paper is establishing a profound connection between the R\'enyi entropies of $\ket{T_{p,p}}$ and number theory. We show that the $k \to \infty$ limit of these R\'enyi entropies can be written in terms of the Witten zeta functions associated with the group $G$ evaluated at positive even integers. We make this connection by proposing a $q$-deformed version of Witten zeta functions using the Chern-Simons theory at level $k$ and studying its large $k$ limit.

It is well known that the Witten zeta functions appear in the calculation of the symplectic volume of the moduli spaces of flat connections on Riemann surfaces of genus $g$, as given in reference \cite{witten1991}. Since we can write the semiclassical limit of the R\'enyi entropies of $\ket{T_{p,p}}$ in terms of classical Witten zeta functions, we can provide a geometric interpretation of the $k \to \infty$ limit of the Rényi and entanglement entropies in terms of these volumes. More broadly, our results point to an intriguing connection between topological entanglement, number-theoretic structures arising from Witten zeta functions, and the geometry of moduli spaces of flat connections.

The paper is organized as follows. In section \ref{sec2}, we discuss the preliminaries related to our setup, which includes giving a brief review of the study of entanglement using the topological machinery, along with a brief discussion on Witten zeta functions. In section \ref{sec3}, we introduce the $q$-deformed version of the Witten zeta function using the Chern-Simons theory at level $k$ and obtain its large $k$ limit. Using this result, we discuss an alternate route of computing the classical Witten zeta functions for any compact group $G$. We provide the SU($N$) group as a concrete example supporting our result and showing detailed steps on how to obtain the Witten zeta functions for the SU($N$) group using this route. We defer the examples for other classical and exceptional Lie groups to Appendix \ref{app2}. In section \ref{sec4}, we show that the R\'enyi entropies for the quantum states $\ket{T_{p,p}}$ for the group $G$ at finite $k$ can be cast in terms of the $q$-deformed Witten zeta functions. We then use the result of section \ref{sec3} to obtain the large $k$ limits of these R\'enyi entropies and show that these entropies converge to finite values in the $k \to \infty$ limit such that the limit can be written in terms of the classical Witten zeta function of $G$ evaluated at positive even integers. We also give a geometric interpretation to the limiting values of entropies by expressing them in terms of the volumes of moduli spaces. We conclude in section \ref{sec5}. For the sake of completeness, we have also analyzed our $q$-deformed Witten zeta function when $q$ is not a root of unity by analytically continuing it to generic complex values of $q$. This analytic continuation for the SU(2) group is presented in the appendix \ref{appB}.
\section{Preliminaries: the topological set-up} \label{sec2}
\subsection{Chern-Simons theory and multi-boundary states}  
The 2+1 dimensional Chern-Simons theory with gauge group $G$ and level $k \in \mathbb{Z}$ is defined on a 3-manifold $M$ with action given by
\begin{equation}
S(A) = \frac{k}{4\pi} \int_M \text{Tr}\left(A \wedge dA + \frac{2}{3} A \wedge A \wedge A \right) ~,
\label{CSaction}
\end{equation}
where $A = A_{\mu}dx^{\mu}$ is a gauge field, which, in this case, is a connection on the trivial $G$-bundle over $M$. The gauge invariant operators in the theory are  Wilson lines. Given an oriented knot $\mathcal{K}$ embedded in $M$, the Wilson line is defined by taking the trace of the holonomy of $A$ around $\mathcal{K}$ :
\begin{equation}
W_{R}(\mathcal{K}) = \text{Tr}_R \,P\exp\left(i\oint_{\mathcal{K}} A \right) ~,
\end{equation} 
where the trace is taken under the representation $R$ of $G$. The computation of the partition function in Chern-Simons theory involves the integration over the infinite-dimensional space of connections. For the bare manifold $M$ without any Wilson line, the partition function is given as
\begin{equation}
Z(M) = \int e^{i S(A)} dA ~,
\end{equation}
where $dA$ is an appropriate quantized measure defined for a connection $A$, and the integration is over all the gauge invariant classes of connections. Since the connection has been integrated out, $Z(M)$ is a topological invariant of $M$.
One can also compute the partition function of $M$ in the presence of knots and links by inserting the appropriate Wilson loop operators in the integral. Given a link $\mathcal{L}$ made of disjoint oriented knot components, i.e. $\mathcal{L} = \mathcal{K}_1 \sqcup \mathcal{K}_2 \sqcup \ldots \sqcup \mathcal{K}_n$, the partition function of $M$ in the presence of $\mathcal{L}$ can be obtained by modifying the path integral as:
\begin{equation}
Z(M; \mathcal{L}) = \int e^{i S(A)}\, W_{R_1}(\mathcal{K}_1)\ldots W_{R_n}(\mathcal{K}_n) \, dA ~.
\end{equation}  
The partition functions defined above are either a vector or a scalar, depending upon whether $M$ is with or without boundary.
\begin{itemize}
\item When $M$ is closed (i.e., without boundary), the partition function $Z(M; \mathcal{L})$ is simply a complex number.
\item When $M$ has a boundary $\Sigma$, the path integral of the theory on $M$ with the Wilson line insertions, and the boundary condition $A\rvert_{\Sigma} = Q$ imposed on $\Sigma$, is interpreted as the wavefunction of a state:
\end{itemize}
\begin{align}
\ket{\Psi} & \equiv Z_Q(M; \mathcal{L}) \nonumber \\ 
& = \int_{A\rvert_{\Sigma} = Q} e^{i S(A)}\, W_{R_1}(\mathcal{K}_1)\ldots W_{R_n}(\mathcal{K}_n) \, dA ~.
\end{align}
The partition function $Z_Q(M; \mathcal{L})$ is a function on the space $F_{\Sigma}$ of gauge equivalence classes on $\Sigma$. With the appropriate measure $dA$ defined for the connections, the space $L^2(F_{\Sigma}, dA) \equiv \mathcal{H}_{\Sigma}$ defines a unique Hilbert space associated with $\Sigma$ and $\ket{\Psi}$ is an element of $\mathcal{H}_{\Sigma}$. Thus, we can associate a unique state with any 3-manifold with a boundary. 

\begin{figure}[htbp]
	\centering
		\includegraphics[width=0.90\textwidth]{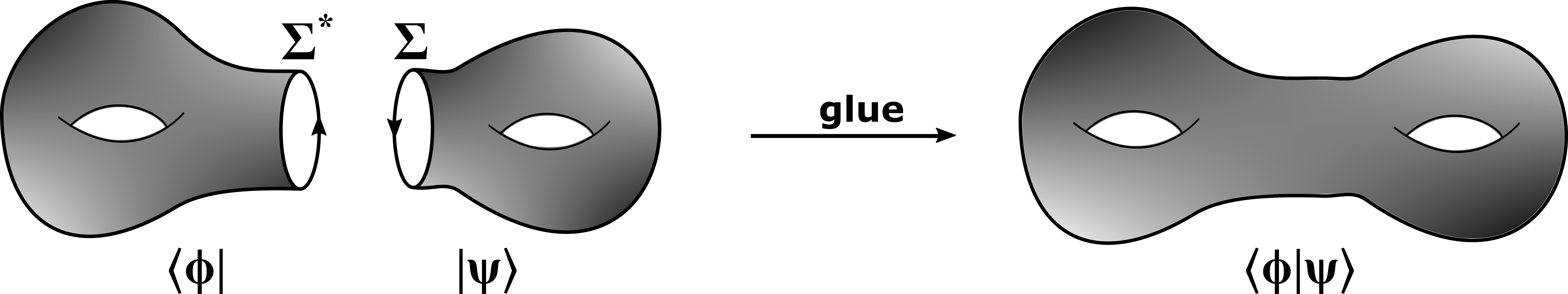}
	\caption{Two manifolds on left with same boundary but opposite orientation. The path integral on these manifolds gives states $\bra{\phi} \in \mathcal{H}_{\Sigma^*}$ and $\ket{\psi} \in \mathcal{H}_{\Sigma}$. The inner product $\bra{\phi}\ket{\psi}$ will be the partition function of the manifold shown in the right obtained by gluing the two manifolds along the common boundary.}
	\label{gluing}
\end{figure}

From the topological point of view, $\ket{\Psi}$ depends on the topology of the manifold $M$ and $\mathcal{H}_{\Sigma}$ depends on the topology of the boundary $\Sigma$. If we consider two topologically different manifolds $M$ and $M'$ with the same boundary $\Sigma = \Sigma'$, then the corresponding partition functions (for a fixed gauge group) give two different states $\ket{\Psi}$ and $\ket{\Psi'}$ in the same Hilbert space $\mathcal{H}_{\Sigma}$. Further note that if we reverse the orientation of the boundary, the associated Hilbert space becomes the dual of the original Hilbert space: 
\begin{equation}
\mathcal{H}_{\Sigma^*} = \mathcal{H}^*_{\Sigma} ~.
\end{equation} 
As a result, there exists a natural pairing, the inner product $\langle \Phi| \Psi \rangle$ for any two states $\ket{\Psi} \in \mathcal{H}_{\Sigma}$ and $\bra{\Phi} \in \mathcal{H}_{\Sigma^*}$. In fact, this technique can be used to compute the partition functions of complicated manifolds by gluing two disconnected pieces along a common boundary whose partition functions are already known. This process is shown in the figure \ref{gluing}.

When the boundary of the manifold $M$ consists of disjoint components, i.e., $\Sigma = \Sigma_1 \sqcup \Sigma_2 \sqcup \ldots \sqcup \Sigma_n$, the Hilbert space associated with $\Sigma$ is the tensor product of Hilbert spaces associated with each component, i.e.
\begin{equation}
\mathcal{H}_{\Sigma} = \mathcal{H}_{\Sigma_1} \otimes \mathcal{H}_{\Sigma_2} \otimes \ldots \otimes \mathcal{H}_{\Sigma_n} ~.
\end{equation} 
Thus, the quantum state $\ket{\Psi}$ associated with $M$ lives in this tensor product of Hilbert spaces, and one can study its entanglement structure by tracing out a subset of the Hilbert spaces. 

In this work, we will consider the link complement manifold 
\begin{equation}
M=S^3 \backslash \mathcal{L} ~,
\end{equation} 
where $\mathcal{L} = \mathcal{K}_1 \sqcup \mathcal{K}_2 \sqcup \ldots \sqcup \mathcal{K}_n$ is a link made up of $n$ number of oriented knots. We denote the associated state as $\ket{\mathcal{L}}$, which can be understood as being defined on $n$ copies of $T^2$. There is a systematic way of computing such states, which was given in \cite{Balasubramanian:2016sro}, and we briefly discuss it here. To obtain the state $\ket{\mathcal{L}}$, first we need to fix the basis of $\mathcal{H}_{T^2}$. As discussed in \cite{Witten:1988hf}, the basis of $\mathcal{H}_{T^2}$ is in one-to-one correspondence with the integrable representations of the affine Lie algebra $\hat{\mathfrak{g}}_k$ at level $k$, where $\mathfrak{g}$ represents the Lie algebra associated with the group $G$. These basis states have a path integral description. Consider a solid torus with a Wilson line carrying an integrable representation $\alpha$ placed in the bulk of a solid torus along its non-contractible cycle. The Chern-Simons path integral on this solid torus will give the basis state $\ket{e_\alpha}$:
\begin{equation}
\begin{array}{c}
\includegraphics[width=0.30\linewidth]{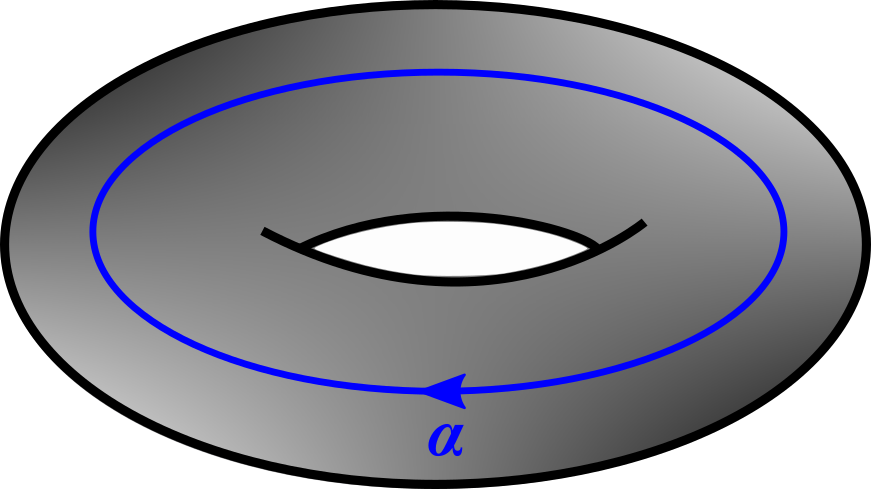}
\end{array} = \ket{e_\alpha} ~.
\end{equation}
The collection of all such states, where $\alpha$ runs over the integrable representations, will define an orthonormal basis of $\mathcal{H}_{T^2}$:
\begin{equation}
\text{basis}(\mathcal{H}_{T^2}) = \{\ket{e_R} : \text{$R$ is an integrable representation} \}~.
\end{equation}
In this work, we will restrict $G$ to be a classical or exceptional Lie group. In such cases, there exists only a finite number of integrable representations at a given level $k$, which makes the Hilbert space a finite-dimensional vector space. We shall denote a representation of $\mathfrak{g}$ as $[a_1, a_2,\ldots, a_r]$ with $a_i$ being the Dynkin labels of its highest weight and $r$ being the rank of the algebra. The integrable representations satisfy the  following conditions: 
\begin{equation}
\phi_1\,a_1 + \phi_2\, a_2 +\ldots + \phi_r\, a_r \leq k \quad;\quad a_i \geq 0 ~,
\end{equation}
where $\phi_i$ are the comarks or the dual Kac labels of $\mathfrak{g}$. We will use the symbol $\mathcal{I}_k$ to denote the set of all the integrable highest weight representations at level $k$. With the Hilbert space fixed, we can expand the state as:
\begin{equation}
\ket{\mathcal{L}} = \sum_{R_1 \, \in \, \mathcal{I}_k} \ldots \sum_{R_n \, \in \, \mathcal{I}_k} C_{R_1,\ldots,R_n} \ket{e_{R_1},\ldots, e_{R_n}} ~,
\end{equation}
where $C_{R_1,\ldots,R_n}$ are complex expansion coefficients. These coefficients can be computed using the following procedure. Take $n$ number of solid tori, with Wilson lines in the representation $R_i$ placed in the bulk of the $i^{\text{th}}$ solid torus, such that their boundaries are oppositely oriented compared to that of $S^3 \backslash \mathcal{L}$. The state associated with this collection of $n$ solid tori will be $\bra{e_{R_1},\ldots, e_{R_n}}$. Taking the inner product of $\bra{e_{R_1},\ldots, e_{R_n}}$ with $\ket{\mathcal{L}}$ is equivalent to gluing in the collection of solid tori along the boundary of the link complement $S^3 \backslash \mathcal{L}$ such that the $i^{\text{th}}$ solid tori in the collection is glued to the $i^{\text{th}}$ toroidal boundary of $S^3 \backslash \mathcal{L}$ as shown below: 
\begin{equation}
\begin{array}{c}
\includegraphics[width=0.65\linewidth]{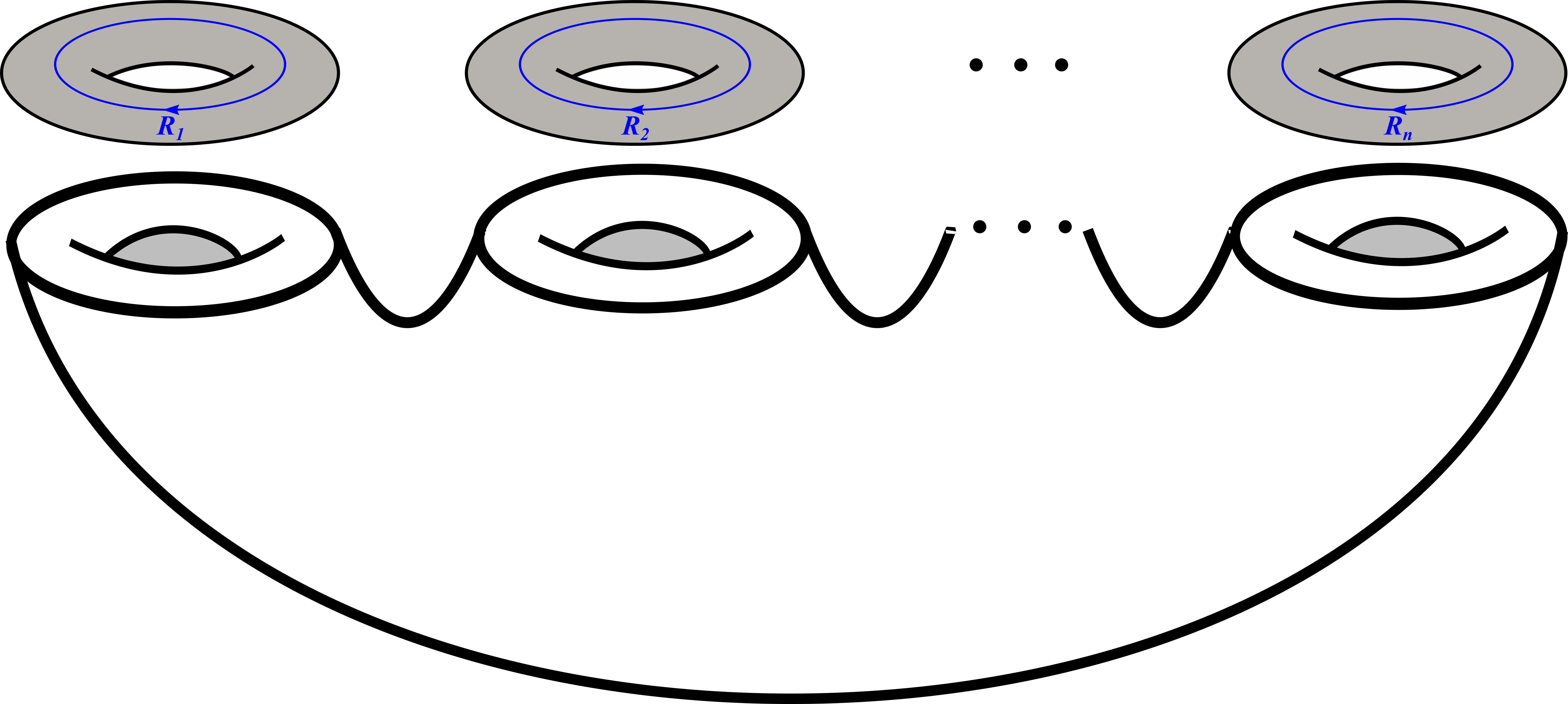}
\end{array} ~.
\end{equation}
This gluing results in $S^3$ with the link $\mathcal{L}$ embedded in it such that the components of $\mathcal{L}$ carry representations $R_1, R_2, \ldots, R_n$. Thus, we get,
\begin{equation}
\bra{e_{R_1},\ldots, e_{R_n}}\ket{\mathcal{L}} = Z(S^3; \mathcal{L}[R_1,\ldots,R_n]) ~.
\end{equation}
Hence, the link state can be written as
\begin{equation}
\ket{\mathcal{L}} = \sum_{R_1 \, \in \, \mathcal{I}_k} \ldots \sum_{R_n \, \in \, \mathcal{I}_k} Z(S^3; \mathcal{L}[R_1,\ldots,R_n]) \ket{e_{R_1},\ldots, e_{R_n}} ~.
\label{linkstate}
\end{equation}
The reduced density matrix can be obtained by tracing out a subset of the Hilbert spaces, and the entropies can be computed.
\subsection{Modular transformation matrix and quantum dimension}  
\begin{figure}[h]
\centerline{\includegraphics[width=2.5in]{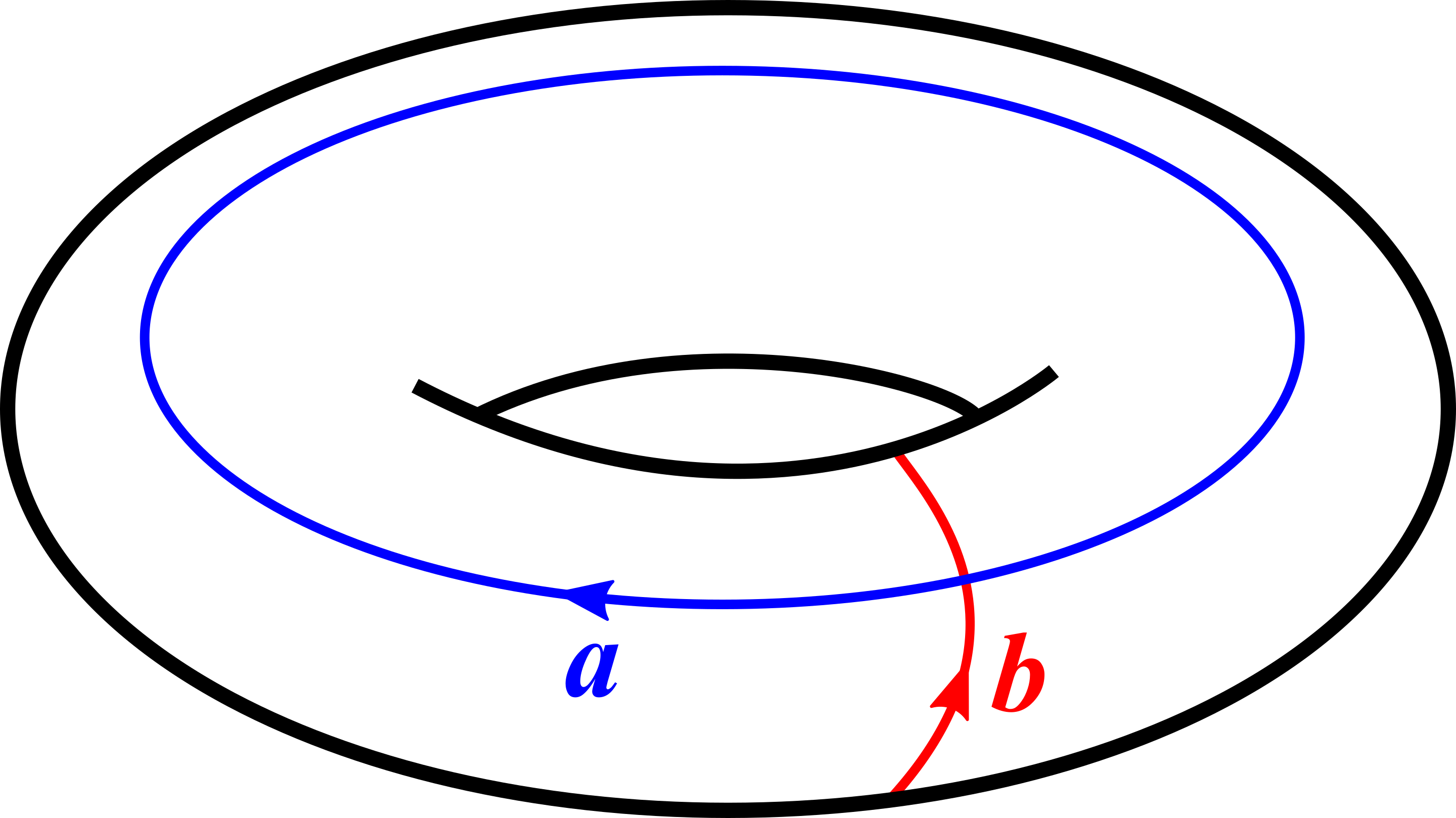}}
\caption[]{The two cycles of a torus which form its homology basis.}
\label{Heegaard}
\end{figure}

The mapping class group MCG($\Sigma$) acts naturally on the Hilbert space $\mathcal{H}_{\Sigma}$ associated with $\Sigma$. In the present case, we are interested in $\Sigma = T^2$ for which the mapping class group is the modular group: 
\begin{equation}
\text{MCG}(T^2) = \text{SL}(2, \mathbb{Z}) ~.
\end{equation}
The unitary representation of SL(2, $\mathbb{Z}$) has two generators $\mathcal{S}$ and $\mathcal{T}$ which act as diffeomorphism operators  for $T^2$ by acting on its homology basis (which consists of the two cycles $a$ and $b$ as shown in the figure \ref{Heegaard}) as following:
\begin{equation}
\mathcal{S}: (a,b) \longrightarrow (b,-a) \quad;\quad  \mathcal{T}: (a,b) \longrightarrow (a,a+b) ~.
\end{equation}
These generators satisfy $\mathcal{S}^2: (a,b) \longrightarrow (-a,-b)$ and $(\mathcal{S}\mathcal{T})^3: (a,b) \longrightarrow (-a,-b)$, which simply reverses the orientations of the two cycles of the torus. Thus we have the relation $\mathcal{S}^2 = (\mathcal{S} \mathcal{T})^3 = \mathcal{C}$, where $\mathcal{C}$ is the charge conjugation matrix (given as $\mathcal{C}_{XY} = \delta_{X \bar{Y}}$) which obeys $\mathcal{C}^2 = I$. These operators can be given a matrix form in the orthonormal basis of $\mathcal{H}_{T^2}$, which are commonly known as modular $\mathcal{S}$ and $\mathcal{T}$ matrices. The elements of these matrices can be explicitly computed for classical and exceptional Lie groups (see, for example \cite{francesco2012conformal}). 

The quantum dimension of an integrable representation $R$ of the affine Lie algebra $\hat{\mathfrak{g}}_k$ at level $k$ can be computed from the modular $\mathcal{S}$ matrix (see \cite{Coquereaux:2010dw} for some explicit computations) and is given as:
\begin{equation}
\text{dim}_q(R) = \frac{\mathcal{S}_{0R}}{\mathcal{S}_{00}} ~.
\label{qdim-S0R/S00}
\end{equation}
This formula is the quantum version of the Weyl dimensional formula, along with the choice of the root of unity $q=\exp(\frac{2 \pi i}{k+y})$, where $y$ is the dual Coxeter number of $G$.
\subsection{Witten zeta functions}  
The prototypical zeta function is the Riemann zeta function 
\begin{equation}
\zeta(s) = \sum_{a=1}^{\infty} a^{-s} ~.
\end{equation} 
A new type of zeta functions was defined in \cite{witten1991quantum}, which were used to express the volume of the moduli space of flat connections on Riemann surfaces. These are called the Witten zeta function and have been an object of interest in both Mathematics and Physics for quite some time, and are formally defined for the Lie group $G$ as:
\begin{equation}
\zeta_G(s) = \sum_{R} \frac{1}{(\text{dim $R$})^{s}} ~,
\label{WZClassical}
\end{equation}
where $R$ labels an irreducible representation of $\mathfrak{g}$. Their dimensions can be calculated by the well-known Weyl formula. For example, for SU(2) and SU(3) groups, we will have:
\begin{align}
\zeta_{\text{SU(2)}}(s) &= \sum_{a=0}^{\infty} \frac{1}{(a+1)^s} \nonumber \\ \zeta_{\text{SU(3)}}(s) &= \sum_{a_2=0}^{\infty} \sum_{a_1=0}^{\infty} \frac{2^s}{(a_1+1)^s (a_2+1)^s (a_1+a_2+2)^s} ~.
\end{align}
In the next section, we will introduce a $q$-deformed version of the Witten zeta functions that can be defined using the Chern-Simons theory at level $k$.
\section{The $q$-deformed Witten zeta function} \label{sec3}
Note that there can be many ways to $q$-deform a Witten zeta function. However, in this section, we would like to introduce a specific $q$-deformed version which naturally appears in the Rényi entropy calculation of certain states in the Chern-Simons theory, as we will see in section \ref{sec4}. Our definition is as follows.
\begin{mdframed}[style=sid2]
\textbf{Definition:} We define the $q$-deformed Witten zeta function as:
\begin{equation}
\zeta_{G}(s;q) \equiv \sum_{R} \frac{1}{(\text{dim}_q \,R)^s} ~.
\label{qDefZetaGen}
\end{equation}
\end{mdframed}
Here, we introduce deformation by simply replacing each classical dimension in \eqref{WZClassical} with its quantum analogue. This takes us into the regime of quantum groups $U_q(\mathfrak{g})$ where the quantum dimension for a representation is defined as:
\begin{equation}
\text{dim}_q\,R= \prod_{\alpha \in \Phi_+} \frac{[(R + \rho, \alpha)]_q}{[(\rho, \alpha)]_q} ~,
\end{equation}
where $\Phi_+$ is the set of positive roots and $\rho$ is the Weyl vector. This is the quantum analog of the classical Weyl dimension formula, where we have replaced each number by a $q$-number defined as:
\begin{equation}
[n]_q = \frac{q^{n/2}-q^{-n/2}}{q^{1/2}-q^{-1/2}} ~.
\end{equation}  
Although the $q$-deformed Witten zeta function defined in \eqref{qDefZetaGen} is an interesting object from a number theory point of view, it lacks a physical intuition, especially if we keep $q$ to be an arbitrary parameter. The physical relevance of this definition can be demonstrated if we take $q$ to be a specific root of unity (i.e., $|q|=1$), which will be the focus of this paper. 
\subsection{When $q=q_0$ is a special root of unity}
Our motivation is to relate the $q$-deformed Witten zeta function with Chern-Simons theory at level $k$. Note that when we consider $q$ to be a root of unity, the sum in equation \eqref{qDefZetaGen} is over only a finite number of representations. In particular, if we consider $q$ to be the primitive $(k+y)$-th root of unity:
\begin{equation}
q = \exp\left(\frac{2 \pi i}{k+y}\right) \equiv q_0 ~,
\label{qRootofUnity}
\end{equation}
where $y$ is the dual Coxeter number of $G$, the number of possible representations is restricted to $R \in \mathcal{I}_k$. Here $\mathcal{I}_k$ is the set of integrable highest weight representations at level $k$ that we have introduced earlier. With this choice of $q$, we will get:
\begin{equation}
\zeta_{G}(s;q_0) \equiv \sum_{R\in \mathcal{I}_k} \frac{1}{(\text{dim}_{q_0} \,R)^s} = \sum_{R\in \mathcal{I}_k} \frac{(\mathcal{S}_{00})^s}{(\mathcal{S}_{0R})^s} ~.
\label{qDefZeta}
\end{equation}
This is the object that naturally appears in the entropy calculations. Another interesting exercise is to study its $q_0\to 1$ limit, which we shall obtain next.
\subsection{$q_0 \to 1$ limit of $q_0$-deformed zeta function}
The $q_0\to 1$ limit of $\zeta_{G}(s;q_0)$, which is equivalent to the semiclassical limit $k \to \infty$ on the Chern-Simons side, is given as follows: 
\begin{mdframed}[style=sid]
\textbf{Result.} \emph{The $q_0\to 1$ limit of the $q_0$-deformed Witten zeta function is an integer multiple of the classical Witten zeta function. More precisely, the following limit holds for any Lie group with $\text{Re}(s) \geq 2$:}
\begin{equation}
\lim_{q_0\to 1}\,\zeta_{G}(s;\,q_0) = \lim_{k \to \infty}\,\left[ \sum_{R\, \in \, \mathcal{I}_k}\frac{1}{(\text{dim}_{q_0} \,R)^s} \right] = |Z_G|\, \zeta_{G}(s) ~.
\label{Conj1}
\end{equation}
\end{mdframed}
Here $Z_G$ denotes the center of the group $G$ whose values for various Lie groups are given below:
\begin{equation}
\begin{tabular}{|c|c|c|c|c|c|c|c|c|c|}
\hline
 & SU($N$) & SO($2N+1$) & Sp($2N$) & SO($2N$) & $G_2$  & $F_4$ & $E_6$ & $E_7$ & $E_8$ \\ \hline
$|Z_G|$ & $N$ & 2 & 2 & 4 & 1  & 1 & 3 & 2 & 1 \\ \hline
\end{tabular}
\label{ConstantGroup}
\end{equation}
The emergence of the factor $|Z_G|$ in \eqref{Conj1} can be explained as follows. For a given group, there will be $|Z_G|$ number of abelian anyons having $q$-dimension 1, and they correspond to the center elements. The fusion with these abelian anyons preserves the quantum dimension, i.e., $\text{dim}_q (g\cdot R) = \text{dim}_q(R) \,\,, \forall\, g\in Z_G$. As a result, the integrable representations are mapped into $|Z_G|$ number of orbits, and the sum in \eqref{Conj1} can be written as: 
\begin{equation}
\sum_{R\, \in \, \mathcal{I}_k} \frac{1}{(\text{dim}_{q_0} \,R)^s} = \sum_{\text{orbits}} \,\sum_{R\, \in \, \text{orbit}} \frac{1}{(\text{dim}_{q_0} \,R)^s}
\end{equation}
When $k$ is large, each orbit will have the same size and will eventually include every irreducible representation. The set of the quantum dimensions in all the orbits remains the same, and since in $k\to \infty$ limit, the quantum dimension converges to the classical dimension, we will get $|Z_G|$ number of copies of classical Witten zeta functions. So we will obtain:
\begin{equation}
\lim_{k \to \infty} \left[\sum_{R\, \in \, \mathcal{I}_k} \frac{1}{(\text{dim}_{q_0} \,R)^s} \right] = |Z_G| \sum_{R} \frac{1}{(\text{dim} \,R)^s} = |Z_G| \, \zeta_G(s) ~.
\end{equation}
For example, consider the SU($N$) group, where the orbits can be elegantly obtained. The action of the center $\mathbb{Z}_N$ in this case will be a cyclic permutation of the weights in the extended Dynkin diagram. If $[a_0,a_1,a_2,\ldots,a_{N-1}]$ is the affine highest weight with $a_0 = k-(a_1+\ldots+a_{N-1})$ and $a_0 \geq 0$, then the orbits will be defined by the cyclic permutation of Dynkin labels in $[a_0,a_1,a_2,\ldots,a_{N-1}]$. The quantum dimension for weights in each permutation will remain the same. 

Take for example SU(2) group with affine highest weight $[a_0,a_1]$ where $a_0=k-a_1$ and $a_0 \geq 0$. In this case, the cyclic permutation $\mathbb{Z}_2$ on the extended Dynkin labels will give $[a_0,a_1]$ and $[a_1,a_0]$ under the action of identity and non-identity elements of $\mathbb{Z}_2$, respectively. Denoting $a_0=a$ and $a_1 = (k-a)$, the full set of integrable representations will split into two orbits. The first orbit will contain the set of representations $[a]$ of the SU(2) group with index $a$ running appropriately. The second orbit will contain the set of representations $[k-a]$ of the SU(2) group. Since $\text{dim}_{q_0}\, [a] = \text{dim}_{q_0}\, [k-a]$, the quantum dimensions of corresponding representations in the two orbits will be the same. Thus, the $q$-Witten zeta function will split into two equal summations. Note that for $k=\text{even}$, the representation $[k/2]$ will be a fixed point. So, we take half of its contribution in the first orbit and half in the second orbit. However, for large $k$, the fixed points will not matter. For finite $k$, we write the two summations below explicitly:
\begin{equation}
 \zeta_{\text{SU(2)}}(s;q_0) = 
\begin{cases}
 \displaystyle S_1+S_2 \quad,\quad & \text{for $k=$ odd} \\
 \displaystyle S_3+S_4 \quad,\quad & \text{for $k=$ even}
\end{cases}
\end{equation}
where the four sums appearing in this equation are given as:
\begin{align}
S_1 &= \sum_{a=0}^{(k-1)/2} \frac{1}{(\text{dim}_{q_0} [a])^s} \quad;\quad S_2 = \sum_{a=(k+1)/2}^{k} \frac{1}{(\text{dim}_{q_0} [a])^s} \nonumber \\ 
S_3 &= \sum_{a=0}^{(k/2)-1} \frac{1}{(\text{dim}_{q_0} [a])^s} + \frac{1}{2}\frac{1}{(\text{dim}_{q_0} [k/2])^s} \quad;\quad S_4 = \frac{1}{2}\frac{1}{(\text{dim}_{q_0} [k/2])^s} + \sum_{a=(k/2)+1}^{k} \frac{1}{(\text{dim}_{q_0} [a])^s} 
\end{align}
Due to the symmetry $\text{dim}_{q_0}\, [a] = \text{dim}_{q_0}\, [k-a]$, we will have $S_1=S_2$ and $S_3=S_4$. In the $q_0 \to 1$ limit or $k \to \infty$ limit, the summations $S_1$ and $S_3$ both converge to $\zeta_{\text{SU}(2)}(s)$ giving a factor of 2 in \eqref{Conj1}.

For the SU(3) group, the highest affine weights in the extended Dynkin diagram are $[a_0,a_1,a_2]$ where $a_0 = k-a_1-a_2$ and $a_0 \geq 0$. The center $\mathbb{Z}_3$ will give the cyclic permutations $[a_0,a_1,a_2]$, $[a_1,a_2,a_0]$ and $[a_2,a_0,a_1]$ corresponding to the three elements of $\mathbb{Z}_3$. As a result, we will get three orbits defined by representations $[a_1,a_2]$, $[a_2,k-a_1-a_2]$, and $[k-a_1-a_2,a_1]$ of SU(3) respectively. It can be checked that the quantum dimensions of these SU(3) representations are the same:   
\begin{equation*}
\text{dim}_{q_0}[a_1,a_2] = \text{dim}_{q_0}[a_2,k-a_1-a_2] = \text{dim}_{q_0}[k-a_1-a_2,a_1] ~. 
\end{equation*}
As a result, in the large $k$ limit, we will get equal contributions from the three orbits, with each orbit contributing a factor of $\zeta_{\text{SU(3)}}(s)$.

The result \eqref{Conj1} is important both from a Physics and a Mathematics point of view. In Chern-Simons theory, this result helps us to obtain the large $k$ limit of the R\'enyi entropies of certain states, as we shall see in the section \ref{sec4}. From the number theory point of view, the result \eqref{Conj1} gives a new definition to the classical Witten zeta functions and facilitates their computation by obtaining the limit on the left-hand side of \eqref{Conj1}. In the remainder of this section, we will discuss the number-theoretical importance of \eqref{Conj1}.
\subsection{Number-theoretical importance}
The result \eqref{Conj1} can be rewritten as: 
\begin{equation}
\zeta_{G}(s) = \frac{1}{|Z_G|}\lim_{k \to \infty}\,\left[ \sum_{R\, \in \, \mathcal{I}_k}\frac{1}{(\text{dim}_{q_0} \,R)^s} \right]  ~.
\label{WZfromConj}
\end{equation}
This equation provides a new definition of the classical Witten zeta function and facilitates its computation by obtaining the limit on the right-hand side of \eqref{WZfromConj}. Here we provide some insights into computing this limit. Let us first analyze the sum
\begin{equation}
\sum_{R\, \in \, \mathcal{I}_k} \frac{1}{(\text{dim}_{q_0}\,R)^{s}} = (\mathcal{S}_{00})^s \, \sum_{R\, \in \, \mathcal{I}_k} \frac{1}{(\mathcal{S}_{0R})^{s}} ~.
\label{EqLHSRHS}
\end{equation}
Now we have the following two critical observations:
\begin{itemize}
    \item The naive numerical calculations suggest that the left-hand side of \eqref{EqLHSRHS} converges to a finite value as $k \to \infty$ for any $s \in \mathbb{C}$ with $\text{Re}(s) \geq 2$. 
    \item The matrix element $\mathcal{S}_{00}$ is a decreasing function in $k$  whose asymptotics is given by a power law:
\begin{equation}
\mathcal{S}_{00} \, \sim \, \frac{\alpha(G)}{k^{(\text{dim $G$})/2}} + \mathcal{O}\left(\frac{1}{k^{(\text{dim $G$}+2)/2}}\right) ~,
\label{S00AsympGen}
\end{equation}
where `$\text{dim $G$}$' denotes the dimension of the group and $\alpha(G)$ is some group dependent constant.
\end{itemize}
Based on the above two observations, we conclude that the sum of the inverse powers of $\mathcal{S}_{0R}$ must diverge with a power law similar to that of $(\mathcal{S}_{00})^s$. Hence, we will have: 
\begin{equation}
\sum_{R\, \in \, \mathcal{I}_k} \frac{1}{(\mathcal{S}_{0R})^{s}} \, \sim \, \beta_s(G)\, k^{s(\text{dim $G$})/2} ~,
\label{S0Rasymp}
\end{equation}
where $\beta_s(G)$ is a group dependent constant which is independent of $k$. Hence, we will get the following limit:
\begin{equation}
\lim_{q_0 \to 1} \zeta_{G}(s;q_0) = \alpha^s\, \beta_s ~.
\label{QWZLimit}
\end{equation}
This enables us to write the value of classical Witten zeta functions by using the equation \eqref{WZfromConj} and can be written as:
\begin{equation}
\boxed{\zeta_{G}(s) = \frac{\alpha^s(G)\, \beta_s(G)}{|Z_G|}} ~.
\label{zetageneral}
\end{equation}

\noindent \textbf{\textcolor{red}{Note:}} The constant $\alpha$ in the above equation is a transcendental number due to the presence of the factor of $\pi$, as we shall see later. However, the value (and hence rationality or irrationality) of $\beta_s$ is not known for arbitrary values of $s$. This translates into the problem of determining whether these zeta functions are irrational or not.\footnote{For example, it is a longstanding problem in number theory to prove the irrationality of the Riemann zeta functions $\zeta_{\text{SU(2)}}(s)$ when $s \geq 5$ is an odd integer.} It would be interesting to study the number-theoretic properties of the constants $\beta_s$ closely, which may give more insight into the structure of these zeta functions. We leave this aspect as a potential future direction. 

When $s=2n$ is a positive even integer, it is a well-known result that the sum of the inverse powers of $\mathcal{S}_{0R}$ over all integrable representations is a polynomial in $k$ with rational coefficients. This makes $\beta_{2n}$ a rational number which can be computed on a case-by-case basis. Thus, the exact expression for classical Witten zeta functions at positive even integers can be written down using the \eqref{zetageneral}. 

We have done several computations and verifications for all classical and exceptional Lie groups. To keep this section concise, we briefly discuss the SU($N$) group, which helps readers understand how to use equation \eqref{zetageneral}. We will also present some concrete values of SU($N$) classical Witten zeta functions for low ranks. For Lie groups other than SU($N$), the relevant details are presented in Appendix~\ref{app2} to avoid cluttering the following section.
\subsection{An example: SU($N$) group}
An irreducible representation for SU($N$) group is denoted as $R = [a_1,a_2,\ldots,a_{N-1}]$, where $a_i$ are the Dynkin labels of the highest weight associated with the representation $R$. For a given level $k$, the set of integrable representations is given by:
\begin{equation}
\mathcal{I}_k = \left\{[a_1, \ldots, a_{N-1}] \,:\, a_{\mu} \geq 0 \,\,\, ; \,\,\, a_1 +\ldots + a_{N-1} \leq k \right\} ~.
\end{equation} 
The matrix element $\mathcal{S}_{0R}$ for an integrable representation $R$ is given as
\begin{equation}
\mathcal{S}_{0R} = \frac{2^{N(N-1)/2}}{\sqrt{N}\, (k+N)^{\frac{N-1}{2}}}\, \left(\prod _{i=1}^{N-1} \sin(\frac{\pi N-\pi i+\pi \ell_i}{k+N})\right)\,\left(\prod_{i=1}^{N-2} \prod_{j=i+1}^{N-1} \sin(\frac{\pi j-\pi i+\pi \ell_i-\pi \ell_j}{k+N})\right) ~,
\end{equation}
where the integers $\ell_i$ are defined as:
\begin{equation}
\ell_i = \sum_{j=i}^{N-1} a_j  ~.
\end{equation}
The element $\mathcal{S}_{00}$ of the $\mathcal{S}$ matrix is given as
\begin{equation}
\mathcal{S}_{00} = \frac{2^{N(N-1)/2}}{\sqrt{N}\, (k+N)^{\frac{N-1}{2}}}\, \left(\prod _{i=1}^{N-1} \sin(\frac{\pi N-\pi i}{k+N})\right)\,\left(\prod_{i=1}^{N-2} \prod_{j=i+1}^{N-1} \sin(\frac{\pi j-\pi i}{k+N})\right) ~.
\end{equation}
The large $k$ expansion of $\mathcal{S}_{00}$ is given as:
\begin{equation}
\mathcal{S}_{00} \sim \frac{(2\pi)^{N(N-1)/2} \, G(N+1)}{\sqrt{N}}\, \frac{1}{k^{(N^2-1)/2}} \,+\cdots ~,
\end{equation}
where `$G$' denotes the Barnes $G$-function whose value at positive integers is:
\begin{equation}
G(N+1) = \prod_{i=1}^{N-1} i! ~.
\end{equation} 
Following the earlier discussion, we would expect the following large $k$ behavior:
\begin{equation}
\sum_{R\, \in \, \mathcal{I}_k} (\mathcal{S}_{0R})^{-s} \, \sim \, \beta_s\, k^{s(N^2-1)/2} ~,
\label{betaSUN}
\end{equation}
where $\beta_s$ is some constant. Thus, the equation \eqref{zetageneral} will give the following formula of the Witten zeta function for the SU($N$) group for $\text{Re}(s) \geq 2$: 
\begin{equation}
\boxed{\zeta_{\text{SU}(N)}(s) = 
\frac{\pi^{sN(N-1)/2}}{N}\left(\frac{2^{N(N-1)/2}\, G(N+1)}{N^{1/2}}\right)^s\beta_s } ~.
\label{ConjSUN}
\end{equation}
For arbitrary $s$, one can numerically obtain $\beta_s$ and use the above equation to calculate the numerical values of $\zeta_{\text{SU}(N)}(s)$ up to desired precision.

For $s=2n$, where $n \geq 1$ is an integer, the $\beta_{2n}$ can be analytically obtained, which turn out to be rational numbers. To do this, we first notice that: 
\begin{equation}
\sum_{R\, \in \, \mathcal{I}_k} \frac{1}{(\mathcal{S}_{0R})^{2n}} = \mathcal{A}_{n,N}(k) ~,
\label{powersofS0RSUN}
\end{equation}
where $\mathcal{A}_{n,\,N}(k)$ is a polynomial of degree $n(N^2-1)$ in the variable $k$ with rational coefficients which has the following form: 
\begin{equation}
\mathcal{A}_{n,\,N}(k) = (k+1)(k+2)\ldots(k+N-1)(k+N)^{nN-n}(k+N+1)\ldots(k+2N-1)\, \sum_{i=0}^{nN^2-nN-2N+2} C_i\, k^i ~,
\label{}
\end{equation}
where $C_i \in \mathbb{Q}_{+}$ are positive rational coefficients. Let us denote the leading order coefficient (i.e., the coefficient of $k^{n(N^2-1)}$ term) of the polynomial $\mathcal{A}_{n,\,N}$ as $C_{\text{lead}}(\mathcal{A}_{n,\,N})$. As a result, the constant $\beta_{2n}$ will be a rational number and can be read off as: 
\begin{equation}
\beta_{2n}  = C_{\text{lead}}(\mathcal{A}_{n,\,N}) ~.
\end{equation}
Thus, the zeta function for the SU($N$) group at positive even integers can be given as:
\begin{equation}
\boxed{\zeta_{\text{SU}(N)}(2n) = \frac{(2\pi)^{nN(N-1)}\, G(N+1)^{2n}}{N^{n+1}}\, C_{\text{lead}}(\mathcal{A}_{n,N})} ~.
\label{zetasuN}
\end{equation}
As an example, consider the SU(3) group and $n=1$. We will have the following summation:
\begin{equation}
\sum_{R\, \in \, \mathcal{I}_k} \frac{1}{(\mathcal{S}_{0R})^{2}} = \sum_{a_1=0}^k \sum_{a_2=0}^k \sum_{i=0}^k \frac{\delta_{a_1+a_2,\,i}}{(\mathcal{S}_{0R})^{2}} 
\end{equation}
where the delta function is used to monitor the integrability condition $a_1+a_2\leq k$ for the SU(3) group. This formula can be easily incorporated into any computational software. The value of this sum will come out to be a positive integer for any value of $k$. Performing this sum for each of the values $k=0,1,2,3,\ldots$, we will get a sequence of positive integers $\{1,9,45,166,504,1332,3168,6930,14157,27313,\ldots \}$. This sequence comes from a degree 8 polynomial in $k$ which can be written as:
\begin{align*}
\mathcal{A}_{1,3}&=(k+1) (k+2) (k+3)^2 (k+4) (k+5) \left(\frac{k^2}{20160}+\frac{k}{3360}+\frac{1}{360}\right)
\end{align*}
The leading order coefficient of this polynomial is: 
\begin{equation*}
C_{\text{lead}}(\mathcal{A}_{1,3}) = \frac{1}{20160} ~.
\end{equation*}
Hence, from our formula \eqref{zetasuN}, we can obtain the zeta function as:
\begin{equation*}
\zeta_{\text{SU}(3)}(2) = \frac{4 \pi ^6}{2835} = \frac{2^2}{3^4\cdot 5^1\cdot 7^1}\pi ^6 ~.
\end{equation*}
Using this technique, we can compute the classical Witten zeta functions of the SU($N$) group at positive even integers. In the table \ref{LeadingSUNTable}, we tabulate the values of $\zeta_{\text{SU}(N)}(2n)$ for some low values of $n$ and $N$.
\captionsetup{width=16cm}
\begin{longtable}{|c||L{15cm}|}
			\caption[Values of Witten zeta function at positive even integers for SU($N$) group]{Values of Witten zeta function at positive even integers for SU($N$) group calculated using \eqref{zetasuN}.}
			\label{LeadingSUNTable} \\ \hline \rowcolor{Gray}
			$n$ & $\zeta_{\text{SU}(2)}(2n)/\pi^{2n}$ \\ \hline \rowcolor{Redd}
			$1$ & \footnotesize{$1/(2^1\cdot 3^1)$} \\ \hline\rowcolor{Bluee}
			$2$ & \footnotesize{$1/(2^1\cdot 3^2\cdot 5^1)$} \\ \hline\rowcolor{Redd}
			$3$ & \footnotesize{$1/(3^3\cdot 5^1\cdot 7^1)$} \\ \hline\rowcolor{Bluee}
			$4$ & \footnotesize{$1/(2^1\cdot 3^3\cdot 5^2\cdot 7^1)$} \\ \hline\rowcolor{Redd}
			$5$ & \footnotesize{$1/(3^5\cdot 5^1\cdot 7^1\cdot 11^1)$} \\ \hline \rowcolor{Bluee}
			$6$ & \footnotesize{$691^1/(3^6\cdot 5^3\cdot 7^2\cdot 11^1\cdot 13^1)$} \\ \hline \rowcolor{Redd}
			$7$ & \footnotesize{$2^1/(3^6\cdot 5^2\cdot 7^1\cdot 11^1\cdot 13^1)$} \\ \hline \rowcolor{Bluee}
			$8$ & \footnotesize{$3617^1/(2^1\cdot 3^7\cdot 5^4\cdot 7^2\cdot 11^1\cdot 13^1\cdot 17^1)$} \\ \hline \rowcolor{Redd}
			$9$ & \footnotesize{$43867^1/(3^9\cdot 5^3\cdot 7^3\cdot 11^1\cdot 13^1\cdot 17^1\cdot 19^1)$} \\ \hline\rowcolor{Bluee}
			$10$ & \footnotesize{$(283^1\cdot 617^1)/(3^9\cdot 5^5\cdot 7^2\cdot 11^2\cdot 13^1\cdot 17^1\cdot 19^1)$} \\\hline \hline \rowcolor{Gray}
			$n$ & $\zeta_{\text{SU}(3)}(2n)/\pi^{6n}$ \\ \hline\rowcolor{Redd}
			$1$ & \footnotesize{$2^2/(3^4\cdot 5^1\cdot 7^1)$} \\ \hline \rowcolor{Bluee}
			$2$ & \footnotesize{$(2^4\cdot 19^1)/(3^7\cdot 5^3\cdot 7^1\cdot 11^1\cdot 13^1)$} \\ \hline \rowcolor{Redd}
			$3$ & \footnotesize{$(2^7\cdot 1031^1)/(3^{10}\cdot 5^3\cdot 7^3\cdot 11^1\cdot 13^1\cdot 17^1\cdot 19^1)$} \\ \hline \rowcolor{Bluee}
			$4$ & \footnotesize{$(2^8\cdot 43^1\cdot 751^1)/(3^{11}\cdot 5^6\cdot 7^4\cdot 11^1\cdot 13^1\cdot 17^1\cdot 19^1\cdot 23^1)$} \\ \hline \rowcolor{Redd}
			$5$ & \footnotesize{$(2^{13}\cdot 27739097^1)/(3^{16}\cdot 5^6\cdot 7^4\cdot 11^3\cdot 13^1\cdot 17^1\cdot 19^1\cdot 23^1\cdot 29^1\cdot 31^1)$} \\ \hline \rowcolor{Bluee}
			$6$ & \footnotesize{$(2^{13}\cdot 29835840687589^1)/(3^{19}\cdot 5^9\cdot 7^6\cdot 11^3\cdot 13^3\cdot 17^1\cdot 19^1\cdot 23^1\cdot 29^1\cdot 31^1\cdot 37^1)$} \\ \hline \rowcolor{Redd}
			$7$ & \footnotesize{$(2^{17}\cdot 89^1\cdot 127^1\cdot 6353243297^1)/(3^{20}\cdot 5^9\cdot 7^7\cdot 11^3\cdot 13^3\cdot 17^1\cdot 19^1\cdot 23^1\cdot 29^1\cdot 31^1\cdot 37^1\cdot 41^1\cdot 43^1)$} \\ \hline \rowcolor{Bluee}
			$8$ & \footnotesize{$(2^{16}\cdot 221137132669842886663^1)/(3^{23}\cdot 5^{12}\cdot 7^8\cdot 11^4\cdot 13^4\cdot 17^3\cdot 19^1\cdot 23^1\cdot 29^1\cdot 37^1\cdot 41^1\cdot 43^1\cdot 47^1)$} \\ \hline \rowcolor{Redd}
			$9$ & \footnotesize{$(2^{21}\cdot 400607^1\cdot 401418649^1\cdot 45096770501^1)/(3^{28}\cdot 5^{12}\cdot 7^9\cdot 11^4\cdot 13^4\cdot 17^3\cdot 19^3\cdot 23^1\cdot 29^1\cdot 31^1\cdot 37^1\cdot 41^1\cdot 43^1\cdot 47^1\cdot 53^1)$} \\ \hline \rowcolor{Bluee}
			$10$ & \footnotesize{$(2^{23}\cdot 67^1\cdot 1815418673^1\cdot 17723407173073733123^1)/(3^{29}\cdot 5^{15}\cdot 7^{10}\cdot 11^6\cdot 13^4\cdot 17^3\cdot 19^3\cdot 23^1\cdot 29^1\cdot 31^1\cdot 37^1\cdot 41^1\cdot 43^1\cdot 47^1\cdot 53^1\cdot 59^1\cdot 61^1)$} \\ \hline \rowcolor{Gray}
			$n$ & $\zeta_{\text{SU}(4)}(2n)/\pi^{12n}$ \\ \hline \rowcolor{Redd}
			$1$ & \footnotesize{$(2^2\cdot 23^1)/(3^4\cdot 5^3\cdot 7^2\cdot 11^1\cdot 13^1)$} \\ \hline \rowcolor{Bluee}
			$2$ & \footnotesize{$(2^6\cdot 14081^1)/(3^8\cdot 5^6\cdot 7^2\cdot 11^2\cdot 13^2\cdot 17^1\cdot 19^1\cdot 23^1)$} \\ \hline \rowcolor{Redd}
			$3$ & \footnotesize{$(2^{11}\cdot 757409^1\cdot 23283173^1)/(3^{12}\cdot 5^9\cdot 7^6\cdot 11^3\cdot 13^3\cdot 17^2\cdot 19^2\cdot 23^1\cdot 29^1\cdot 31^1\cdot 37^1)$} \\ \hline \rowcolor{Bluee}
			$4$ & \footnotesize{$(2^{14}\cdot 1021^1\cdot 5529809^1\cdot 754075957^1)/(3^{14}\cdot 5^{12}\cdot 7^8\cdot 11^3\cdot 13^3\cdot 17^3\cdot 19^2\cdot 23^2\cdot 29^1\cdot 31^1\cdot 37^1\cdot 41^1\cdot 43^1\cdot 47^1)$} \\ \hline \rowcolor{Redd}
			$5$ & \footnotesize{$(2^{21}\cdot 116763209^1\cdot 1872391681^1\cdot 3187203549787^1)/(3^{20}\cdot 5^{15}\cdot 7^{10}\cdot 11^6\cdot 13^3\cdot 17^3\cdot 19^3\cdot 23^2\cdot 29^2\cdot 31^2\cdot 37^1\cdot 41^1\cdot 43^1\cdot 47^1\cdot 53^1\cdot 59^1\cdot 61^1)$} \\ \hline \rowcolor{Bluee}
			$6$ & \footnotesize{$(2^{23}\cdot 1798397149^1\cdot 5509496891^1\cdot 6127205846988571484743^1)/(3^{24}\cdot 5^{18}\cdot 7^{12}\cdot 11^6\cdot 13^6\cdot 17^3\cdot 19^3\cdot 23^3\cdot 29^2\cdot 31^2\cdot 37^2\cdot 41^1\cdot 43^1\cdot 47^1\cdot 53^1\cdot 59^1\cdot 61^1\cdot 67^1\cdot 71^1\cdot 73^1)$} \\ \hline \rowcolor{Redd}
			$7$ & \footnotesize{$(2^{31}\cdot 107060512957308326131930505315555844558595160011^1)/(3^{27}\cdot 5^{21}\cdot 7^{14}\cdot 11^7\cdot 13^7\cdot 17^3\cdot 19^2\cdot 23^3\cdot 29^3\cdot 31^2\cdot 37^2\cdot 41^2\cdot 43^2\cdot 47^1\cdot 53^1\cdot 59^1\cdot 61^1\cdot 67^1\cdot 71^1\cdot 73^1\cdot 79^1\cdot 83^1)$} \\ \hline \rowcolor{Bluee}
			$8$ & \footnotesize{$(2^{30}\cdot 77855522117^1\cdot 33600567284007472260739^1\cdot 81145847646624374989395609791^1)/(3^{31}\cdot 5^{24}\cdot 7^{16}\cdot 11^9\cdot 13^8\cdot 17^6\cdot 19^3\cdot 23^3\cdot 29^3\cdot 31^3\cdot 37^2\cdot 41^2\cdot 43^2\cdot 47^2\cdot 53^1\cdot 59^1\cdot 61^1\cdot 67^1\cdot 71^1\cdot 73^1\cdot 79^1\cdot 83^1\cdot 89^1\cdot 97^1)$} \\ \hline \rowcolor{Redd}
			$9$ & \footnotesize{$(2^{37}\cdot 331^1\cdot 10368540971^1\cdot 40653670834097855534893^1\cdot 406079879527767338511975408843427482112103^1)/(3^{36}\cdot 5^{27}\cdot 7^{18}\cdot 11^9\cdot 13^9\cdot 17^6\cdot 19^6\cdot 23^3\cdot 29^3\cdot 31^3\cdot 37^3\cdot 41^2\cdot 43^2\cdot 47^2\cdot 53^2\cdot 59^1\cdot 61^1\cdot 67^1\cdot 71^1\cdot 73^1\cdot 79^1\cdot 83^1\cdot 89^1\cdot 97^1\cdot 101^1\cdot 103^1\cdot 107^1\cdot 109^1)$} \\ \hline\rowcolor{Bluee}
			$10$ & \footnotesize{$(2^{41}\cdot 321961^1\cdot 431010253879^1\cdot 316394055331853^1\cdot 217274379906777968833^1\cdot 52096269477671207981197145179883^1)/(3^{39}\cdot 5^{30}\cdot 7^{20}\cdot 11^{12}\cdot 13^8\cdot 17^7\cdot 19^6\cdot 23^3\cdot 29^3\cdot 31^3\cdot 37^2\cdot 41^3\cdot 43^2\cdot 47^2\cdot 53^2\cdot 59^2\cdot 61^2\cdot 67^1\cdot 71^1\cdot 73^1\cdot 79^1\cdot 83^1\cdot 89^1\cdot 97^1\cdot 101^1\cdot 103^1\cdot 107^1\cdot 109^1\cdot 113^1)$} \\ \hline \rowcolor{Gray}
			$n$ & $\zeta_{\text{SU}(5)}(2n)/\pi^{20n}$ \\ \hline\rowcolor{Redd}
			$1$ & \footnotesize{$2^{10}/(3^6\cdot 5^6\cdot 7^3\cdot 11^2\cdot 17^1)$} \\\hline \rowcolor{Bluee}
			$2$ & \footnotesize{$(2^{20}\cdot 1523^1\cdot 2625375581^1)/(3^{12}\cdot 5^{11}\cdot 7^6\cdot 11^4\cdot 13^3\cdot 17^2\cdot 19^2\cdot 23^1\cdot 29^1\cdot 31^1\cdot 37^1\cdot 41^1)$} \\\hline \rowcolor{Redd}
			$3$ & \footnotesize{$(2^{31}\cdot 30677^1\cdot 2082905565627654787323001^1)/(3^{18}\cdot 5^{16}\cdot 7^{10}\cdot 11^6\cdot 13^5\cdot 17^3\cdot 19^3\cdot 23^2\cdot 29^2\cdot 31^2\cdot 37^1\cdot 41^1\cdot 43^1\cdot 47^1\cdot 53^1\cdot 59^1\cdot 61^1)$} \\ \hline \rowcolor{Bluee}
			$4$ & \footnotesize{$(2^{40}\cdot 85081^1\cdot 728520415874861^1\cdot 1361779882876127669651^1)/(3^{22}\cdot 5^{21}\cdot 7^{13}\cdot 11^6\cdot 13^6\cdot 17^5\cdot 19^4\cdot 23^3\cdot 29^2\cdot 31^2\cdot 37^2\cdot 41^2\cdot 43^1\cdot 47^1\cdot 53^1\cdot 59^1\cdot 61^1\cdot 67^1\cdot 71^1\cdot 73^1\cdot 79^1)$} \\ \hline \rowcolor{Redd}
			$5$ & \footnotesize{$(2^{51}\cdot 2143^1\cdot 6751027^1\cdot 430667831149^1\cdot 201223346979560452521803194127591413^1)/(3^{30}\cdot 5^{26}\cdot 7^{16}\cdot 11^{10}\cdot 13^5\cdot 17^6\cdot 19^5\cdot 23^4\cdot 29^2\cdot 31^3\cdot 37^2\cdot 41^2\cdot 43^2\cdot 47^2\cdot 53^1\cdot 59^1\cdot 61^1\cdot 67^1\cdot 71^1\cdot 73^1\cdot 79^1\cdot 83^1\cdot 89^1\cdot 97^1\cdot 101^1)$} \\ \hline \rowcolor{Bluee}
			$6$ & \footnotesize{$(2^{61}\cdot 12165631^1\cdot 879445620034002539535907927568397845963873954198598074696163224141259900991^1)/(3^{36}\cdot 5^{30}\cdot 7^{19}\cdot 11^{12}\cdot 13^{10}\cdot 17^6\cdot 19^6\cdot 23^5\cdot 29^4\cdot 31^4\cdot 37^3\cdot 41^2\cdot 43^2\cdot 47^2\cdot 53^2\cdot 59^2\cdot 61^2\cdot 67^1\cdot 71^1\cdot 73^1\cdot 79^1\cdot 83^1\cdot 89^1\cdot 97^1\cdot 101^1\cdot 103^1\cdot 107^1\cdot 109^1\cdot 113^1)$} \\ \hline \hline \rowcolor{Gray}
			$n$ & $\zeta_{\text{SU}(6)}(2n)/\pi^{30n}$ \\ \hline \rowcolor{Redd}
$1$ & \footnotesize{$(2^{17}\cdot 46511^1)/(3^{10}\cdot 5^5\cdot 7^5\cdot 11^3\cdot 13^2\cdot 17^1\cdot 19^1\cdot 23^1\cdot 29^1\cdot 31^1)$} \\ \hline \rowcolor{Bluee}
$2$ & \footnotesize{$(2^{35}\cdot 2809087^1\cdot 1366804622087788067^1)/(3^{19}\cdot 5^{11}\cdot 7^{10}\cdot 11^6\cdot 13^5\cdot 17^3\cdot 19^3\cdot 23^2\cdot 29^2\cdot 31^2\cdot 37^1\cdot 41^1\cdot 43^1\cdot 47^1\cdot 53^1\cdot 59^1\cdot 61^1)$} \\ \hline
\end{longtable}
This method can be used to obtain the exact analytic values of the zeta functions for all classical and exceptional Lie groups at positive even integers, some of which are presented in the Appendix~\ref{app2}. This highlights an important number-theoretic application of \eqref{Conj1}.

We will now discuss how the \eqref{Conj1} is applicable and important in the context of topological entanglement of quantum states in Chern-Simons theory.
\section{R\'enyi entropies associated with state $\ket{S^3  \backslash T_{p,p}}$ for torus links $T_{p,p}$} \label{sec4}
Let us consider the link complement $S^3\backslash \mathcal{L}$ where $\mathcal{L}$ is the torus link $T_{p,p}$. The torus link $T_{p,p}$ consists of $p$ circles such that any two circles are linked together exactly once (in other words, the linking number between any two circles is 1). To visualize this, we have presented some of these links in Table \ref{Tpplinkstable}.   
\begin{table}[htb]
\begin{center}
$\begin{array}{|c|c|c|} \hline
\rowcolor{Gray}
\text{Link} & \text{Diagram} & \text{Link drawn on surface of a torus} \\ \hline
T_{2,2} & {\begin{array}{c}
\includegraphics[width=0.20\linewidth]{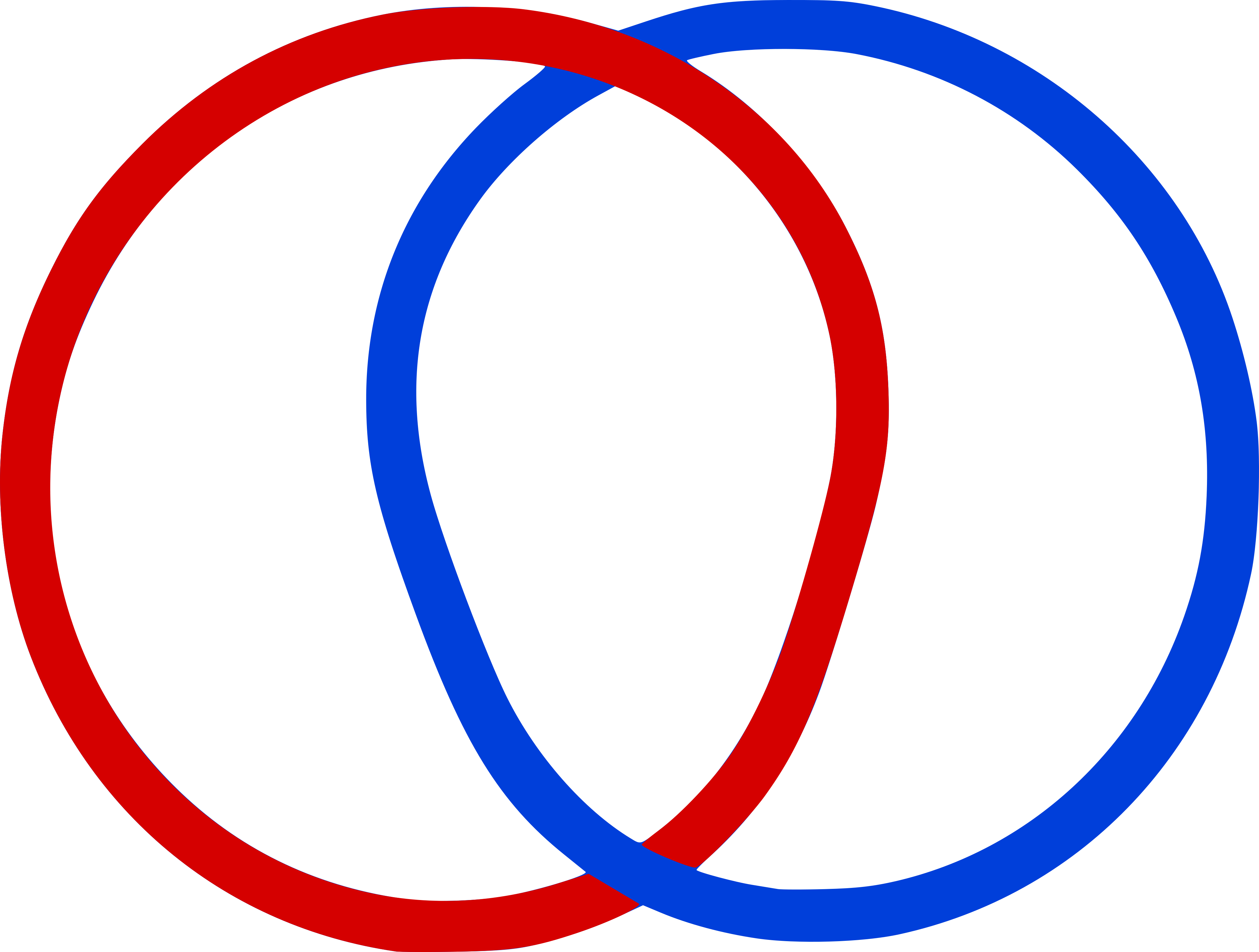}
\end{array}} & {\begin{array}{c}
\includegraphics[width=0.25\linewidth]{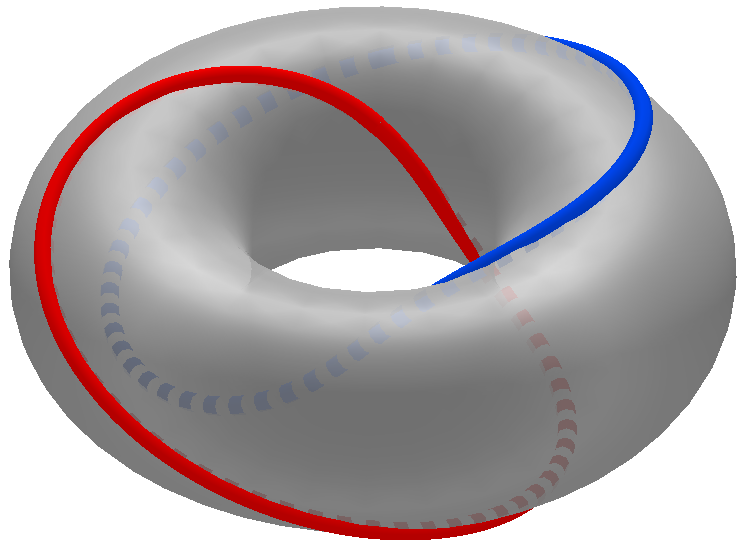}
\end{array}} \\
T_{3,3} & {\begin{array}{c}
\includegraphics[width=0.20\linewidth]{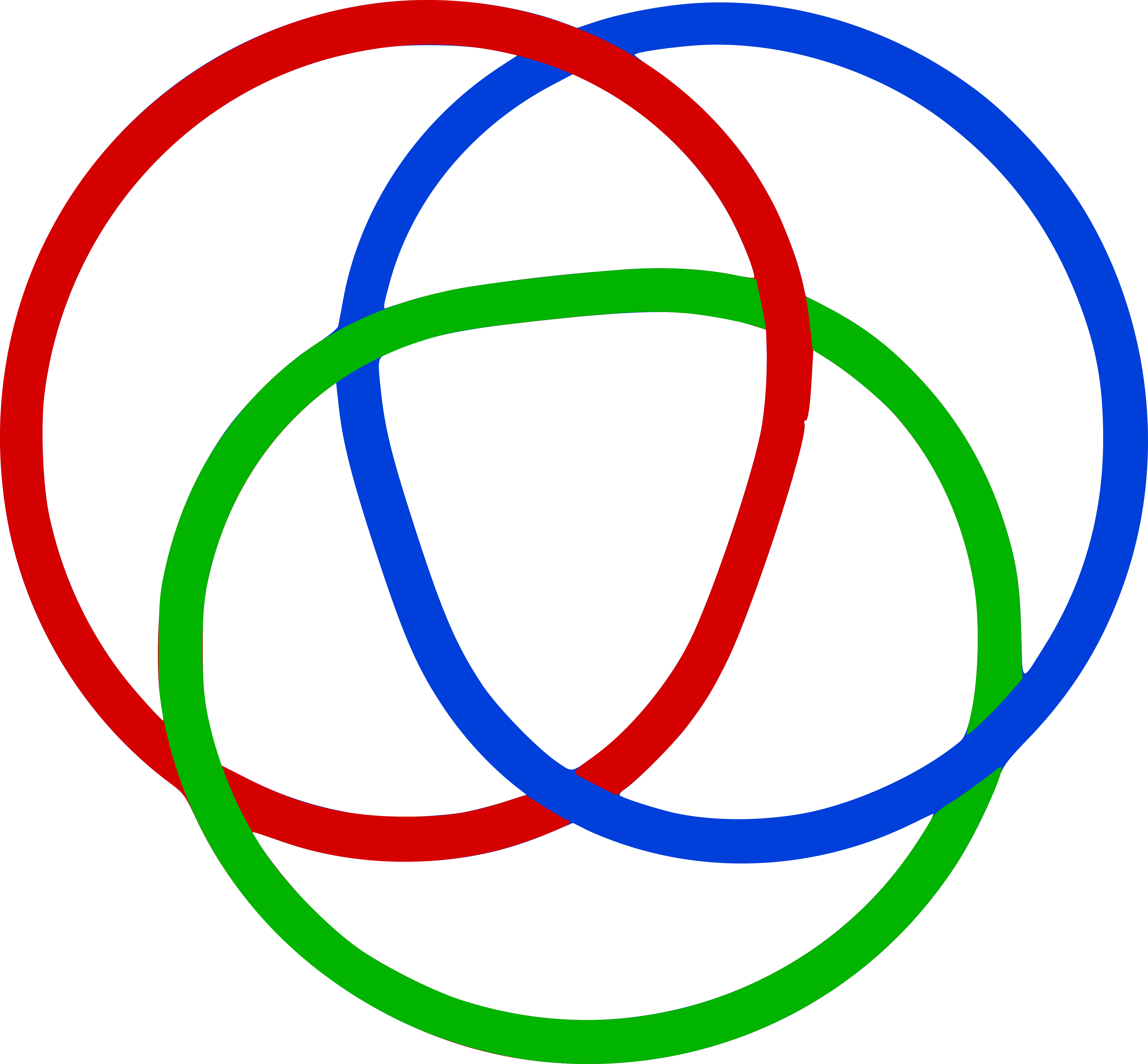}
\end{array}} & {\begin{array}{c}
\includegraphics[width=0.25\linewidth]{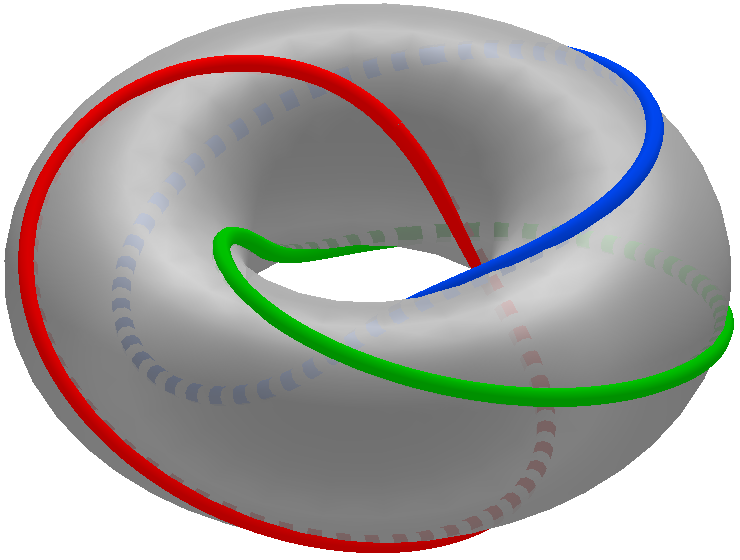}
\end{array}} \\
T_{4,4} & {\begin{array}{c}
\includegraphics[width=0.20\linewidth]{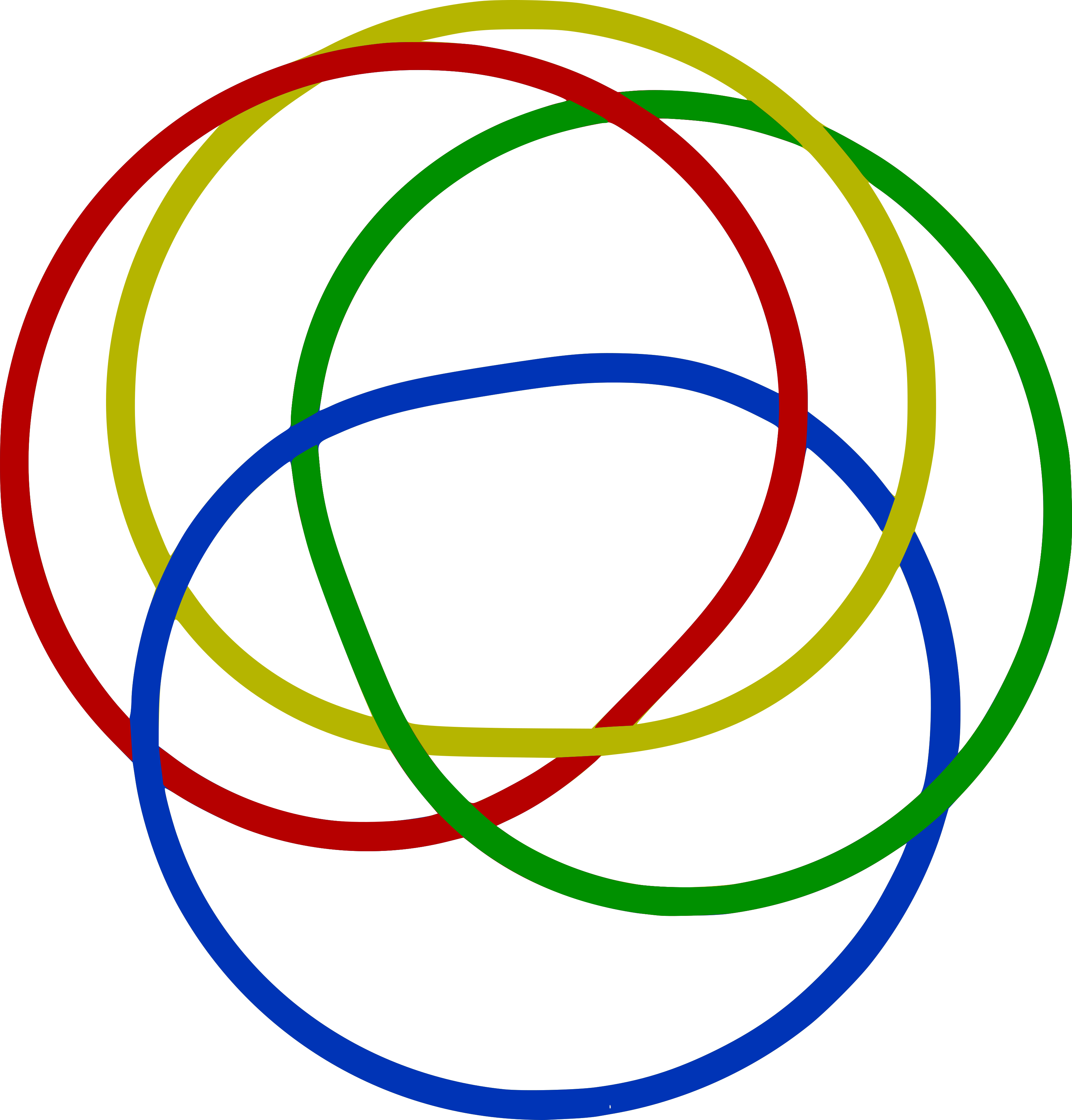}
\end{array}} & {\begin{array}{c}
\includegraphics[width=0.25\linewidth]{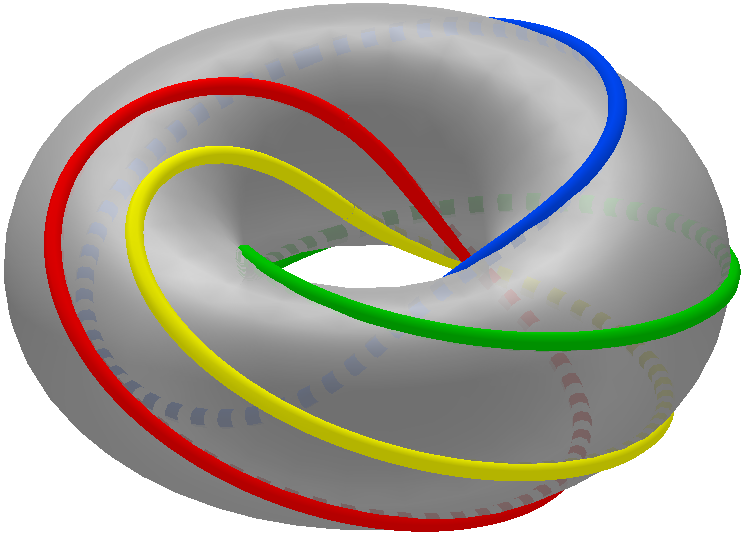}
\end{array}} \\
 \hline
\end{array} $ 
\end{center}
\caption{Diagrammatic presentation of torus links of type $T_{p,p}$}
\label{Tpplinkstable}
\end{table}

Following the discussion of section 2, the state associated with $S^3 \backslash T_{p,p}$ can be given as:
\begin{equation}
\ket{S^3  \backslash T_{p,p}} \equiv \ket{T_{p,p}} = \sum_{R_1 \, \in \, \mathcal{I}_k} \ldots \sum_{R_p \, \in \, \mathcal{I}_k} Z(S^3; T_{p,p}[R_1,\ldots,R_p]) \ket{e_{R_1},\ldots, e_{R_p}} ~,
\label{stateTpp}
\end{equation}
where $Z(S^3; T_{p,p}[R_1,\ldots,R_p])$ is the partition function of $S^3$ in presence of the link $T_{p,p}$ where the $p$ circles carry integrable representations $R_1, \ldots, R_p$ of the gauge group $G$. This partition function can be written in terms of the elements of the $\mathcal{S}$ and $\mathcal{T}$ matrices as \cite{Stevan:2010jh,Brini:2011wi}:  
\begin{equation}
Z(S^3; T_{p,p}[R_1,\ldots,R_p]) = \sum_{X,\,Y\,\in \, \mathcal{I}_k}  \frac{\mathcal{S}_{XY}^* \,\mathcal{S}_{0 Y}}{(\mathcal{S}_{0 X})^{p-1}}\, (\mathcal{T}_{YY}) \left(\prod_{i=1}^p \mathcal{S}_{R_i X}\right)  ~.
\label{Z-torus}
\end{equation}
Thus, the state $\ket{T_{p,p}}$ can be obtained by substituting the partition function \eqref{Z-torus} into the \eqref{stateTpp}. Since the entanglement properties of a state do not change under a local unitary transformation of the basis states, we can further rewrite the state by making the following transformation in the $i^{\text{th}}$ basis state:
\begin{equation}
\ket{e_{R_i}} = \sum_{Y_i} \mathcal{S}_{R_i Y_i}^{*}\ket{f_{Y_i}}  ~.
\label{}
\end{equation}
In the new basis $\{\ket{f_{Y_1}},\ket{f_{Y_2}},\ldots,\ket{f_{Y_p}}\}$, the expansion coefficients of the state $\ket{T_{p,p}}$ will change, and it will be given as:  
\begin{equation}
\ket{T_{p,p}} = \sum_{Y_1,\ldots,Y_p} \, \sum_{R_1,\ldots,R_p}\, \sum_{X,Y}   \frac{\mathcal{S}_{XY}^* \,\mathcal{S}_{0 Y}}{(\mathcal{S}_{0 X})^{p-1}}\, (\mathcal{T}_{YY}) \left(\prod_{i=1}^p \mathcal{S}_{R_i X} \mathcal{S}_{R_i Y_i}^{*}\right) \ket{f_{Y_1},\ldots, f_{Y_p}} ~.
\end{equation}
Using the symmetric and unitary property of $\mathcal{S}$ matrix, we get:
\begin{align}
\ket{T_{p,p}} &= \sum_{Y_1,\ldots,Y_p}\, \sum_{X,Y}   \frac{\mathcal{S}_{XY}^* \,\mathcal{S}_{0 Y}}{(\mathcal{S}_{0 X})^{p-1}}\, \mathcal{T}_{YY} \left(\prod_{i=1}^p \delta_{X,Y_i} \right) \ket{f_{Y_1},\ldots, f_{Y_p}} \nonumber \\
&= \sum_{X,\,Y}   \frac{\mathcal{S}_{XY}^* \,\mathcal{S}_{0 Y}}{(\mathcal{S}_{0 X})^{p-1}}\, \mathcal{T}_{YY}  \ket{f_{X},\ldots, f_{X}} ~.
\end{align}
This state can be written in a compact form as:
\begin{equation}
\ket{T_{p,p}} = \sum_{R\in \mathcal{I}_k}  \frac{(\mathcal{S}\mathcal{T}\mathcal{S}^*)_{0R}}{(\mathcal{S}_{0R})^{p-1}}\, \ket{f_{R},\ldots, f_{R}} ~.
\end{equation}
There is further computational simplification possible due to the identities involving the $\mathcal{S}$ and $\mathcal{T}$ matrices. We note the following:
\begin{alignat}{2}
\mathcal{S} \mathcal{T} \mathcal{S}^{*}  &= \mathcal{C} \mathcal{T}^{-1} \mathcal{S}^{-1} \mathcal{T}^{-1} \mathcal{S}^{-1} \mathcal{S}^* \quad && (\text{using } (\mathcal{S}\mathcal{T})^3 = \mathcal{C}) \nonumber \\
&= \mathcal{C} \mathcal{T}^{*} \mathcal{S}^{*} \mathcal{T}^{*} \mathcal{S}^{-1} \mathcal{S}^{-1} \quad && (\text{using $\mathcal{S}^{-1} = \mathcal{S}^*$ and $\mathcal{T}^{-1} = \mathcal{T}^*$}) \nonumber \\
&= \mathcal{C} \mathcal{T}^{*} \mathcal{S}^{*} \mathcal{T}^{*} \mathcal{C} \quad && (\text{using $\mathcal{S}^{2} = \mathcal{C}$ and $\mathcal{C}^{-1} = \mathcal{C}$}) ~.
\end{alignat} 
Here, the equality in the first line is due to $(\mathcal{S}\mathcal{T})^3 = \mathcal{C}$. The second equality uses $\mathcal{S}^{-1} = \mathcal{S}^*$ and $\mathcal{T}^{-1} = \mathcal{T}^*$), and the third equality is because $\mathcal{S}^{2} = \mathcal{C}$ and $\mathcal{C}^{-1} = \mathcal{C}$. Finally, we arrive at the following matrix element:
\begin{equation}
(\mathcal{S} \mathcal{T} \mathcal{S}^{*})_{0R} = (\mathcal{C} \mathcal{T}^{*} \mathcal{S}^{*} \mathcal{T}^{*} \mathcal{C})_{0R} = \mathcal{C}_{00} \mathcal{T}^{*}_{00}\, \mathcal{S}^{*}_{0\bar{R}} \mathcal{T}^{*}_{\bar{R}\bar{R}}\, \mathcal{C}_{\bar{R}R} = \mathcal{T}^{*}_{00}\, \mathcal{S}^{*}_{0\bar{R}} \mathcal{T}^{*}_{\bar{R}\bar{R}} = \mathcal{T}^{*}_{00}\, \mathcal{S}_{0R} \mathcal{T}^{*}_{\bar{R}\bar{R}} ~,
\end{equation} 
where we used $\mathcal{T}_{XY} = \mathcal{T}_{XX}\delta_{X Y}$ and $\mathcal{C}_{XY} = \delta_{X \bar{Y}}$ with $\bar{Y}$ denoting the conjugate of representation $Y$. Also $\mathcal{S}_{0R} = \mathcal{S}_{0\bar{R}} = (\mathcal{S}_{0 \bar{R}})^*$. With this, we will have:
\begin{equation}
 \ket{T_{p,p}} = \sum_{R\in \mathcal{I}_k} \mathcal{T}^{*}_{00}\mathcal{T}^{*}_{\bar{R}\bar{R}} \, (\mathcal{S}_{0R})^{2-p} \, \ket{f_{R},\ldots, f_{R}}  ~.  
\end{equation}
This state is not normalized. So we must divide it by $\bra{T_{p,p}}\ket{T_{p,p}}^{1/2}$. Using the fact that $\abs{\mathcal{T}_{RR}}^2=1$ for all representations $R$, we will obtain:
\begin{equation}
\braket{T_{p,p}} = \sum_{R\in \mathcal{I}_k} (\mathcal{S}_{0R})^{4-2p} ~.
\end{equation} 
To compute the entanglement measures associated with $\ket{T_{p,p}}$, we bi-partition the total Hilbert space into two Hilbert spaces $\mathcal{H}_A$ and $\mathcal{H}_B$ where $\mathcal{H}_A$ is a tensor product of $p_1$ Hilbert spaces and $\mathcal{H}_B$ is the tensor product of remaining $(p-p_1)$ Hilbert spaces. Tracing out $\mathcal{H}_B$ will give the reduced density matrix $\rho$ acting on $\mathcal{H}_A$ which will be a diagonal matrix of order $=\text{dim}\,\mathcal{H}_A =  \abs{\mathcal{I}_k}^{p_1}$. We find that it has only $\abs{\mathcal{I}_k}$ number of non-vanishing eigenvalues which are given as,
\begin{equation}
\lambda_{R} = \frac{(\mathcal{S}_{0R})^{4-2p}}{\braket{T_{p,p}}} \quad;\quad R \, \in \, \mathcal{I}_k ~.
\label{eigen-toruslink}
\end{equation}
The R\'enyi entropy with index $m$ is given as
\begin{equation}
\mathcal{R}_m = \frac{1}{1-m} \ln\left( \sum_{R\in \mathcal{I}_k} \lambda_{R}^m \right)  ~.
\label{RenyiGen}
\end{equation}
In the following subsections, we will express these Rényi entropies in terms of the $q_0$-deformed Witten zeta function that we defined earlier, and then investigate its large $k$ limit.
\subsection{R\'enyi entropy in terms of $q_0$-deformed Witten zeta function}
At finite $k$ values, the eigenvalues can be written as follows:
\begin{equation}
\lambda_{R} = \frac{(\mathcal{S}_{0R})^{4-2p}}{\sum_{R\in \mathcal{I}_k} (\mathcal{S}_{0R})^{4-2p}} = \frac{(\mathcal{S}_{0R}/\mathcal{S}_{00})^{4-2p}}{\sum_{R\in \mathcal{I}_k} (\mathcal{S}_{0R}/\mathcal{S}_{00})^{4-2p}} = \frac{(\text{dim}_q\,R)^{4-2p}}{\sum_{R\in \mathcal{I}_k}(\text{dim}_q\,R)^{4-2p}}  ~.
\label{eigen-qdim}
\end{equation}
Hence, the R\'enyi entropy can be written in terms of $q_0$-deformed Witten zeta functions as:
\begin{equation}
\mathcal{R}_m = \frac{1}{1-m} \ln \zeta_G(2pm-4m\,;\,q_0) - \frac{m}{1-m} \ln \zeta_G(2p-4\,;\,q_0) ~.
\label{Renyifinitek}
\end{equation}
The entanglement entropy (EE) can be simply obtained by taking the $m \to 1$ limit of the R\'enyi entropy $\mathcal{R}_m$. Since we have already obtained the $q_0 \to 1$ limit of the $q_0$-deformed Witten zeta functions in \eqref{Conj1}, we can obtain the $k \to \infty$ limit of these Rényi entropies.
\subsection{Large $k$ limit of R\'enyi entropies and Witten zeta functions}
\begin{figure}[htbp]
	\centering
		\includegraphics[width=0.95\textwidth]{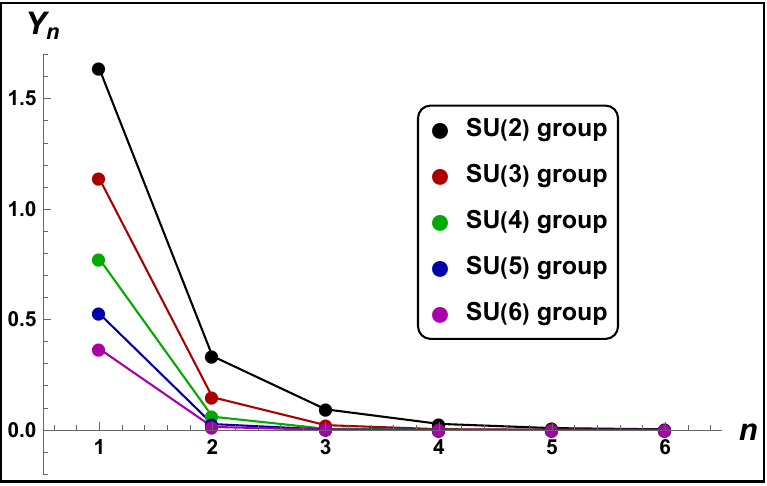}
	\caption{The plot showing the variation of $Y_n$ vs $n$ for various SU($N$) groups. The function $Y_n$ is defined in \eqref{FunctionYn}.}
	\label{PlotYn}
\end{figure}

\begin{figure}[htbp]
	\centering
		\includegraphics[width=0.95\textwidth]{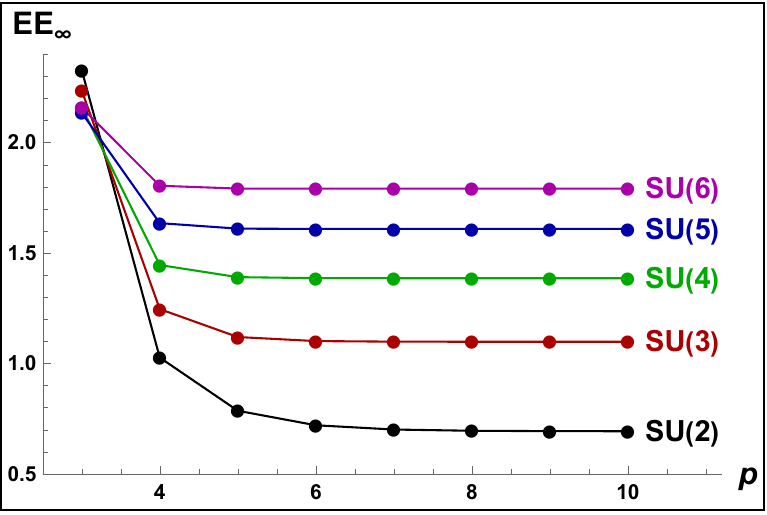}
	\caption{The plot showing the variation of $\text{EE}_{\infty}$ vs $p$ for various SU($N$) groups. The values converge to $\ln N$ respectively, as $p$ becomes large.}
	\label{PlotEEInfinity}
\end{figure}

We find that the  Rényi entropies converge in the $k \to \infty$ limit. We will denote the limiting values by the following notation:
\begin{equation*}
\mathcal{R}_m^{\infty} \equiv   \lim_{k \to \infty}\mathcal{R}_m  \quad;\quad \text{EE}_{\infty} \equiv   \lim_{k \to \infty}\text{EE} ~.
\end{equation*}
The large $k$ limit of the Rényi entropies can be obtained by using the \eqref{Conj1} and substituting it in the \eqref{Renyifinitek}. We have the following result.
\begin{mdframed}[style=sid]
\textbf{Result.} \emph{The large $k$ limit of the R\'enyi entropy for the state $\ket{T_{p,p}}$ where $p\geq3$, for the gauge group $G$, will be given as:}
\begin{equation}
\mathcal{R}_m^{\infty} = \ln |Z_G| + \frac{1}{1-m} \ln \left[ \frac{\zeta_{G}(2pm-4m)}{(\zeta_{G}(2p-4))^m} \right] ~.
\label{ResultLargekRE}
\end{equation}
\end{mdframed}
Thus, we see that the Rényi entropies converge in the $k \to \infty$ limit and can be written in terms of the Witten zeta functions. We can also obtain the limiting value of the entanglement entropy by taking the $m \to 1$ limit of the above expression, and we obtain the following result:

\begin{mdframed}[style=sid]
\textbf{Result.} \emph{The large $k$ limit of the entanglement entropy for the state $\ket{T_{p,p}}$ where $p\geq3$, for the gauge group $G$, will be given as:}
\begin{equation}
\text{EE}_{\infty} = \ln |Z_G|  + \ln  \zeta_{G}(2p-4) +(2p-4) \frac{\zeta_{G}'(2p-4)}{\zeta_{G}(2p-4)} ~,
\label{ResultLargekEE}
\end{equation}
\end{mdframed}
where $\zeta_{G}'(n)$ denotes the derivative of the Witten zeta function evaluated at $n$, that is:
\begin{equation}
\zeta_{G}'(n) \equiv \left. \frac{d \zeta_{G}(s)}{ds} \right\vert_{s=n} = -\sum_{R} \frac{\ln (\text{dim $R$})}{(\text{dim $R$})^{n}} ~.
\end{equation}
So far, these are the general results and are applicable to any Lie group. Let us study one example in detail. In the remainder of this section, we will do the numerical analysis of the $\text{EE}_{\infty}$ for the SU($N$) group. To get more insight into $\text{EE}_{\infty}$, let us rewrite it as follows. 
\begin{equation}
\text{EE}_{\infty} = \ln N  + Y_{p-2}
\end{equation}
where we have defined $Y_n$ (with $n=1,2,3,\ldots$) as:
\begin{equation}
Y_n \equiv \ln  \zeta_{\text{SU}(N)}(2n) -(2n) \frac{\zeta_{\text{SU}(N)}'(2n)}{\zeta_{\text{SU}(N)}(2n)} ~.
\label{FunctionYn}
\end{equation}
Let us analyze $Y_n$ closely. Note that although we can calculate $\zeta_{\text{SU}(N)}(2n)$ using \eqref{zetasuN}, computing $\zeta_{\text{SU}(N)}'(2n)$ by simply taking its derivative will not work because we do not have a close form expression of $C_{\text{lead}}(\mathcal{A}_{n,N})$ in $n$. Further, if we use the modular $\mathcal{S}$ matrix definition of \eqref{powersofS0RSUN} and take the derivative on both sides of \eqref{powersofS0RSUN}, we see that $d\mathcal{A}_{n,N}/dn$ will no longer be a polynomial with rational coefficients because of the presence of logarithmic terms. Nevertheless, it will be an interesting problem to study $\zeta_{\text{SU}(N)}'(2n)$ using \eqref{powersofS0RSUN} to see if new number theoretic properties of derivatives of zeta functions can be obtained. We leave this as a possible future study. Since we do not have an analytic expression of $Y_n$, we present a numerical plot of $Y_n$ vs $n$ for various SU($N$) groups in the figure \ref{PlotYn}. We notice that $Y_n$ is always positive and exponentially decreases to 0 for large values of $n$. The convergence rate becomes even faster as the rank of the group increases.
This tells us that for a given SU($N$) group, the $\text{EE}_{\infty}$ decreases as $p$ increases and saturates to the value $\ln \,N$ for large values of $p$. We present the plot of $\text{EE}_{\infty}$ vs $p$ in figure \ref{PlotEEInfinity} which illustrates this pattern.

\subsection{Large $k$ limit of entropies and volume of moduli space of flat connections}
Our definition of $q$-deformed Witten zeta functions in \eqref{qDefZetaGen} and the expression of the Rényi entropy \eqref{Renyifinitek} in terms of these $q$-deformed zeta functions lead us to express the large $k$ limit of the Rényi entropy in terms of the symplectic volume of the moduli space of flat connections on Riemann surfaces. Let us explore this in detail. 

Consider $\Sigma_g$ to be a Riemann surface with genus $g$ and $\mathcal{M}_g$ be the moduli space of flat connections on $\Sigma_g$. This moduli space is defined in \cite{Verlinde:1988sn} as follows:
\begin{equation}
\mathcal{M}_g = \text{Hom}(\pi_1(\Sigma_g), G)/G ~.
\end{equation}
This moduli space is a manifold with real dimensions $\text{dim }\mathcal{M}_g = (2g-2)\,\text{dim }G$ and admits a natural symplectic structure with the following symplectic volume:
\begin{equation}
\text{vol}(\mathcal{M}_g) = \int_{\mathcal{M}_g} \frac{\omega^d}{d!} \quad;\quad d = \frac{\text{dim }\mathcal{M}_g}{2} ~.
\end{equation} 
Here $\omega$ is the Atiyah–Bott symplectic form and the symplectic volume is the top exterior power of the Atiyah–Bott form integrated over the moduli space. Witten, in \cite{witten1991}, gave a standard technique to obtain such symplectic volumes by exploiting the Verlinde / Chern–Simons asymptotics, and can be computed as the following limit:
\begin{equation}
\text{vol}(\mathcal{M}_g) = \lim_{k \to \infty}  k^{-d} \, \text{dim }\mathcal{H}_{\Sigma_g} ~,
\label{VolumeWitten}
\end{equation} 
where $\mathcal{H}_{\Sigma_g}$ is the Hilbert space associated with the Riemann surface $\Sigma_g$. According to the Verlinde formula, the dimension of this Hilbert space is given as:
\begin{equation}
\text{dim }\mathcal{H}_{\Sigma_g} = \sum_{R\, \in \, \mathcal{I}_k} \frac{1}{(\mathcal{S}_{0R})^{2g-2}}
\end{equation} 
We can write this as follows: 
\begin{equation}
\text{dim }\mathcal{H}_{\Sigma_g} = (\mathcal{S}_{00})^{2-2g} \sum_{R\, \in \, \mathcal{I}_k}  \frac{1}{(\text{dim}_{q_0}\,R)^{2g-2}}
\end{equation} 
Now, we invoke our result \eqref{Conj1} and the $\mathcal{S}_{00}$ asymptotic \eqref{S00AsympGen} to extract the limit and get the following expression of the volume:
\begin{equation}
\boxed{ \text{vol}(\mathcal{M}_g) = \alpha(G)^{2-2g}\, |Z_G|\, \zeta_{G}(2g-2) } ~.
\label{VolumeGeneral}
\end{equation} 
Here $\alpha(G)$ is the group-dependent constant as mentioned in \eqref{S00AsympGen}. For example, for the SU($N$) group, the expression of volume will become:
\begin{equation}
\boxed{ \text{vol}(\mathcal{M}_g) = \frac{ N^g }{G(N+1)^{2g-2}}\, \frac{\zeta_{\text{SU}(N)}(2g-2)}{(2\pi)^{(g-1)N(N-1)}} } ~.
\label{VolumeSUN}
\end{equation} 
For other classical and exceptional Lie groups, the expression of $\alpha(G)$ can be obtained from the $\mathcal{S}_{00}$ asymptotics given in the Appendix \ref{app2} and the symplectic volume can be calculated using \eqref{VolumeGeneral}. 

Now that we have an expression for the volume of the moduli spaces in terms of Witten zeta functions, we can recast our large $k$ results of Rényi entropy of \eqref{ResultLargekRE} in terms of the volume of the moduli spaces of flat connections. For this, we notice that: 
\begin{equation}
\frac{\text{vol}(\mathcal{M}_{pm-2m+1})}{[\text{vol}(\mathcal{M}_{p-1})]^m} = |Z_G|^{1-m} \, \frac{\zeta_{G}(2pm-4m)}{[\zeta_{G}(2p-4)]^m} ~.
\end{equation}
With this identification, we are ready with our result.

\begin{mdframed}[style=sid]
\textbf{Result:} 
\emph{The large $k$ limit of the R\'enyi entropy for the state $\ket{T_{p,p}}$ where $p\geq3$, for the gauge group $G$, will be given as:}
\begin{equation}
\mathcal{R}_m^{\infty} = \frac{1}{1-m} \ln \left[ \frac{\text{vol}(\mathcal{M}_{pm-2m+1})}{\text{vol}(\mathcal{M}_{p-1})^m} \right] ~.
\label{LargekREandVolume}
\end{equation}
\end{mdframed}
This formula closely resembles the structure of a Rényi entropy and is reminiscent of the replica-trick construction. Although we do not have a mathematically rigorous explanation of the above result from a topological point of view, we can intuitively see why a genus with a combination $(pm-2m+1)$ enters into this formula. Recall from \eqref{VolumeWitten} that $\text{dim }\mathcal{H}_{\Sigma_g} \sim k^d \, \text{vol}(\mathcal{M}_g)$ with $d=(g-1)\,\text{dim }G$. As a result, in the large $k$ limit, we will have:
\begin{equation}
\frac{\text{dim }\mathcal{H}_{\Sigma_{pm-2m+1}}}{[\text{dim }\mathcal{H}_{\Sigma_{p-1}}]^m} \,\, \longrightarrow \,\,  \frac{\text{vol}(\mathcal{M}_{pm-2m+1})}{[\text{vol}(\mathcal{M}_{p-1})]^m} ~.
\end{equation}
We also know that $Z(\Sigma_g \times S^1) = \text{dim }\mathcal{H}_{\Sigma_g}$. Therefore, we have:
\begin{equation}
\mathcal{R}_m^{\infty} = \frac{1}{1-m} \ln \left[ \frac{Z(\Sigma_{pm-2m+1} \times S^1)}{[Z(\Sigma_{p-1} \times S^1)]^m} \right] ~.
\end{equation}
We would like to make an analogy of this result with the result of (1+1)-dimensional Chern-Simons theory that was obtained in \cite{Dwivedi:2020jyx}. In this reference, the entanglement structure of the quantum states $\ket{\Sigma_{g,n}}$ corresponding to Riemann surfaces of genus $g$ with $n$ disks removed (or $n$ boundaries) was studied in the context of (1+1)-dimensional Chern-Simons theory. These states live in the $n$-fold tensor product of $\mathcal{H}_{S^1}$. According to the results of \cite{Dwivedi:2020jyx}, if we consider $S^2$ with $p$ disks removed, i.e., if we take the state $\ket{\Sigma_{0,p}}$, then the Rényi entropy for this state using the Replica trick was obtained to be \cite{Dwivedi:2020jyx}:  
\begin{equation}
\mathcal{R}_m^{(1+1)} = \frac{1}{1-m} \ln \left[ \frac{Z(\Sigma_{pm-2m+1})}{[Z(\Sigma_{p-1}]^m} \right] ~,
\label{2dCSResult}
\end{equation}
where $Z(\Sigma_g)$ is the (1+1) Chern-Simons partition function of the closed 2d Riemann surface $\Sigma_g$. On the other hand, in the (2+1)-dimensional picture, we have:  
\begin{equation}
\mathcal{R}_m^{(2+1)}  \,\, \xrightarrow[]{k \to \infty} \,\, \frac{1}{1-m} \ln \left[ \frac{Z(\Sigma_{pm-2m+1} \times S^1)}{[Z(\Sigma_{p-1} \times S^1)]^m} \right] ~.
\label{3dCSLargekResult}
\end{equation}
The similarity between \eqref{2dCSResult} and \eqref{3dCSLargekResult} is profound, though the exact machinery of why this is happening in the case of $\ket{T_{p,p}}$ associated with $S^3/T_{p,p}$ is not clear to us, and we leave this question for a future investigation. 

From the Rényi entropy result \eqref{LargekREandVolume}, we can also write the large $k$ limit of the entanglement entropy in terms of the volume by taking the $m \to 1$ limit.\\
\begin{mdframed}[style=sid]
\textbf{Result:} 
\emph{The large $k$ limit of the entanglement entropy for the state $\ket{T_{p,p}}$ where $p\geq3$, for the gauge group $G$, will be given as:}
\begin{equation}
\text{EE}_{\infty} = \ln \text{vol}(\mathcal{M}_{p-1})  - (p-2) \frac{\text{vol}'(\mathcal{M}_{p-1})}{\text{vol}(\mathcal{M}_{p-1})} ~,
\label{LargekEEandVolume}
\end{equation}
\end{mdframed}
where our notation is:
\begin{equation}
\text{vol}'(\mathcal{M}_{p-1}) \equiv \left. \frac{d \, \text{vol}(\mathcal{M}_g)}{dg} \right\vert_{g=p-1} ~,
\end{equation}
which calculates the derivative of the symplectic volume of the moduli space with respect to the genus evaluated at $g=(p-1)$. We can also write $\text{EE}_{\infty}$ as a derivative of an overall function of $g$ evaluated at $g=(p-1)$ as given below:
\begin{equation}
\text{EE}_{\infty} = \frac{d}{dg}\left[(p-2)^2\,\frac{\ln \text{vol}(\mathcal{M}_{g})}{1-g} \right]  ~,
\end{equation}
where the derivative is evaluated at genus $g=(p-1)$. We can write this as: 
\begin{equation}
\boxed{\text{EE}_{\infty} = \left. \frac{d\, \Phi(g)}{dg} \right\vert_{g=p-1}}   ~.
\label{EEGeometry}
\end{equation}
Geometrically, it means that the $\text{EE}_{\infty}$ measures the rate of change of the `normalized logarithmic volume' $\Phi(g)$ with genus $g$, at the point $g=(p-1)$, where the function $\Phi(g)$ is given as: 
\begin{equation}
\Phi(g) =  (p-2)^2\,\frac{\ln \text{vol}(\mathcal{M}_{g})}{1-g} ~.
\end{equation}
One can also obtain the large $k$ limit of the minimum Rényi entropy by taking the $m \to \infty$ limit of the \eqref{ResultLargekRE}. Note that as $s\to \infty$, we will have $\zeta_G(s) \to 1$. This happens because $(\text{dim}\,R)^{-s} \to 0$, unless $R$ is the trivial representation for which the dimension is 1. Hence $\ln \zeta_G(2pm-4m) \to 0$ as $m \to \infty$. With this realization, it is now trivial to take the $m \to \infty$ limit of the \eqref{ResultLargekRE}, which simply gives $\ln |Z_G| + \ln \zeta_{G}(2p-4)$. We can then use \eqref{VolumeGeneral} to convert the zeta function into the volume term, and we will have the following result:\\ \\
\textbf{Result:} 
\emph{The large $k$ limit of the minimum Rényi entropy for the state $\ket{T_{p,p}}$ where $p\geq3$, for the gauge group $G$, will be given as:}
\begin{equation}
\mathcal{R}_{\text{min}}^{\infty} = \ln \left[\alpha(G)^{2p-4}\, \text{vol}(\mathcal{M}_{p-1}) \right] ~.
\label{LargekREminandVolume}
\end{equation}
Thus, we see that the minimum Rényi entropy cleanly captures the volume of the moduli space, with the following expression: 
\begin{equation}
\boxed{ \exp[\mathcal{R}_{\text{min}}^{\infty}] = \alpha(G)^{2p-4}\, \text{vol}(\mathcal{M}_{p-1}) } ~.
\label{expREmin}
\end{equation}
\section{\label{sec5}Conclusion}
In this work, we delve into topological entanglement and its connection with number theory. The case study here is a very small class of links, which are torus links of type $T_{p,p}$, which consists of $p$ circles such that every circle is linked once with the other circles. The Chern-Simons path integral is used to obtain the quantum state $\ket{T_{p,p}}$ associated with the link complement manifold $S^3 \backslash T_{p,p}$. We compute the R\'enyi entropies of this state for various Lie groups and Chern-Simons levels. We see from \eqref{Renyifinitek} that these R\'enyi entropies can be written in terms of the $q$-deformed Witten zeta function, which we introduce in \eqref{qDefZetaGen}. The parameter $q$ for this case must be a specific root of unity, denoted by $q_0$ and given in \eqref{qRootofUnity}.

Our primary focus was to find what happens to the entropies in the semiclassical limit of $k \to \infty$. To do this, we first analyzed the $q_0$-deformed Witten zeta functions and studied their values in the limit of $q_0 \to 1$ for all the classical and exceptional Lie groups. We find in \eqref{Conj1} that this limit will give $|Z_G|$ times the classical Witten zeta function associated with that particular group. This result itself is interesting from the number theory point of view, as it provides an alternate method of determining the Witten zeta functions. We have also checked (without going into the intricacies, if any) that \eqref{Conj1} is also true for any complex number $s$ with $\text{Re}(s) \geq 2$. Apart from this, we hope that this result may also be useful to number theorists in gaining more insight into the zeta functions.

Using the result \eqref{Conj1}, we find that the R\'enyi entropies for the state $\ket{T_{p,p}}$ converge to a finite number in the limit $k \to \infty$ for any gauge group. This limiting value can be written in terms of the Witten zeta function associated with that particular group, which is evident from \eqref{ResultLargekRE}. Also, the entanglement entropy converges to a finite number in the limit $k \to \infty$, and its limiting value can be written in terms of the Witten zeta function and its derivatives as given in \eqref{ResultLargekEE}. 

It is clear from the results that the torus links provide an elegant topological setup where the entanglement measures have interesting number-theoretic properties. In this work, we restricted ourselves to torus links $T_{p,p}$ for which the entropy turned out to be a function of quantum dimensions. However, it will be interesting to investigate other classes of torus links where the entanglement spectrum will be a complicated function of modular $\mathcal{S}$ and $\mathcal{T}$ matrices. For example, see \cite{Dwivedi:2020rlo, Caputa:2024qkk} for the large $k$ asymptotics of R\'enyi entropies associated with $T_{p,pn}$ torus links for the SU(2) group. Thus, it will be worthwhile to study the large $k$ topological entanglement for these torus links for generic Lie groups. We believe that the results of such studies can be later developed into exact mathematical statements and that these observations may bring some new insight into modern mathematical physics applications. 

Although the theme of this work was to analyze the $q$-deformed Witten zeta function when $q$ was a specific root of unity, we have also studied one example where $q$ is not a root of unity. In Appendix \ref{appB}, we have studied $\zeta_{\text{SU}(2)}(2;q)$ and obtained its value by analytically continuing $q$ to complex values such that $|q| \neq 1$. We first show that $\zeta_{\text{SU}(2)}(2;q)$ can be written in terms of the $q$-polygamma function for positive real values of $q$. The analytic continuation of $\zeta_{\text{SU}(2)}(2;q)$ is then inherited from the analytic continuation of $q$-polygamma functions resulting in \eqref{zetawhenqneq1}. We have verified the correctness of \eqref{zetawhenqneq1} by doing numerical calculations for various complex values of $q$. It would be interesting to explore $\zeta_{G}(s;q)$ for other Lie groups as well and explore it by analytically continuing the values of $s$ and $q$. We leave it as a future problem.

Understanding the semiclassical ($k \to \infty$) behavior of the entanglement measures in (2+1)-dimensional Chern-Simons link states is of particular interest because of its conceptual similarity to the philosophy behind the volume conjecture. In the volume conjecture, the asymptotic limits of quantum invariants are expected to encode classical geometric quantities associated with the link complement. Therefore, one can expect that the large $k$ limit of the entanglement measures may also encode some geometrical information. However, since the entanglement measures of link states are weighted sums over various quantum invariants, it is not obvious which geometry naturally emerges in the large $k$ limit. In the present work, we identify an instance in which such a geometric interpretation becomes apparent. For the family of torus links $T_{p,p}$, we show that the large $k$ limit of the Rényi entropies of the associated link states can be expressed in terms of the symplectic volumes of moduli spaces of flat connections on Riemann surfaces. In this sense, the entanglement structure of these link states admits a direct interpretation in terms of classical phase-space volumes arising from the moduli space of flat connections. This is clear from \eqref{LargekREandVolume}, \eqref{LargekEEandVolume}, and \eqref{LargekREminandVolume}. The $\mathcal{R}_m^{\infty}$ in \eqref{LargekREandVolume} involves the ratio of two volumes. The $\text{EE}_{\infty}$ in \eqref{LargekEEandVolume} involves a derivative of the volume. In fact, as mentioned in \eqref{EEGeometry}, the  $\text{EE}_{\infty}$ can be geometrically viewed as the rate of change of some kind of `normalized logarithmic volume' with respect to genus, evaluated at the point $g=(p-1)$. The minimum Rényi entropy has a direct and clean interpretation in terms of the volume, which is evident from \eqref{expREmin}. These results tell us that the classical phase-space volumes arising from the moduli space of flat connections are the objects of interest, at least for this family of torus links.  

Although the torus links $T_{p,p}$ represent a particularly tractable class of examples, the appearance of moduli-space volumes in the semiclassical limit suggests a broader geometric mechanism underlying entanglement in topological quantum field theories. It would be interesting to investigate whether analogous relations hold for more general torus links $T_{p,q}$, where additional structure may arise from the richer topology of the link complement. An even more intriguing direction is the case of the notorious class of hyperbolic links, whose complements admit hyperbolic geometry, and one might hope for connections between entanglement measures and hyperbolic volumes, in closer spirit to the volume conjecture. Establishing such relations could provide new insight into how classical geometric structures emerge from entanglement properties of quantum states in Chern–Simons theory.
\vspace{0.3in}

\noindent \textbf{Acknowledgment}\\
The author is supported by the “SERB Start-Up Research Grant SRG/2023/001023”.
\appendix

\section{Analytic continuation of $\zeta_{\text{SU(2)}}(2;q)$} \label{appB}
Here we study the simplest case of $\zeta_{\text{SU(2)}}(2;q)$ and obtain its analytic continuation for $|q| \neq 1$. Using our definition \eqref{qDefZetaGen}, we will have:
\begin{align}
\zeta_{\text{SU(2)}}(2;q) &= \sum_{a=0}^\infty \left(\frac{q^{1/2}-q^{-1/2}}{q^{(a+1)/2}-q^{-(a+1)/2}} \right)^2 \nonumber \\ 
&= \sum_{a=0}^{\infty} \frac{(1-q)^2\, q^a}{\left(1-q^{a+1}\right)^2}     ~.
\end{align}
When $|q| \neq 1$, we will show that this sum can be expressed in terms of the $q$-polygamma function. For this, we first define the $q$-digamma function, which is the $q$-deformed version of the digamma function. It is formally defined as:
\begin{equation}
\psi_q(z) = \frac{d}{dz} \ln \Gamma_q(z) ~,
\end{equation}
where $\Gamma_q(z)$ is the $q$-gamma function. The $q$-digamma function can be defined for positive real values of $q$ as follows \cite{mansour2009some}:
\begin{equation}
 \psi_q(z) = 
\begin{cases}
  -\ln(1-q) + \ln q \displaystyle \sum_{n=1}^{\infty} \dfrac{q^{nz}}{1-q^n}  \\
  -\ln(q-1) + \ln q \left(z-\dfrac{1}{2} - \displaystyle \sum_{n=1}^{\infty} \dfrac{q^{-nz}}{1-q^{-n}} \right)
\end{cases}
\label{q-digamma-defn}
\end{equation}
The upper case in the above equation is valid for $0<q<1$ while the second case is valid for $q>1$. Note that these definitions can be analytically continued to complex values of $q$ through the analytic continuation of $\Gamma_q(z)$. Thus, the above two regions will be analytically continued to $|q|<1$ and $|q|>1$, and we can use the respective definitions (where we chose the principal branch of the log function). Here the domain is $z \in \mathbb{C}$ such that $z\neq 0,-1,-2,-3,\ldots$. Note that the $q$-digamma function is not defined for $|q|=1$, and we have already discussed the $|q|=1$ case in section \ref{sec3}. Taking the derivatives of $\psi_q(z)$ will give us the $q$-polygamma functions which are defined as:
\begin{equation}
\psi^{(m)}_q(z) = \frac{d^m}{dz^m} \psi_q(z) = \frac{d^{m+1}}{dz^{m+1}} \ln \Gamma_q(z) ~.
\end{equation}
Using the series form written in \eqref{q-digamma-defn}, we can write the $q$-polygamma functions as:
\begin{equation}
 \psi^{(m)}_q(z) = 
\begin{cases}
  (\ln q)^{m+1} \displaystyle \sum_{n=1}^{\infty} \dfrac{n^m\,q^{nz}}{1-q^n} \\[0.5cm]
  \ln q \left(\dfrac{d^m z}{dz^m}\right) + (-\ln q)^{m+1} \displaystyle \sum_{n=1}^{\infty} \dfrac{n^m\,q^{-nz}}{1-q^{-n}} 
\end{cases}
\label{q-polygamma-defn}
\end{equation}
where the first line is valid for $0<q<1$ while the second is valid for $q>1$. Just like before, these definitions can be analytically continued to  $|q|<1$ and $|q|>1$, respectively. Let us now show how the $\zeta_{\text{SU(2)}}(2;q)$ can be written in terms of $\psi^{(m)}_q(z)$. For $|q|<1$, and $a>0$, we note that the following Taylor series converges:
\begin{equation}
\frac{1}{(1-q^{a+1})^2} = \sum_{n=0}^{\infty} (n+1)\, q^{n(a+1)} ~.
\end{equation}
So for $|q|<1$, we will have the following summation:
\begin{align}
\zeta_{\text{SU(2)}}(2;q) &=  \sum_{a=0}^{\infty} \sum_{n=0}^{\infty} (1-q)^2 (n+1)\,q^a \, q^{n(a+1)} \nonumber \\
&=(1-q)^2\sum_{n=0}^{\infty} \frac{(n+1)\,q^n}{1-q^{n+1}} \nonumber \\ 
&=(1-q)^2\sum_{n=1}^{\infty} \frac{n\,q^{n-1}}{1-q^n} ~.
\end{align}
We can immediately see from \eqref{q-polygamma-defn} that:
\begin{equation}
\zeta_{\text{SU(2)}}(2;q) = \frac{(1-q)^2}{q\,(\ln q)^2}\, \psi_q^{(1)}(1)  ~.
\end{equation}
Let us now consider the case $|q|>1$. For $a>0$, we note that the following Laurent series converges:
\begin{equation}
\frac{1}{(1-q^{a+1})^2} = \sum_{n=0}^{\infty} \frac{n}{q^{(n+1)(a+1)}} =  \sum_{n=1}^{\infty} \frac{n}{q^{(n+1)(a+1)}} ~. \nonumber
\end{equation}
Thus, we get for $|q|>1$ the following summation:
\begin{align}
\zeta_{\text{SU(2)}}(2;q) &=  \sum_{a=0}^{\infty} \sum_{n=1}^{\infty} (1-q)^2\frac{n\,q^a}{q^{(n+1)(a+1)}} \nonumber \\
&=\frac{(1-q)^2}{q}\sum_{n=1}^{\infty} \frac{n\,q^{-n}}{1-q^{-n}} ~.
\end{align}
Comparing it with \eqref{q-polygamma-defn}, we see that:
\begin{equation}
\zeta_{\text{SU(2)}}(2;q) = \frac{(q-1)^2}{q\,(\ln q)^2}\left( \psi_q^{(1)}(1) - \ln q\right)  ~.
\end{equation}
Thus, for $|q|\neq 1$, we will have the following values:
\begin{equation}
 \zeta_{\text{SU(2)}}(2;q) = 
\begin{cases}
 \displaystyle \frac{(1-q)^2}{q\,(\ln q)^2}\, \psi_q^{(1)}(1) \quad,\quad & \text{if } |q|<1 \\[0.5cm]
 \displaystyle \frac{(1-q)^2}{q\,(\ln q)^2}\, \psi_q^{(1)}(1) - \frac{(1-q)^2}{q\ln q} \quad,\quad & \text{if } |q|>1
\end{cases} ~.
\label{zetawhenqneq1}
\end{equation}
This completes the analytic continuation for $|q|\neq 1$.
\section{Results for the remaining Lie groups} \label{app2}
We have already presented the results for the SU($N$) group in section 3. Here we will present the results for the remaining Lie groups.
\subsection{SO($2N+1$) group}
An irreducible representation for SO($2N+1$) group is denoted as $R = [a_1,a_2,\ldots,a_{N}]$, where $a_i$ are the Dynkin labels of the highest weight associated with the representation $R$. For a given level $k$, the set of integrable representations is given by:
\begin{equation}
\mathcal{I}_k =\begin{cases}\left\{[a_1, \ldots, a_{N}] \, : \, a_{\mu} \geq 0 \,\,\, ; \,\,\, a_1+2a_2+\ldots+2a_{N-1}+a_N \leq k \right\}  & ,\, \quad N>1 \\[0.4cm] \left\{[a_1] \,:\, 0 \leq a_1 \leq 2k \right\} & ,\, \quad N=1 \end{cases}
  ~.
\end{equation} 
The matrix element $\mathcal{S}_{0R}$ for an integrable representation $R$ is given as\footnote{Note that as pointed out in \cite{Mlawer:1990uv}, the formula \eqref{S-matrix-BN} for $N=1$, gives the $\mathcal{S}$ matrix element of SO(3)$_{2k}$.}:
\begin{equation}
\mathcal{S}_{0R} = \begin{cases} (-1)^{N(N-1)/2} \dfrac{2^{N-1}}{(k+2N-1)^{N/2}}\, \text{det}\, M  & ,\, \quad N>1 \\ \dfrac{1}{\sqrt{k+1}}\, \sin \left[\dfrac{\pi(2\ell_1+1)}{2(k+1)}\right] & ,\, \quad N=1 \end{cases}~,
\label{S-matrix-BN}
\end{equation} 
where $M$ is an order $N$ matrix with elements given as,
\begin{equation}
M_{ij} = \sin\left(\pi \frac{(2N-2i+1)(2N+2\ell_j-2j+1)}{2(k+2N-1)} \right)
\end{equation}
with $1 \leq i,j \leq N$ and the half-integers $\ell_i$ are given as,
\begin{equation}
\ell_i = \begin{cases} a_{N}/2 + \sum_{j=i}^{N-1} a_j & ,\, \quad 1\leq i \leq N-1 \\ a_N/2 & ,\, \quad i=N \end{cases}  ~.
\end{equation}
The large $k$ expansion of $\mathcal{S}_{00}$ is given as:
\begin{equation}
\mathcal{S}_{00} \sim \frac{2^{(N-1)(2N+1)}\, \pi^{N^2}\, G(N+1)\, G(N+\frac{3}{2})}{\pi^{(N+1)/2}\,G(\frac{1}{2})}\, \frac{1}{k^{(2N^2+N)/2}} \,+ \cdots ~,
\end{equation}
where the Barnes $G$-function can be computed at half-integers using the following:
\begin{equation}
G(z+1) = \Gamma(z) G(z) \quad;\quad G(1/2) = 2^{1/24}\, e^{3\zeta'(-1)/2}\, \pi^{-1/4} ~.
\end{equation}
Following the earlier discussion, we would expect the following large $k$ behavior:
\begin{equation}
\sum_{R\, \in \, \mathcal{I}_k} (\mathcal{S}_{0R})^{-s} \, \sim \, \beta_s\, k^{s(2N^2+N)/2} ~,
\label{betaSO2N+1}
\end{equation}
where $\beta_s$ is some constant. Thus, the equation \eqref{zetageneral} will give the following formula of the Witten zeta function for the $G=$SO($2N+1$) group for $\text{Re}(s) \geq 2$: 
\begin{equation}
\boxed{\zeta_{\text{SO}(2N+1)}(s) = \frac{\pi^{sN^2}}{2} \left( \frac{2^{(N-1)(2N+1)}\, G(N+1)\, G(N+\frac{3}{2})}{\pi^{(N+1)/2}\,G(\frac{1}{2})} \right)^s \beta_s}
 ~.
\label{ConjSO2N+1}
\end{equation}
The term inside the parentheses on the RHS of \eqref{ConjSO2N+1} is a positive integer for $N\geq2$. For arbitrary $s$, one can numerically obtain $\beta_s$ and use the above equation to calculate the numerical values of $\zeta_{\text{SO}(2N+1)}(s)$ up to desired precision.

For $s= 2n$, where $n \geq 1$ is an integer, we can calculate the RHS of \eqref{ConjSO2N+1} by explicitly computing the constants $\beta_{2n}$ which turn out to be rational numbers. To do this, we first notice that: 
\begin{equation}
\sum_{R\, \in \, \mathcal{I}_k} \frac{1}{(\mathcal{S}_{0R})^{2n}} = \begin{cases}
\mathcal{B}_{n,\,N}^{\text{even}}(k)  & ,\, \quad k=\text{even} \\ \mathcal{B}_{n,\,N}^{\text{odd}}(k) & ,\, \quad k=\text{odd} \end{cases} ~,
\label{powersofS0RBN}
\end{equation}
where $\mathcal{B}_{n,\,N}^{\text{even}}$ and $\mathcal{B}_{n,\,N}^{\text{even}}$ are polynomials of degree $nN(2N+1)$ in the variable $k$ with rational coefficients.\footnote{Recall that the level $k$ for SO(3) group is even, so we only consider $\mathcal{B}_{n,\,1}^{\text{even}}$ for the SO(3) group.} The leading order coefficient (i.e., the coefficient of $k^{nN(2N+1)}$ term) of the polynomials $\mathcal{B}_{n,\,N}^{\text{even}}$ and $\mathcal{B}_{n,\,N}^{\text{odd}}$ are equal which we denote as $C_{\text{lead}}(\mathcal{B}_{n,\,N})$. As a result, the constant $\beta_{2n}$ will be a rational number and can be read off as: 
\begin{equation}
\beta_{2n}  = C_{\text{lead}}(\mathcal{B}_{n,\,N}) ~.
\end{equation}
Thus, the zeta function for SO($2N+1$) group at positive even integers can be given as:
\begin{equation}
\boxed{\zeta_{\text{SO}(2N+1)}(2n) = \pi^{2nN^2} \left( \frac{2^{2n(N-1)(2N+1)}\, G(N+1)^{2n}\, G(N+\frac{3}{2})^{2n}}{2\,\pi^{n(N+1)}\,G(\frac{1}{2})^{2n}} \right)\, C_{\text{lead}}(\mathcal{B}_{n,\,N})} ~.
\label{zetaBN}
\end{equation}
We can obtain the leading order coefficients $C_{\text{lead}}(\mathcal{B}_{n,\,N})$ for any given $n$ and $N$ using the technique discussed for the SU($N$) group. This will enable us to compute $\zeta_{\text{SO}(2N+1)}(2n)$ via the \eqref{zetaBN}. We present the table \ref{ZetaBNTable} where we tabulate the values of $\zeta_{\text{SO}(2N+1)}(2n)$ for some low values of $n$ and $N$.
\captionsetup{width=16cm}
\begin{longtable}{|c||L{15cm}|}
			\caption[Values of Witten zeta function at positive even integers for SO($2N+1$) group]{Values of zeta functions at positive even integers for SO($2N+1$) group calculated using \eqref{zetaBN}.}
			\label{ZetaBNTable} \\ \hline \rowcolor{Gray}
			$n$ & $\zeta_{\text{SO}(3)}(2n)$ comes out to be the same as $\zeta_{\text{SU}(2)}(2n)$ given in Table \ref{LeadingSUNTable} \\ \hhline{|=#=|}
   \rowcolor{Gray}
			$n$ & $\zeta_{\text{SO}(5)}(2n)/\pi^{8n}$ \\ \hline
   \rowcolor{Redd}
			$1$ & \footnotesize{$1/(2^4\cdot 3^1\cdot 5^2\cdot 7^1)$} \\ \hline \rowcolor{Bluee}
			$2$ & \footnotesize{$(479^1)/(2^6\cdot 3^2\cdot 5^4\cdot 7^2\cdot 11^1\cdot 13^1\cdot 17^1)$} \\ \hline \rowcolor{Redd}
			$3$ & \footnotesize{$(43^1\cdot 19309^1)/(2^6\cdot 3^5\cdot 5^3\cdot 7^4\cdot 11^2\cdot 13^2\cdot 17^1\cdot 19^1\cdot 23^1)$} \\ \hline \rowcolor{Bluee}
			$4$ & \footnotesize{$(241^1\cdot 64009163^1)/(2^{10}\cdot 3^6\cdot 5^8\cdot 7^4\cdot 11^2\cdot 13^1\cdot 17^2\cdot 19^1\cdot 23^1\cdot 29^1\cdot 31^1)$} \\ \hline \rowcolor{Redd}
			$5$ & \footnotesize{$(457^1\cdot 254030104653331^1)/(2^{10}\cdot 3^8\cdot 5^{10}\cdot 7^5\cdot 11^4\cdot 13^2\cdot 17^2\cdot 19^2\cdot 23^1\cdot 29^1\cdot 31^1\cdot 37^1\cdot 41^1)$} \\ \hline \rowcolor{Bluee}
			$6$ & \footnotesize{$(347^1\cdot 388237^1\cdot 4066121^1\cdot 162492760783^1)/(2^{12}\cdot 3^{11}\cdot 5^{12}\cdot 7^8\cdot 11^4\cdot 13^4\cdot 17^2\cdot 19^2\cdot 23^2\cdot 29^1\cdot 31^1\cdot 37^1\cdot 41^1\cdot 43^1\cdot 47^1)$} \\ \hline \rowcolor{Redd}
			$7$ & \footnotesize{$(61^1\cdot 127^1\cdot 281^1\cdot 5464238483528872797439^1)/(2^{12}\cdot 3^{12}\cdot 5^{14}\cdot 7^8\cdot 11^5\cdot 13^4\cdot 17^2\cdot 19^2\cdot 23^2\cdot 29^2\cdot 31^1\cdot 37^1\cdot 41^1\cdot 43^1\cdot 47^1\cdot 53^1)$} \\ \hline \rowcolor{Bluee}
			$8$ & \footnotesize{$(10177^1\cdot 62278508825405753685348611625373^1)/(2^{18}\cdot 3^{12}\cdot 5^{15}\cdot 7^{10}\cdot 11^5\cdot 13^4\cdot 17^4\cdot 19^2\cdot 23^2\cdot 29^2\cdot 31^2\cdot 37^1\cdot 41^1\cdot 43^1\cdot 47^1\cdot 53^1\cdot 59^1\cdot 61^1)$} \\ \hline \rowcolor{Redd}
			$9$ & \footnotesize{$(52447819^1\cdot 1341387759113^1\cdot 529792839498988256125642914121^1)/(2^{18}\cdot 3^{17}\cdot 5^{18}\cdot 7^{12}\cdot 11^6\cdot 13^6\cdot 17^4\cdot 19^4\cdot 23^1\cdot 29^2\cdot 31^2\cdot 37^2\cdot 41^1\cdot 43^1\cdot 47^1\cdot 53^1\cdot 59^1\cdot 61^1\cdot 67^1\cdot 71^1\cdot 73^1)$} \\ \hline \rowcolor{Bluee}
			$10$ & \footnotesize{$(397^1\cdot 4602863^1\cdot 3070587686057^1\cdot 1485486855888271^1\cdot 2407308907771100657^1)/(2^{20}\cdot 3^{16}\cdot 5^{20}\cdot 7^{12}\cdot 11^8\cdot 13^6\cdot 17^5\cdot 19^4\cdot 23^2\cdot 29^2\cdot 31^2\cdot 37^2\cdot 41^2\cdot 43^1\cdot 47^1\cdot 53^1\cdot 59^1\cdot 61^1\cdot 67^1\cdot 71^1\cdot 73^1\cdot 79^1)$} \\ \hhline{|=#=|} \rowcolor{Gray}
			$n$ & $\zeta_{\text{SO}(7)}(2n)/\pi^{18n}$ \\ \hline \rowcolor{Redd}
			$1$ & \footnotesize{$(19^1)/(2^4\cdot 3^5\cdot 5^1\cdot 7^3\cdot 11^1\cdot 13^1\cdot 17^1)$} \\ \hline \rowcolor{Bluee}
			$2$ & \footnotesize{$(307^1\cdot 267743941589^1)/(2^6\cdot 3^{10}\cdot 5^4\cdot 7^6\cdot 11^3\cdot 13^3\cdot 17^2\cdot 19^2\cdot 23^1\cdot 29^1\cdot 31^1\cdot 37^1)$}  \\ \hline
\end{longtable}

\subsection{Sp($2N$) group}
An irreducible representation for Sp($2N$) group is denoted as $R = [a_1,a_2,\ldots,a_{N}]$, where $a_i$ are the Dynkin labels of the highest weight associated with the representation $R$. For a given level $k$, the set of integrable representations is given by:
\begin{equation}
\mathcal{I}_k = \left\{[a_1, \ldots, a_{N}] \, : \, a_{\mu} \geq 0 \,\,\, ; \,\,\, a_1+a_2+\ldots+a_N \leq k \right\} 
  ~.
\end{equation} 
The matrix element $\mathcal{S}_{0R}$ for an integrable representation $R$ is given as:
\begin{equation}
\mathcal{S}_{0R} = (-1)^{N(N-1)/2} \dfrac{2^{N/2}}{(k+N+1)^{N/2}}\, \text{det}\, M ~,
\label{S-matrix-CN}
\end{equation} 
where $M$ is a $N\times N$ matrix with elements given as,
\begin{equation}
M_{ij} = \sin\left(\pi \frac{(N+1-i)(N+\ell_j-j+1)}{k+N+1} \right)
\end{equation}
with $1 \leq i,j \leq N$ and the integers $\ell_i$ are given as,
\begin{equation}
\ell_i = \sum_{j=i}^{N} a_j ~.
\end{equation}
The large $k$ expansion of $\mathcal{S}_{00}$ is given as:
\begin{equation}
\mathcal{S}_{00} \sim \frac{2^{N(2N+1)/2}\, \pi^{N^2}\, G(N+1)\, G(N+\frac{3}{2})}{\pi^{(N+1)/2}\,G(\frac{1}{2})}\, \frac{1}{k^{(2N^2+N)/2}} \,+ \cdots ~,
\end{equation}
So, we would expect the following large $k$ behavior:
\begin{equation}
\sum_{R\, \in \, \mathcal{I}_k} (\mathcal{S}_{0R})^{-s} \, \sim \, \beta_s\, k^{s(2N^2+N)/2} ~,
\label{betaSp2N}
\end{equation}
Thus, the equation \eqref{zetageneral} will give the following formula of the Witten zeta function for the $G=$Sp($2N$) group for $\text{Re}(s) \geq 2$:
\begin{equation}
\boxed{\zeta_{\text{Sp}(2N)}(s) = \frac{\pi^{sN^2}}{2} \left(\frac{2^{N(2N+1)/2}\, G(N+1)\, G(N+\frac{3}{2})}{\pi^{(N+1)/2}\,G(\frac{1}{2})} \right)^s \beta_s }
 ~.
\label{ConjSp2N}
\end{equation}
For arbitrary $s$, one can numerically obtain $\beta_s$ and use the above equation to calculate the numerical values of $\zeta_{\text{Sp}(2N)}(s)$ up to desired precision.

For $s=2n$, where $n \geq 1$ is an integer, the $\beta_{2n}$ can be analytically obtained, which turn out to be rational numbers. To do this, we first notice that:
\begin{equation}
\sum_{R\, \in \, \mathcal{I}_k} \frac{1}{(\mathcal{S}_{0R})^{2n}} = 
\mathcal{C}_{n,\,N}(k)  ~,
\label{powersofS0RCN}
\end{equation}
where $\mathcal{C}_{n,\,N}$ is a  polynomial of degree $nN(2N+1)$ in the variable $k$ with rational coefficients which has the following form: 
\begin{equation}
\mathcal{C}_{n,\,N}(k) = (k+1)(k+2)\ldots(k+N)(k+N+1)^{nN}(k+N+2)\ldots(k+2N+1)\, \sum_{i=0}^{2nN^2-2N} C_i\, k^i ~,
\label{}
\end{equation}
where $C_i \in \mathbb{Q}_{+}$ are positive rational coefficients. We denote the leading order coefficient (i.e., the coefficient of $k^{nN(2N+1)}$ term) of the polynomial $\mathcal{C}_{n,\,N}$ as $C_{\text{lead}}(\mathcal{C}_{n,\,N})$. As a result, the constant $\beta_{2n}$ will be a rational number and can be read off as: 
\begin{equation}
\beta_{2n}  = C_{\text{lead}}(\mathcal{C}_{n,\,N}) ~.
\end{equation}
Thus, the zeta function for Sp($2N$) group at positive even integers can be given as:
\begin{equation}
\boxed{\zeta_{\text{Sp}(2N)}(2n) = \pi^{2nN^2} \left( \frac{2^{nN(2N+1)}\, G(N+1)^{2n}\, G(N+\frac{3}{2})^{2n}}{2\,\pi^{n(N+1)}\,G(\frac{1}{2})^{2n}} \right)\, C_{\text{lead}}(\mathcal{C}_{n,\,N})} ~.
\label{zetaCN}
\end{equation}
We can obtain the leading order coefficients $C_{\text{lead}}(\mathcal{C}_{n,\,N})$ for any given $n$ and $N$ using the earlier techniques which will enable us to compute $\zeta_{\text{Sp}(2N)}(2n)$ via the \eqref{zetaCN}. We present the table \ref{ZetaCNTable} where we tabulate the values of $\zeta_{\text{Sp}(2N)}(2n)$ for some low values of $n$ and $N$.
\captionsetup{width=16cm}
\begin{longtable}{|c||L{15cm}|}
			\caption[Values of Witten zeta function at positive even integers for Sp($2N$) group]{Values of zeta functions at positive even integers for Sp($2N$) group calculated using \eqref{zetaCN}.}
			\label{ZetaCNTable} \\ \hline \rowcolor{Gray}
			$n$ & $\zeta_{\text{Sp}(2)}(2n)$ comes out to be the same as $\zeta_{\text{SU}(2)}(2n)$ given in Table \ref{LeadingSUNTable} \\ \hhline{|=#=|} \rowcolor{Gray}
   $n$ & $\zeta_{\text{Sp}(4)}(2n)$ comes out to be the same as $\zeta_{\text{SO}(5)}(2n)$ given in Table \ref{ZetaBNTable} \\ \hhline{|=#=|}
   \rowcolor{Gray}
			$n$ & $\zeta_{\text{Sp}(6)}(2n)/\pi^{18n}$ \\ \hline
   \rowcolor{Redd}
   $1$ & \footnotesize{$(19^1)/(2^4\cdot 3^5\cdot 5^1\cdot 7^3\cdot 11^1\cdot 13^1\cdot 17^1)$} \\ \hline \rowcolor{Bluee}
			$2$ & \footnotesize{$(104701^1\cdot 3140775089^1)/(2^8\cdot 3^{10}\cdot 5^4\cdot 7^6\cdot 11^3\cdot 13^3\cdot 17^2\cdot 19^2\cdot 23^1\cdot 29^1\cdot 31^1\cdot 37^1)$} \\ \hline \rowcolor{Redd}
			$3$ & \footnotesize{$(3774593^1\cdot 20951970345196831001^1)/(2^9\cdot 3^{15}\cdot 5^6\cdot 7^9\cdot 11^5\cdot 13^4\cdot 17^3\cdot 19^3\cdot 23^2\cdot 29^1\cdot 31^1\cdot 37^1\cdot 41^1\cdot 43^1\cdot 47^1\cdot 53^1)$} \\ \hline \rowcolor{Bluee}
			$4$ & \footnotesize{$(757^1\cdot 769^1\cdot 16651^1\cdot 2343331477562563285766267904404545351^1)/(2^{16}\cdot 3^{19}\cdot 5^9\cdot 7^{12}\cdot 11^6\cdot 13^6\cdot 17^4\cdot 19^4\cdot 23^3\cdot 29^2\cdot 31^2\cdot 37^2\cdot 41^1\cdot 43^1\cdot 47^1\cdot 53^1\cdot 59^1\cdot 61^1\cdot 67^1\cdot 71^1\cdot 73^1)$} \\ \hhline{|=#=|} \rowcolor{Gray}
			$n$ & $\zeta_{\text{Sp}(8)}(2n)/\pi^{32n}$ \\ \hline \rowcolor{Redd}
			$1$ & \footnotesize{$(839^1\cdot 3181^1)/(2^{11}\cdot 3^6\cdot 5^3\cdot 7^2\cdot 11^2\cdot 13^2\cdot 17^2\cdot 19^1\cdot 23^1\cdot 29^1\cdot 31^1)$} \\ \hline \rowcolor{Bluee}
			$2$ & \footnotesize{$(2856079^1\cdot 12055759^1\cdot 8640780643542223777^1)/(2^{19}\cdot 3^{16}\cdot 5^7\cdot 7^6\cdot 11^6\cdot 13^5\cdot 17^4\cdot 19^3\cdot 23^2\cdot 29^2\cdot 31^2\cdot 37^1\cdot 41^1\cdot 43^1\cdot 47^1\cdot 53^1\cdot 59^1\cdot 61^1)$}  \\ \hhline{|=#=|} \rowcolor{Gray}
			$n$ & $\zeta_{\text{Sp}(10)}(2n)/\pi^{50n}$ \\ \hline \rowcolor{Redd}
			$1$ & \footnotesize{$(5395475362754527^1)/(2^9\cdot 3^7\cdot 5^6\cdot 7^4\cdot 11^5\cdot 13^3\cdot 17^3\cdot 19^2\cdot 23^2\cdot 29^1\cdot 31^1\cdot 37^1\cdot 41^1\cdot 43^1\cdot 47^1) $} \\ \hline
\end{longtable}
\subsection{SO($2N$) group}
An irreducible representation for SO($2N$) group is denoted as $R = [a_1,a_2,\ldots,a_{N}]$, where $a_i$ are the Dynkin labels of the highest weight associated with the representation $R$. For a given level $k$, the set of integrable representations is given by:
\begin{equation}
\mathcal{I}_k =\begin{cases} \left\{[a_1, \ldots, a_{N}] \, : \, a_{\mu} \geq 0 \,\,\, ; \,\,\, a_1+2a_2+\ldots+2a_{N-2}+a_{N-1}+a_N \leq k \right\}  & ,\, \quad N \geq 4  \\[0.3cm] 
\left\{[a_1, a_2,a_3] \,:\, a_{\mu} \geq 0 \,\,\, ; \,\,\, a_1+a_2+a_3 \leq k \right\} & ,\, \quad N=3 \\[0.3cm] 
\left\{[a_1, a_2] \, : \, 0\leq a_{1} \leq k \,\,\, ; \,\,\, 0\leq a_{2} \leq k \right\}  & ,\, \quad N=2 \end{cases}
  ~.
\end{equation} 
The matrix element $\mathcal{S}_{0R}$ for an integrable representation $R$ is given as:
\begin{equation}
\mathcal{S}_{0R} = (-1)^{N(N-1)/2} \frac{2^{N-2}}{(k+2N-2)^{N/2}}\,(\text{det}\, M_1 + i^N \text{det}\, M_2) ~,
\label{S-matrix-DN}
\end{equation} 
where $M_1$ and $M_2$ are $N \times N$ matrices with elements given as,
\begin{equation}
(M_1)_{ij} = \cos\left(2\pi \frac{(N-i)(N+\ell_j-j)}{k+2N-2} \right) \quad;\quad (M_2)_{ij} = \sin\left(2\pi \frac{(N-i)(N+\ell_j-j)}{k+2N-2} \right)
\end{equation}
with $1 \leq i,j \leq N$. The variables $\ell_i$ are given as,
\begin{equation}
\ell_i = \begin{cases} (a_{N}+a_{N-1})/2 + \sum_{j=i}^{N-2} a_j &,\quad 1\leq i \leq N-2 \\ (a_{N}+a_{N-1})/2 &,\quad i=N-1 \\ (a_{N}-a_{N-1})/2 &,\quad i=N \end{cases}~.
\end{equation}
The large $k$ expansion of $\mathcal{S}_{00}$ is given as:
\begin{equation}
\mathcal{S}_{00} \sim \frac{2^{N (2N-3)}\, \pi^{N^2}\, G(N+1)\, G(N+\frac{1}{2})}{\pi^{3N/2}\,G(\frac{1}{2})}\, \frac{1}{k^{(2N^2-N)/2}} \,+ \cdots ~,
\end{equation}
Thus we would expect the following large $k$ behavior:
\begin{equation}
\sum_{R\, \in \, \mathcal{I}_k} (\mathcal{S}_{0R})^{-s} \, \sim \, \beta_s\, k^{s(2N^2-N)/2} ~.
\label{betaSO2N}
\end{equation}
Thus, the equation \eqref{zetageneral} will give the following formula of the Witten zeta function for the $G=$SO($2N$) group for $\text{Re}(s) \geq 2$:
\begin{equation}
\zeta_{\text{SO}(2N)}(s) = \frac{\pi^{s(N^2-N)}}{4} \left(\frac{2^{N(2N-3)}\, G(N+1)\, G(N+\frac{1}{2})}{\pi^{N/2}\,G(\frac{1}{2})} \right)^s \beta_s
 ~.
\label{ConjSO2N}
\end{equation}
For arbitrary $s$, one can numerically obtain $\beta_s$ and use the above equation to calculate the numerical values of $\zeta_{\text{SO}(2N)}(s)$ up to desired precision.
 
For $s= 2n$, where $n \geq 1$ is an integer, we can calculate the RHS of \eqref{ConjSO2N} by explicitly computing the constants $\beta_{2n}$ which turn out to be rational numbers. To do this, we first notice that: 
\begin{equation}
\sum_{R\, \in \, \mathcal{I}_k} \frac{1}{(\mathcal{S}_{0R})^{2n}} = 
\mathcal{D}_{n,\,N}(k)  ~,
\label{powersofS0RDN}
\end{equation}
where $\mathcal{D}_{n,\,N}$ is a  polynomial of degree $nN(2N-1)$ in the variable $k$ with rational coefficients. We denote the leading order coefficient (i.e. the coefficient of $k^{nN(2N-1)}$ term) of the polynomial $\mathcal{D}_{n,\,N}$ as $C_{\text{lead}}(\mathcal{D}_{n,\,N})$. As a result, the constant $\beta_{2n}$ will be a rational number and can be read off as: 
\begin{equation}
\beta_{2n}  = C_{\text{lead}}(\mathcal{D}_{n,\,N}) ~.
\end{equation}
Thus, the zeta function for SO($2N$) group at positive even integers can be given as:
\begin{equation}
\boxed{\zeta_{\text{SO}(2N)}(2n) = \pi^{2nN(N-1)} \left( \frac{2^{2nN(2N-3)}\, G(N+1)^{2n}\, G(N+\frac{1}{2})^{2n}}{4\pi^{nN}\,G(\frac{1}{2})^{2n}} \right)\, C_{\text{lead}}(\mathcal{D}_{n,\,N})} ~.
\label{zetaDN}
\end{equation}
We can obtain the leading order coefficients $C_{\text{lead}}(\mathcal{D}_{n,\,N})$ for any given $n$ and $N$ using the earlier techniques and can compute $\zeta_{\text{SO}(2N)}(2n)$ via the \eqref{zetaDN}. We present the table \ref{ZetaDNTable} where we tabulate the values of $\zeta_{\text{SO}(2N)}(2n)$ for some low values of $n$ and $N$.
\captionsetup{width=16cm}
\begin{longtable}{|c||L{15cm}|}
			\caption[Values of Witten zeta function at positive even integers for SO($2N$) group]{Values of zeta functions at positive even integers for SO($2N$) group calculated using \eqref{zetaDN}.}
			\label{ZetaDNTable} \\ \hline \rowcolor{Gray}
			$n$ & $\zeta_{\text{SO}(4)}(2n)$ comes out to be the same as $\left(\zeta_{\text{SU}(2)}(2n)\right)^2$ \\ \hhline{|=#=|} \rowcolor{Gray}
     \rowcolor{Gray}
			$n$ & $\zeta_{\text{SO}(6)}(2n)/\pi^{12n}$ \\ \hline
   \rowcolor{Redd}
   $1$ & \footnotesize{$(2^2\cdot 23^1)/(3^4\cdot 5^3\cdot 7^2\cdot 11^1\cdot 13^1)$} \\ \hline \rowcolor{Bluee}
			$2$ & \footnotesize{$(2^6\cdot 14081^1)/(3^8\cdot 5^6\cdot 7^2\cdot 11^2\cdot 13^2\cdot 17^1\cdot 19^1\cdot 23^1)$} \\ \hline \rowcolor{Redd}
			$3$ & \footnotesize{$(2^{11}\cdot 757409^1\cdot 23283173^1)/(3^{12}\cdot 5^9\cdot 7^6\cdot 11^3\cdot 13^3\cdot 17^2\cdot 19^2\cdot 23^1\cdot 29^1\cdot 31^1\cdot 37^1)$} \\ \hline \rowcolor{Bluee}
			$4$ & \footnotesize{$(2^{14}\cdot 1021^1\cdot 5529809^1\cdot 754075957^1)/(3^{14}\cdot 5^{12}\cdot 7^8\cdot 11^3\cdot 13^3\cdot 17^3\cdot 19^2\cdot 23^2\cdot 29^1\cdot 31^1\cdot 37^1\cdot 41^1\cdot 43^1\cdot 47^1)$} \\  \hline
\end{longtable}
\subsection{$G_2$ group}
An irreducible representation for $G_2$ group is denoted as $R = [a_1,a_2]$, where $a_i$ are the Dynkin labels of the highest weight associated with the representation $R$. For a given level $k$, the set of integrable representations is given by:
\begin{equation}
\mathcal{I}_k = \left\{[a_1, a_{2}] \,:\, a_1,a_2 \geq 0 \,\,\, ; \,\,\, 2a_1+a_2 \leq k \right\} ~.
\end{equation} 
The matrix element $\mathcal{S}_{0R}$ for an integrable representation $R$ is given as:
\begin{equation}
\mathcal{S}_{0R} = \frac{64 \sin \left(\frac{\pi  \left(a_1+1\right)}{k+4}\right) \sin \left(\frac{\pi  \left(a_2+1\right)}{3 (k+4)}\right) \sin \left(\frac{\pi  \left(a_1+a_2+2\right)}{k+4}\right) \sin \left(\frac{\pi  \left(2 a_1+a_2+3\right)}{k+4}\right) \sin \left(\frac{\pi  \left(3 a_1+a_2+4\right)}{3 (k+4)}\right) \sin \left(\frac{\pi  \left(3 a_1+2 a_2+5\right)}{3 (k+4)}\right)}{\sqrt{3}\, (k+4)} ~.
\label{S-matrix-G2}
\end{equation} 
The large $k$ expansion of $\mathcal{S}_{00}$ is given as:
\begin{equation}
\mathcal{S}_{00} \sim \left(\frac{2560\, \pi ^6}{9 \sqrt{3}}\right) \frac{1}{k^7} + \cdots ~.
\end{equation}
Thus we would expect the following large $k$ behavior:
\begin{equation}
\sum_{R\, \in \, \mathcal{I}_k} (\mathcal{S}_{0R})^{-s} \, \sim \, \beta_s\, k^{7s} ~.
\label{betaG2}
\end{equation}
Thus, the equation \eqref{zetageneral} will give the following formula of the Witten zeta function for the $G_2$ group for $\text{Re}(s) \geq 2$:
\begin{equation}
\boxed{\zeta_{G_2}(s) = \pi^{6s}\left(\frac{2560}{9 \sqrt{3}}\right)^s \beta_s }
 ~.
\label{ConjG2}
\end{equation}
For arbitrary $s$, one can numerically obtain $\beta_s$ and use the above equation to calculate the numerical values of $\zeta_{G_2}(s)$ up to desired precision.

For $s= 2n$, where $n \geq 1$ is an integer, we can calculate the RHS of \eqref{ConjG2} by explicitly computing the constants $\beta_{2n}$ which turn out to be rational numbers. To do this, we first notice that: 
\begin{equation}
\sum_{R\, \in \, \mathcal{I}_k} \frac{1}{(\mathcal{S}_{0R})^{2n}} = 
\begin{cases} 
 \mathcal{G}_n^{\text{even}}(k) &,\,\, \text{for $k=$ even}  \\[0.2cm]
\mathcal{G}_n^{\text{odd}}(k) &,\,\, \text{for $k=$ odd} 
\end{cases}  ~,
\label{powersofS0RG2}
\end{equation}
where $\mathcal{G}_n^{\text{even}}$ and $\mathcal{G}_n^{\text{odd}}$ are  polynomials of degree $14n$ in the variable $k$ with rational coefficients. The leading order coefficient (i.e. the coefficient of $k^{14n}$ term) is the same for both the polynomials. So we denote the leading order coefficient as $C_{\text{lead}}(\mathcal{G}_{n})$. As a result, the constant $\beta_{2n}$ will be a rational number and can be read off as: 
\begin{equation}
\beta_{2n}  = C_{\text{lead}}(\mathcal{G}_{n}) ~.
\end{equation}
Thus, the zeta function for $G_2$ group at positive even integers can be given as:
\begin{equation}
\boxed{\zeta_{G_2}(2n) = \pi^{12n} \left(\frac{6553600^n}{243^n}\right) \beta_{2n}} ~.
\label{zetaG2}
\end{equation}
We can obtain the leading order coefficients $C_{\text{lead}}(\mathcal{G}_n)$ for any given $n$ using the earlier techniques and can compute $\zeta_{G_2}(2n)$ via the \eqref{zetaG2}. We present the table \ref{zetaG2Table} where we tabulate the values of $\zeta_{G_2}(2n)$ for some low values of $n$.
\captionsetup{width=16cm}
\begin{longtable}{|c||L{15cm}|}
			\caption[Values of Witten zeta function at positive even integers for $G_2$ group]{Values of Witten zeta function at positive even integers for $G_2$ group computed using \eqref{zetaG2}.}
			\label{zetaG2Table} \\ \hline \rowcolor{Gray}
			$n$ & $\zeta_{G_2}(2n)/\pi^{12n}$ \\ \hhline{|=#=|} \rowcolor{Redd}
			$1$ & \footnotesize{$(2^2\cdot 5 \cdot 23 )/(3^{10}\cdot 7^2\cdot 11^1\cdot 13^1)$} \\ \hline\rowcolor{Bluee}
			$2$ & \footnotesize{$(2^4\cdot 8165653)/(3^{19}\cdot 5^1\cdot 7^4\cdot 11^2\cdot 13^2\cdot 17^1\cdot 23^1)$} \\\hline \rowcolor{Redd}
			$3$ & \footnotesize{$(2^7\cdot 55940539974690617)/(3^{30}\cdot 7^6\cdot 11^3\cdot 13^3\cdot 17^2\cdot 19^2\cdot 23^1\cdot 29^1\cdot 31^1\cdot 37^1)$} \\\hline \rowcolor{Bluee}
			$4$ & \footnotesize{$(2^8\cdot 55487 \cdot 853287535121366387947 )/(3^{38}\cdot 5^3\cdot 7^7\cdot 11^4\cdot 13^4\cdot 17^2\cdot 19^2\cdot 23^2\cdot 29^1\cdot 31^1\cdot 37^1\cdot 41^1\cdot 43^1\cdot 47^1)$} \\\hline \rowcolor{Redd}
			$5$ & \footnotesize{$(2^{13}\cdot 12225989 \cdot 222465917 \cdot 9587024790875491112749 )/(3^{49}\cdot 5^1\cdot 7^{10}\cdot 11^6\cdot 13^4\cdot 17^3\cdot 19^3\cdot 23^2\cdot 29^2\cdot 31^2\cdot 37^1\cdot 41^1\cdot 43^1\cdot 47^1\cdot 53^1\cdot 59^1\cdot 61^1)$} \\ \hline\rowcolor{Bluee}
			$6$ & \footnotesize{$(2^{13}\cdot 4057 \cdot 40591 \cdot 140423 \cdot 33320646926928967 \cdot 1068094980951542245514899 )/(3^{59}\cdot 5^5\cdot 7^{12}\cdot 11^6\cdot 13^6\cdot 17^4\cdot 19^4\cdot 23^3\cdot 29^2\cdot 31^2\cdot 37^2\cdot 41^1\cdot 43^1\cdot 47^1\cdot 53^1\cdot 59^1\cdot 61^1\cdot 67^1\cdot 71^1\cdot 73^1)$} \\\hline \rowcolor{Redd}
			$7$ & \footnotesize{$(2^{17}\cdot 147661499\cdot 72106219887426503049471001180353212291045796427220341)/(3^{69}\cdot 5^3\cdot 7^{14}\cdot 11^7\cdot 13^7\cdot 17^4\cdot 19^4\cdot 23^3\cdot 29^2\cdot 31^2\cdot 37^2\cdot 41^2\cdot 43^2\cdot 47^1\cdot 53^1\cdot 59^1\cdot 61^1\cdot 67^1\cdot 71^1\cdot 73^1\cdot 79^1\cdot 83^1)$}
			\\ \hline
\end{longtable}
\subsection{$F_4$ group}
An irreducible representation for $F_4$ group is denoted as $R = [a_1,a_2,a_3,a_4]$, where $a_i$ are the Dynkin labels of the highest weight associated with the representation $R$. For a given level $k$, the set of integrable representations is given by:
\begin{equation}
\mathcal{I}_k = \left\{[a_1, a_{2}] \,:\, a_1,a_2,a_3,a_4 \geq 0 \,\,\, ; \,\,\, 2a_1+3a_2+2a_3+a_4 \leq k \right\} ~.
\end{equation} 
The matrix element $\mathcal{S}_{0R}$ for an integrable representation $R$ is given as:
\begingroup
\allowdisplaybreaks
\begin{align}
\mathcal{S}_{0R} &=  \frac{8388608}{(k+9)^2} \sin\left[\frac{\pi  \left(a_1+1\right)}{k+9}\right] \sin\left[\frac{\pi\left(a_2+1\right)}{k+9}\right] \sin\left[\frac{\pi\left(a_3+1\right)}{2(k+9)}\right] \sin\left[\frac{\pi  \left(a_4+1\right)}{2(k+9)}\right] \sin\left[\frac{\pi  \left(a_0+8\right)}{k+9}\right] \nonumber  \\
& \times \sin\left[\frac{\pi\left(a_1+a_2+2\right)}{k+9}\right] \sin\left[\frac{\pi\left(a_2+a_3+2\right)}{k+9}\right] \sin\left[\frac{\pi  \left(a_3+a_4+2\right)}{2 (k+9)}\right] \nonumber \\
& \times \sin\left[\frac{\pi\left(a_1+a_2+a_3+3\right)}{k+9}\right] \sin \left[\frac{\pi\left(a_2+a_3+a_4+3\right)}{k+9}\right] \sin \left[\frac{\pi  \left(2 a_2+a_3+3\right)}{2 (k+9)}\right] \nonumber \\
& \times \sin\left[\frac{\pi\left(a_1+a_2+a_3+a_4+4\right)}{k+9}\right] \sin \left[\frac{\pi \left(a_1+2 a_2+a_3+4\right)}{k+9}\right] \sin\left[\frac{\pi  \left(2 a_2+a_3+a_4+4\right)}{2 (k+9)}\right] \nonumber \\
& \times  \sin\left[\frac{\pi  \left(a_0-a_1-a_2-a_3+5\right)}{k+9}\right] \sin \left[\frac{\pi  \left(a_0-a_2-a_3-a_4+5\right)}{2 (k+9)}\right] \sin \left[\frac{\pi  \left(a_0-2 a_1-a_2+5\right)}{2 (k+9)}\right] \nonumber \\
& \times \sin \left[\frac{\pi  \left(a_0-a_1-a_2+6\right)}{k+9}\right] \sin \left[\frac{\pi  \left(a_0-a_2-a_3+6\right)}{2 (k+9)}\right] \sin \left[\frac{\pi  \left(a_0-a_2+7\right)}{2 (k+9)}\right] \sin \left[\frac{\pi  \left(a_0-a_1+7\right)}{k+9}\right] \nonumber \\
& \times  \sin \left[\frac{\pi  \left(a_0+a_2+9\right)}{2 (k+9)}\right] \sin \left[\frac{\pi  \left(a_0+a_2+a_3+10\right)}{2 (k+9)}\right] \sin \left[\frac{\pi  \left(a_0+a_2+a_3+a_4+11\right)}{2 (k+9)}\right] ~,
\label{S-matrix-F4}
\end{align} 
\endgroup
where we have defined $a_0 = 2 a_1+3 a_2+2 a_3+a_4$. The large $k$ expansion of $\mathcal{S}_{00}$ is given as:
\begin{equation}
\mathcal{S}_{00} \sim \frac{49442161950720000 \,\pi ^{24}}{k^{26}} \,+\, \cdots ~.
\end{equation}
Thus, we would expect the following large $k$ behavior:
\begin{equation}
\sum_{R\, \in \, \mathcal{I}_k} (\mathcal{S}_{0R})^{-s} \, \sim \, \beta_s\, k^{26s} ~.
\label{betaF4}
\end{equation}
Thus, the equation \eqref{zetageneral} will give the following formula of the Witten zeta function for the $F_4$ group for $\text{Re}(s) \geq 2$: 
\begin{equation}
\boxed{\zeta_{F_4}(s) = \pi^{24s}\, 49442161950720000^s \beta_s }
 ~.
\label{ConjF4}
\end{equation}
For arbitrary $s$, one can numerically obtain $\beta_s$ and use the above equation to calculate the numerical values of $\zeta_{F_4}(s)$ up to desired precision.

For $s= 2n$, where $n \geq 1$ is an integer, we expect 
\begin{equation}
\sum_{R\, \in \, \mathcal{I}_k} \frac{1}{(\mathcal{S}_{0R})^{2n}} = 
\mathcal{F}_n(k) ~,
\label{powersofS0RF4}
\end{equation}
where $\mathcal{F}_n$ would be a polynomial of degree $52n$ in the variable $k$ with rational coefficients. Hence, the constant $\beta_{2n}$ will be a rational number which will be the leading order coefficient of $\mathcal{F}_n$: 
\begin{equation}
\beta_{2n}  = C_{\text{lead}}(\mathcal{F}_{n}) ~.
\end{equation} 
Thus, the zeta function at positive even integers can be given as:
\begin{equation}
\boxed{\zeta_{F_4}(2n) = \pi^{48n} \cdot 2^{52n}\cdot 3^{14n}\cdot 5^{8n}\cdot 7^{4n}\cdot 11^{2n} \cdot \beta_{2n}} ~.
\label{zetaF4}
\end{equation} 
We can obtain the leading order coefficients $C_{\text{lead}}(\mathcal{F}_n)$ for any given $n$ using the earlier techniques and can compute $\zeta_{F_4}(2n)$ via the \eqref{zetaF4}.
\subsection{$E_6$ group}
The set of integrable representations is given by:
\begin{equation}
\mathcal{I}_k = \left\{[a_1, a_{2},a_3,a_4,a_5,a_6] \,:\, a_i \geq 0 \,\,\, ; \,\,\, a_1 + 2 a_2 + 3 a_3 + 2 a_4 + a5 + 2 a_6 \leq k \right\} ~.
\end{equation} 
The matrix element $\mathcal{S}_{0R}$ for a representation $R = [a_1, a_{2},a_3,a_4,a_5,a_6]$ is given as, 
\begingroup
\allowdisplaybreaks
\begin{align}
\mathcal{S}_{0R} &= \text{const}\times \textsf{S}_{a_0+11}\,   \textsf{S}_{a_1+1}\,\textsf{S}_{a_2+1}\,\textsf{S}_{a_3+1}\,\textsf{S}_{a_4+1}\,\textsf{S}_{a_5+1}\,\textsf{S}_{a_6+1}\,\textsf{S}_{a_1+a_2+2}\,\textsf{S}_{a_2+a_3+2}\,\textsf{S}_{a_3+a_4+2}\,\textsf{S}_{a_4+a_5+2}\,\textsf{S}_{a_3+a_6+2}   \nonumber \\
& \times  \textsf{S}_{a_1+a_2+a_3+3}\,\textsf{S}_{a_2+a_3+a_4+3}\,\textsf{S}_{a_3+a_4+a_5+3}\,\textsf{S}_{a_2+a_3+a_6+3}\,\textsf{S}_{a_3+a_4+a_6+3}\,\textsf{S}_{a_1+a_2+a_3+a_6+4}\,\textsf{S}_{a_1+a_2+a_3+a_4+4}\nonumber \\
& \times \textsf{S}_{a_2+a_3+a_4+a_5+4}\,\textsf{S}_{a_2+a_3+a_4+a_6+4}\,\textsf{S}_{a_3+a_4+a_5+a_6+4}\,\textsf{S}_{a_1+a_2+a_3+a_4+a_5+5}\,\textsf{S}_{a_1+a_2+a_3+a_4+a_6+5}\,\textsf{S}_{a_2+2 a_3+a_4+a_6+5} \nonumber \\
& \times \textsf{S}_{a_2+a_3+a_4+a_5+a_6+5}\,\textsf{S}_{a_1+a_2+2 a_3+a_4+a_6+6}\,\textsf{S}_{a_1+a_2+a_3+a_4+a_5+a_6+6}\,\textsf{S}_{a_2+2 a_3+a_4+a_5+a_6+6}\,\textsf{S}_{a_0-a_3-a_4-a_5-a_6+7}  \nonumber \\
& \times \textsf{S}_{a_0-a_2-a_3-a_4-a_6+7}\,\textsf{S}_{a_0-a_1-a_2-a_3-a_6+7}\,\textsf{S}_{a_0-a_3-a_4-a_6+8}\,\textsf{S}_{a_0-a_2-a_3-a_6+8}\,\textsf{S}_{a_0-a_3-a_6+9}\,\textsf{S}_{a_0-a_6+10}  ~,
\label{S-matrix-E6}
\end{align} 
\endgroup
where we have defined $a_0 = a_1+2 a_2+3 a_3+2 a_4+a_5+2 a_6$ and
\begin{equation}
 \text{const}= \frac{68719476736}{\sqrt{3}\, (k+12)^3} \quad;\quad  \textsf{S}_x = \sin\left[\frac{\pi\,  x}{k+12}\right] ~.
\end{equation}
The large $k$ expansion of $\mathcal{S}_{00}$ is given as:
\begin{equation}
\mathcal{S}_{00} \sim \frac{535128220927132468405862400000 \sqrt{3}\, \pi ^{36}}{k^{39}} \,+\, \cdots ~.
\end{equation}
Thus, we would expect the following large $k$ behavior:
\begin{equation}
\sum_{R\, \in \, \mathcal{I}_k} (\mathcal{S}_{0R})^{-s} \, \sim \, \beta_s\, k^{39s} ~.
\label{betaE6}
\end{equation}
Thus, the equation \eqref{zetageneral} will give the following formula of the Witten zeta function for the $E_6$ group for $\text{Re}(s) \geq 2$: 
\begin{equation}
\boxed{ \zeta_{E_6}(s) = \frac{\pi^{36s}}{3}\times 535128220927132468405862400000^s \times 3^{s/2} \times \beta_s }
 ~.
\label{ConjE6}
\end{equation}
For arbitrary $s$, one can numerically obtain $\beta_s$ and use the above equation to calculate the numerical values of $\zeta_{E_6}(s)$ up to desired precision.

For $s=2n$, where $n \geq 1$ is an integer, the $\beta_{2n}$ can be analytically obtained, which turn out to be rational numbers. To do this, we first notice that:  
\begin{equation}
\sum_{R\, \in \, \mathcal{I}_k} \frac{1}{(\mathcal{S}_{0R})^{2n}} = 
\mathcal{E}_{n,6}(k) ~,
\label{powersofS0RE6}
\end{equation}
where $\mathcal{E}_{n,6}$ would be a polynomial of degree $78n$ in the variable $k$ with rational coefficients. Hence, the constant $\beta_{2n}$ will be a rational number which will be the leading order coefficient of $\mathcal{E}_{n,6}$: 
\begin{equation}
\beta_{2n}  = C_{\text{lead}}(\mathcal{E}_{n,6}) ~.
\end{equation} 
Thus, the zeta function at positive even integers can be given as:
\begin{equation}
\boxed{\zeta_{E_6}(2n) = \pi^{72n} \times 2^{122n} \times 3^{19n-1}\times 5^{10n}\times 7^{6n}\times 11^{2n} \times \beta_{2n}} ~.
\label{zetaE6}
\end{equation} 
We can obtain the leading order coefficients $C_{\text{lead}}(\mathcal{E}_{n,6})$ for any given $n$ using the earlier techniques and can compute $\zeta_{E_6}(2n)$ via the \eqref{zetaE6}.
\subsection{$E_7$ group}
The set of integrable representations is given by:
\begin{equation}
\mathcal{I}_k = \left\{[a_1, a_{2},a_3,a_4,a_5,a_6,a_7] \,:\, a_i \geq 0 \,\,\, ; \,\,\, 2 a_1+3 a_2+4 a_3+3 a_4+2 a_5+a_6+2 a_7 \leq k \right\} ~.
\end{equation}  
The matrix element $\mathcal{S}_{0R}$ for a representation $R = [a_1, a_{2},a_3,a_4,a_5,a_6,a_7]$ is given as, 
\begingroup
\allowdisplaybreaks
\begin{align}
\mathcal{S}_{0R} &= \text{const}\times \textsf{S}_{a_0+17}\, \textsf{S}_{a_0-a_1+16}\, \textsf{S}_{a_1+1}\, \textsf{S}_{a_0-a_1-a_2+15}\, \textsf{S}_{a_2+1}\, \textsf{S}_{a_1+a_2+2}\, \textsf{S}_{a_0-a_1-a_2-a_3+14}\, \textsf{S}_{a_3+1}\, \textsf{S}_{a_2+a_3+2}\, \textsf{S}_{a_1+a_2+a_3+3} \nonumber \\
&\times \textsf{S}_{a_0-a_1-a_2-a_3-a_4+13}\, \textsf{S}_{a_4+1}\, \textsf{S}_{a_3+a_4+2}\, \textsf{S}_{a_2+a_3+a_4+3}\, \textsf{S}_{a_1+a_2+a_3+a_4+4}\, \textsf{S}_{a_0-a_1-a_2-a_3-a_4-a_5+12}\, \textsf{S}_{a_5+1}\, \textsf{S}_{a_4+a_5+2}\nonumber \\
& \times \textsf{S}_{a_3+a_4+a_5+3}\, \textsf{S}_{a_2+a_3+a_4+a_5+4}\, \textsf{S}_{a_1+a_2+a_3+a_4+a_5+5}\, \textsf{S}_{a_6+1}\, \textsf{S}_{a_5+a_6+2}\, \textsf{S}_{a_4+a_5+a_6+3}\, \textsf{S}_{a_3+a_4+a_5+a_6+4} \, \textsf{S}_{a_7+1}\, \textsf{S}_{a_3+a_7+2}\nonumber \\
& \times \textsf{S}_{a_2+a_3+a_4+a_5+a_6+5}\, \textsf{S}_{a_1+a_2+a_3+a_4+a_5+a_6+6} \textsf{S}_{a_0-a_1-a_2-a_3-a_7+13}\, \textsf{S}_{a_0-a_1-2 a_2-2 a_3-a_4-a_7+10}\, \textsf{S}_{a_1+a_2+a_3+a_7+4} \nonumber \\
& \times \textsf{S}_{a_0-a_1-a_2-2 a_3-a_4-a_7+11}\,\textsf{S}_{a_0-a_1-a_2-a_3-a_4-a_7+12}\, \, \textsf{S}_{a_2+a_3+a_7+3}\, \, \textsf{S}_{a_3+a_4+a_7+3}\, \textsf{S}_{a_2+a_3+a_4+a_7+4}\, \textsf{S}_{a_1+a_2+a_3+a_4+a_7+5} \nonumber\\
& \times \textsf{S}_{a_2+2 a_3+a_4+a_7+5}\, \textsf{S}_{a_1+a_2+2 a_3+a_4+a_7+6}\, \textsf{S}_{a_1+2 a_2+2 a_3+a_4+a_7+7}\, \textsf{S}_{a_3+a_4+a_5+a_7+4}\,\textsf{S}_{a_2+a_3+a_4+a_5+a_7+5} \nonumber\\
& \times \textsf{S}_{a_1+a_2+a_3+a_4+a_5+a_7+6}\, \textsf{S}_{a_2+2 a_3+a_4+a_5+a_7+6}\, \textsf{S}_{a_1+a_2+2 a_3+a_4+a_5+a_7+7}\, \textsf{S}_{a_1+2 a_2+2 a_3+a_4+a_5+a_7+8}\nonumber\\
& \times \textsf{S}_{a_2+2 a_3+2 a_4+a_5+a_7+7}\,\textsf{S}_{a_1+a_2+2 a_3+2 a_4+a_5+a_7+8}\, \textsf{S}_{a_1+2 a_2+2 a_3+2 a_4+a_5+a_7+9}\, \textsf{S}_{a_1+2 a_2+3 a_3+2 a_4+a_5+a_7+10} \nonumber\\
& \times \textsf{S}_{a_3+a_4+a_5+a_6+a_7+5}\, \textsf{S}_{a_2+a_3+a_4+a_5+a_6+a_7+6}\, \textsf{S}_{a_1+a_2+a_3+a_4+a_5+a_6+a_7+7}\, \textsf{S}_{a_2+2 a_3+a_4+a_5+a_6+a_7+7} \nonumber \\
&\times \textsf{S}_{a_1+a_2+2 a_3+a_4+a_5+a_6+a_7+8}\, \textsf{S}_{a_1+2 a_2+2 a_3+a_4+a_5+a_6+a_7+9}\, \textsf{S}_{a_2+2 a_3+2 a_4+a_5+a_6+a_7+8}\, \textsf{S}_{a_1+a_2+2 a_3+2 a_4+a_5+a_6+a_7+9} \nonumber \\
&\times\textsf{S}_{a_1+2 a_2+2 a_3+2 a_4+a_5+a_6+a_7+10} \,\textsf{S}_{a_1+2 a_2+3 a_3+2 a_4+a_5+a_6+a_7+11}\, \textsf{S}_{a_2+2 a_3+2 a_4+2 a_5+a_6+a_7+9} \nonumber \\ &\times \textsf{S}_{a_1+2 a_2+3 a_3+2 a_4+a_5+2 a_7+11}
\label{S-matrix-E7}
\end{align} 
\endgroup
where we have defined $a_0 = 2 a_1+3 a_2+4 a_3+3 a_4+2 a_5+a_6+2 a_7$ and
\begin{equation}
 \text{const}= \frac{4611686018427387904 \sqrt{2}}{(k+18)^{7/2}} \quad;\quad \textsf{S}_x = \sin\left[\frac{\pi\,  x}{k+18}\right] ~.
\end{equation}
The large $k$ expansion of $\mathcal{S}_{00}$ is given as:
\begin{equation}
\mathcal{S}_{00} \sim \frac{\sqrt{2} \times \pi^{63} \times  2^{109} \times 3^{22} \times 5^{10} \times 7^6 \times 11^3\times 13^2\times 17^1}{k^{133/2}} \,+\, \cdots ~.
\end{equation}
Thus, we would expect the following large $k$ behavior:
\begin{equation}
\sum_{R\, \in \, \mathcal{I}_k} (\mathcal{S}_{0R})^{-s} \, \sim \, \beta_s\, k^{133s/2} ~.
\label{betaE7}
\end{equation}
Thus, the equation \eqref{zetageneral} will give the following formula of the Witten zeta function for the $E_7$ group for $\text{Re}(s) \geq 2$: 
\begin{equation}
\boxed{ \zeta_{E_7}(s) = \frac{\pi^{63s}}{2}\times 2^{219s/2} \times 3^{22s} \times 5^{10s} \times 7^{6s} \times 11^{3s} \times 13^{2s} \times 17^s \times \beta_s }
 ~.
\label{ConjE7}
\end{equation}
For arbitrary $s$, one can numerically obtain $\beta_s$ and use the above equation to calculate the numerical values of $\zeta_{E_7}(s)$ up to desired precision.

For $s= 2n$, where $n \geq 1$ is an integer, we expect 
\begin{equation}
\sum_{R\, \in \, \mathcal{I}_k} \frac{1}{(\mathcal{S}_{0R})^{2n}} = 
\mathcal{E}_{n,7}(k) ~,
\label{powersofS0RE7}
\end{equation}
where $\mathcal{E}_{n,7}$ would be a polynomial of degree $133n$ in the variable $k$ with rational coefficients. Hence, the constant $\beta_{2n}$ will be a rational number which will be the leading order coefficient of $\mathcal{E}_{n,7}$: 
\begin{equation}
\beta_{2n}  = C_{\text{lead}}(\mathcal{E}_{n,7}) ~.
\end{equation} 
Thus, the zeta function at positive even integers can be given as:
\begin{equation}
\boxed{\zeta_{E_7}(2n) = \pi^{126n} \times 2^{219n-1} \times 3^{44n} \times 5^{20n} \times 7^{12n} \times 11^{6n} \times 13^{4n} \times 17^{2n} \times \beta_{2n}} ~.
\label{zetaE7}
\end{equation} 
We can obtain the leading order coefficients $C_{\text{lead}}(\mathcal{E}_{n,7})$ for any given $n$ using the earlier techniques and can compute $\zeta_{E_7}(2n)$ via the \eqref{zetaE7}.
\subsection{$E_8$ group}
The set of integrable representations is given by:
\begin{equation}
\mathcal{I}_k = \left\{[a_1, a_{2},a_3,a_4,a_5,a_6,a_7,a_8] \,:\, a_i \geq 0 \,\,\, ; \,\,\, 2 a_1+3 a_2+4 a_3+5 a_4+6 a_5+4 a_6+2 a_7+3 a_8 \leq k \right\} ~.
\end{equation}  
The matrix element $\mathcal{S}_{0R}$ for a representation $R = [a_1, a_{2},a_3,a_4,a_5,a_6,a_7,a_8]$ is given as, 
\begingroup
\allowdisplaybreaks
\begin{align}
\mathcal{S}_{0R} &= \text{const}\times \textsf{S}_1\, \textsf{S}_{a_1+1}\, \textsf{S}_{a_2+1} \, \textsf{S}_{a_1+a_2+2} \, \textsf{S}_{a_3+1} \, \textsf{S}_{a_2+a_3+2} \, \textsf{S}_{a_1+a_2+a_3+3} \, \textsf{S}_{a_4+1} \, \textsf{S}_{a_3+a_4+2} \, \textsf{S}_{a_2+a_3+a_4+3} \, \textsf{S}_{a_1+a_2+a_3+a_4+4} \nonumber \\ 
& \times \textsf{S}_{a_5+1} \, \textsf{S}_{a_4+a_5+2} \, \textsf{S}_{a_3+a_4+a_5+3} \, \textsf{S}_{a_2+a_3+a_4+a_5+4} \, \textsf{S}_{a_1+a_2+a_3+a_4+a_5+5}\, \textsf{S}_{a_0+a_2+2 a_3-2 a_5-a_6+29} \, \textsf{S}_{a_6+1} \, \textsf{S}_{a_5+a_6+2} \nonumber \\ 
& \times \textsf{S}_{a_4+a_5+a_6+3} \, \textsf{S}_{a_3+a_4+a_5+a_6+4} \, \textsf{S}_{a_2+a_3+a_4+a_5+a_6+5} \, \textsf{S}_{a_1+a_2+a_3+a_4+a_5+a_6+6} \, \textsf{S}_{a_0+a_2+2 a_3-3 a_5-2 a_6-a_7+26}\,\textsf{S}_{a_7+1} \nonumber \\ 
& \times \textsf{S}_{a_0+a_3-a_4-3 a_5-2 a_6-a_7+23} \, \textsf{S}_{a_0-a_1+a_3-a_4-3 a_5-2 a_6-a_7+22} \, \textsf{S}_{a_0+a_2+a_3-a_4-3 a_5-2 a_6-a_7+24} \, \textsf{S}_{a_4+a_5+a_6+a_7+4} \nonumber \\ 
& \times \textsf{S}_{a_0+a_2+2 a_3-a_4-3 a_5-2 a_6-a_7+25} \, \textsf{S}_{a_0+a_2+2 a_3-2 a_5-2 a_6-a_7+27} \, \textsf{S}_{a_0+a_2+2 a_3-2 a_5-a_6-a_7+28} \, \textsf{S}_{a_6+a_7+2} \, \textsf{S}_{a_5+a_6+a_7+3}  \nonumber \\ 
& \times \textsf{S}_{a_3+a_4+a_5+a_6+a_7+5} \, \textsf{S}_{a_2+a_3+a_4+a_5+a_6+a_7+6} \, \textsf{S}_{a_1+a_2+a_3+a_4+a_5+a_6+a_7+7} \, \textsf{S}_{a_0-2 a_4-4 a_5-3 a_6-2 a_7-a_8+17} \nonumber \\ 
& \times \textsf{S}_{a_0-2 a_4-4 a_5-3 a_6-a_7-a_8+18} \, \textsf{S}_{a_0-a_1-2 a_4-4 a_5-3 a_6-a_7-a_8+17} \, \textsf{S}_{a_0-2 a_4-4 a_5-2 a_6-a_7-a_8+19} \, \textsf{S}_{\left(a_8+1\right) \left(a_3+a_8+2\right)} \nonumber \\ 
& \times \textsf{S}_{a_0-a_1-2 a_4-4 a_5-2 a_6-a_7-a_8+18} \, \textsf{S}_{a_0-2 a_4-3 a_5-2 a_6-a_7-a_8+20} \, \textsf{S}_{a_0-a_1-2 a_4-3 a_5-2 a_6-a_7-a_8+19} \, \textsf{S}_{a_2+a_3+a_8+3} \, \textsf{S}_{a_3+a_4+a_8+3} \nonumber \\ 
& \times \textsf{S}_{a_0-a_1-a_2-2 a_4-3 a_5-2 a_6-a_7-a_8+18} \, \textsf{S}_{a_0-a_4-3 a_5-2 a_6-a_7-a_8+21} \, \textsf{S}_{a_0-a_1-a_4-3 a_5-2 a_6-a_7-a_8+20} \nonumber \\ 
& \times \textsf{S}_{a_0-a_1-a_2-a_4-3 a_5-2 a_6-a_7-a_8+19} \, \textsf{S}_{a_0+a_3-a_4-3 a_5-2 a_6-a_7-a_8+22} \, \textsf{S}_{a_0-a_1+a_3-a_4-3 a_5-2 a_6-a_7-a_8+21} \nonumber \\ 
& \times \textsf{S}_{a_0+a_2+a_3-a_4-3 a_5-2 a_6-a_7-a_8+23}   \, \textsf{S}_{a_1+a_2+a_3+a_8+4}  \, \textsf{S}_{a_2+a_3+a_4+a_8+4} \, \textsf{S}_{a_1+a_2+a_3+a_4+a_8+5} \, \textsf{S}_{a_2+2 a_3+a_4+a_8+5} \nonumber \\ 
& \times \textsf{S}_{a_1+a_2+2a_3+a_4+a_8+6} \, \textsf{S}_{a_1+2 a_2+2 a_3+a_4+a_8+7} \, \textsf{S}_{a_3+a_4+a_5+a_8+4} \, \textsf{S}_{a_2+a_3+a_4+a_5+a_8+5} \, \textsf{S}_{a_1+a_2+a_3+a_4+a_5+a_8+6} \nonumber \\ 
& \times \textsf{S}_{a_2+2 a_3+a_4+a_5+a_8+6} \, \textsf{S}_{a_1+a_2+2 a_3+a_4+a_5+a_8+7} \, \textsf{S}_{a_1+2 a_2+2 a_3+a_4+a_5+a_8+8} \, \textsf{S}_{a_2+2 a_3+2 a_4+a_5+a_8+7} \nonumber \\ 
& \times \textsf{S}_{a_1+a_2+2 a_3+2 a_4+a_5+a_8+8} \, \textsf{S}_{a_1+2 a_2+2 a_3+2 a_4+a_5+a_8+9} \, \textsf{S}_{a_1+2 a_2+3 a_3+2 a_4+a_5+a_8+10} \, \textsf{S}_{a_3+a_4+a_5+a_6+a_8+5} \nonumber \\ 
& \times \textsf{S}_{a_2+a_3+a_4+a_5+a_6+a_8+6} \, \textsf{S}_{a_1+a_2+a_3+a_4+a_5+a_6+a_8+7} \, \textsf{S}_{a_2+2 a_3+a_4+a_5+a_6+a_8+7} \, \textsf{S}_{a_1+a_2+2 a_3+a_4+a_5+a_6+a_8+8} \nonumber \\ 
& \times \textsf{S}_{a_1+2 a_2+2 a_3+a_4+a_5+a_6+a_8+9} \, \textsf{S}_{a_2+2 a_3+2 a_4+a_5+a_6+a_8+8} \, \textsf{S}_{a_1+a_2+2 a_3+2 a_4+a_5+a_6+a_8+9} \, \textsf{S}_{a_1+2 a_2+2 a_3+2 a_4+a_5+a_6+a_8+10} \nonumber \\ 
& \times \textsf{S}_{a_1+2 a_2+3 a_3+2 a_4+a_5+a_6+a_8+11} \, \textsf{S}_{a_2+2 a_3+2 a_4+2 a_5+a_6+a_8+9} \, \textsf{S}_{a_1+a_2+2 a_3+2 a_4+2 a_5+a_6+a_8+10} \nonumber \\ 
& \times \textsf{S}_{a_1+2 a_2+2 a_3+2 a_4+2 a_5+a_6+a_8+11} \, \textsf{S}_{a_1+2 a_2+3 a_3+2 a_4+2 a_5+a_6+a_8+12} \, \textsf{S}_{a_1+2 a_2+3 a_3+3 a_4+2 a_5+a_6+a_8+13}\nonumber \\ 
& \times \textsf{S}_{a_3+a_4+a_5+a_6+a_7+a_8+6} \, \textsf{S}_{a_2+a_3+a_4+a_5+a_6+a_7+a_8+7} \, \textsf{S}_{a_1+a_2+a_3+a_4+a_5+a_6+a_7+a_8+8} \, \textsf{S}_{a_2+2 a_3+a_4+a_5+a_6+a_7+a_8+8} \nonumber \\ 
& \times \textsf{S}_{a_1+a_2+2 a_3+a_4+a_5+a_6+a_7+a_8+9} \, \textsf{S}_{a_1+2 a_2+2 a_3+a_4+a_5+a_6+a_7+a_8+10} \, \textsf{S}_{a_2+2 a_3+2 a_4+a_5+a_6+a_7+a_8+9} \nonumber \\ 
& \times \textsf{S}_{a_1+a_2+2 a_3+2 a_4+a_5+a_6+a_7+a_8+10} \, \textsf{S}_{a_1+2 a_2+2 a_3+2 a_4+a_5+a_6+a_7+a_8+11} \, \textsf{S}_{a_1+2 a_2+3 a_3+2 a_4+a_5+a_6+a_7+a_8+12} \nonumber \\ 
& \times \textsf{S}_{a_2+2 a_3+2 a_4+2 a_5+a_6+a_7+a_8+10} \, \textsf{S}_{a_1+a_2+2 a_3+2 a_4+2 a_5+a_6+a_7+a_8+11} \, \textsf{S}_{a_1+2 a_2+2 a_3+2 a_4+2 a_5+a_6+a_7+a_8+12} \nonumber \\ 
& \times \textsf{S}_{a_1+2 a_2+3 a_3+2 a_4+2 a_5+a_6+a_7+a_8+13} \, \textsf{S}_{a_1+2 a_2+3 a_3+3 a_4+2 a_5+a_6+a_7+a_8+14} \, \textsf{S}_{a_2+2 a_3+2 a_4+2 a_5+2 a_6+a_7+a_8+11} \nonumber \\ 
& \times \textsf{S}_{a_1+a_2+2 a_3+2 a_4+2 a_5+2 a_6+a_7+a_8+12} \, \textsf{S}_{a_1+2 a_2+2 a_3+2 a_4+2 a_5+2 a_6+a_7+a_8+13} \, \textsf{S}_{a_1+2 a_2+3 a_3+2 a_4+2 a_5+2 a_6+a_7+a_8+14} \nonumber \\ 
& \times \textsf{S}_{a_1+2 a_2+3 a_3+3 a_4+2 a_5+2 a_6+a_7+a_8+15} \, \textsf{S}_{a_1+2 a_2+3 a_3+3 a_4+3 a_5+2 a_6+a_7+a_8+16} \, \textsf{S}_{a_1+2 a_2+3 a_3+2 a_4+a_5+2 a_8+11} \nonumber \\ 
& \times \textsf{S}_{a_1+2 a_2+3 a_3+2 a_4+a_5+a_6+2 a_8+12} \, \textsf{S}_{a_1+2 a_2+3 a_3+2 a_4+2 a_5+a_6+2 a_8+13} \, \textsf{S}_{a_1+2 a_2+3 a_3+3 a_4+2 a_5+a_6+2 a_8+14} \nonumber \\ 
& \times \textsf{S}_{a_1+2 a_2+4 a_3+3 a_4+2 a_5+a_6+2 a_8+15} \, \textsf{S}_{a_1+3 a_2+4 a_3+3 a_4+2 a_5+a_6+2 a_8+16} \, \textsf{S}_{a_1+2 a_2+3 a_3+2 a_4+a_5+a_6+a_7+2 a_8+13} \nonumber \\ 
& \times \textsf{S}_{a_1+2 a_2+3 a_3+2 a_4+2 a_5+a_6+a_7+2 a_8+14} \, \textsf{S}_{a_1+2 a_2+3 a_3+3 a_4+2 a_5+a_6+a_7+2 a_8+15} \, \textsf{S}_{a_1+2 a_2+4 a_3+3 a_4+2 a_5+a_6+a_7+2 a_8+16} \nonumber \\ 
& \times \textsf{S}_{a_1+2 a_2+3 a_3+2 a_4+2 a_5+2 a_6+a_7+2 a_8+15} \, \textsf{S}_{a_1+2 a_2+3 a_3+3 a_4+2 a_5+2 a_6+a_7+2 a_8+16} \nonumber \\ 
& \times \textsf{S}_{a_1+2 a_2+4 a_3+3 a_4+2 a_5+2 a_6+a_7+2 a_8+17} \, \textsf{S}_{a_1+2 a_2+3 a_3+3 a_4+3 a_5+2 a_6+a_7+2 a_8+17}
\label{S-matrix-E8}
\end{align} 
\endgroup
where we have defined $a_0 = 2 a_1+3 a_2+4 a_3+5 a_4+6 a_5+4 a_6+2 a_7+3 a_8$ and
\begin{equation}
 \text{const}= \frac{1329227995784915872903807060280344576}{(k+30)^4} \quad;\quad \textsf{S}_x = \sin\left[\frac{\pi\,  x}{k+30}\right] ~.
\end{equation}
The large $k$ expansion of $\mathcal{S}_{00}$ is given as:
\begin{equation}
\mathcal{S}_{00} \sim \frac{\pi^{120} \times 2^{217} \times 3^{47} \times 5^{21} \times 7^{14} \times 11^8 \times 13^6 \times 17^4 \times 19^3 \times 23^2 \times 29^1}{k^{124}} \,+\, \cdots ~.
\end{equation}
Thus, we would expect the following large $k$ behavior:
\begin{equation}
\sum_{R\, \in \, \mathcal{I}_k} (\mathcal{S}_{0R})^{-s} \, \sim \, \beta_s\, k^{124s} ~.
\label{betaE8}
\end{equation}
Thus, the equation \eqref{zetageneral} will give the following formula of the Witten zeta function for the $E_8$ group for $\text{Re}(s) \geq 2$:
\begin{equation}
\boxed{ \zeta_{E_8}(s) = \pi^{120s} \times 2^{217s} \times 3^{47s} \times 5^{21s} \times 7^{14s} \times 11^{8s} \times 13^{6s} \times 17^{4s} \times 19^{3s} \times 23^{2s} \times 29^s \times \beta_s }
 ~.
\label{ConjE8}
\end{equation}
For arbitrary $s$, one can numerically obtain $\beta_s$ and use the above equation to calculate the numerical values of $\zeta_{E_8}(s)$ up to desired precision.

For $s= 2n$, where $n \geq 1$ is an integer, we expect 
\begin{equation}
\sum_{R\, \in \, \mathcal{I}_k} \frac{1}{(\mathcal{S}_{0R})^{2n}} = 
\mathcal{E}_{n,8}(k) ~,
\label{powersofS0RE8}
\end{equation}
where $\mathcal{E}_{n,8}$ would be a polynomial of degree $248n$ in the variable $k$ with rational coefficients. Hence, the constant $\beta_{2n}$ will be a rational number which will be the leading order coefficient of $\mathcal{E}_{n,8}$: 
\begin{equation}
\beta_{2n}  = C_{\text{lead}}(\mathcal{E}_{n,8}) ~.
\end{equation} 
Thus, the zeta function at positive even integers can be given as:
\begin{equation}
\boxed{\zeta_{E_8}(2n) = \pi^{240n} \times 2^{434n} \times 3^{94n} \times 5^{42n} \times 7^{28n} \times 11^{16n} \times 13^{12n} \times 17^{8n} \times 19^{6n} \times 23^{4n} \times 29^{2n} \times \beta_{2n}}
\label{zetaE8}
\end{equation} 
We can obtain the leading order coefficients $C_{\text{lead}}(\mathcal{E}_{n,8})$ for any given $n$ using the earlier techniques and can compute $\zeta_{E_8}(2n)$ via the \eqref{zetaE8}.
\bibliographystyle{JHEP}
\bibliography{PRDbiblio}

\end{document}